\documentclass[multphys,vecphys]{svmult}

\usepackage{makeidx}     
\usepackage{graphicx}    
\usepackage{multicol}    

\makeindex             

\def\gappeq{\mathrel{\rlap {\raise.5ex\hbox{$>$}}
{\lower.5ex\hbox{$\sim$}}}}

\def\lappeq{\mathrel{\rlap{\raise.5ex\hbox{$<$}}
{\lower.5ex\hbox{$\sim$}}}}

\def\eg{{\em e.g.}}
\def\s{{\,\rm s}}

\def\H{H\hskip-8.5pt/\hskip2pt}

\def\beq{\begin{equation}}
\def\eeq{\end{equation}}

\def\I#1{{\rm Im}\,#1}
\def\Tr{{\rm Tr}\,}
\catcode`\@=11 
\def\coeff#1#2{{\textstyle{#1\over #2}}}

\def\VEV#1{\left\langle #1\right\rangle}

\def\lsim{\mathrel{\mathpalette\@versim<}}
\def\gsim{\mathrel{\mathpalette\@versim>}}
\def\@versim#1#2{\vcenter{\offinterlineskip
    \ialign{$\m@th#1\hfil##\hfil$\crcr#2\crcr\sim\crcr } }}

\def\t1{{\tilde 1}}

\def\GeV{\,{\rm GeV}}

\def\to{\rightarrow}

\def\gappeq{\mathrel{\rlap {\raise.5ex\hbox{$>$}}
{\lower.5ex\hbox{$\sim$}}}}

\def\lappeq{\mathrel{\rlap{\raise.5ex\hbox{$<$}}
{\lower.5ex\hbox{$\sim$}}}}

\begin{document}

\title*{CPT Violation and Decoherence in Quantum Gravity}
\titlerunning{CPT Violation} 
\author{Nick E. Mavromatos}
\authorrunning{N. Mavromatos} 
\institute{King$'$s College London, Department of Physics, Strand, London WC2R 2LS, U.K.
\texttt{Nikolaos.Mavromatos@kcl.ac.uk}}
%
%
\maketitle

In these lectures I review, in as much
pedagogical way as possible, various theoretical
ideas and motivation for violation of CPT invariance 
in some models of Quantum Gravity, 
and discuss the relevant phenomenology. 
Since the subject is vast, I pay particular emphasis
on the CPT Violating decoherence scenario for quantum gravity, 
due to space-time foam. In my opinion this seems 
to be the most likely scenario  to be realised in Nature, 
should quantum gravity 
be responsible for the violation of this symmetry. 
In this context, I also discuss how the CPT Violating decoherence
scenario can explain experimental ``anomalies'' in neutrino data,
such as LSND results, in agreement with the rest of the presently available
data, without enlarging the neutrino sector. 

\section{Introduction and Summary}

Next year, Special Relativity celebrates a century of enormous success, 
having passed many stringent experimental tests, 
in both its classical 
and quantum versions 
(relativistic quantum field theories in flat space times).
Unfortunately, the same is not true for its curved-space counterpart,
General Relativity. 
A consistently quantized theory of gravity, that is  a dynamical 
theory of curved 
geometries themselves, still remains a mystery.
Despite the enormous effort invested 
for this purpose on behalf of the scientic community 
over the past ninety years, Quantum Gravity\index{quantum gravity} 
is still far from being understood as a physical theory.

Of course, 
elegant and mathematically consistent models, 
such as string\index{string} or, better, brane\index{brane} 
theory\cite{strings}, 
have been developed to a great detail 
from a mathematical viewpoint. Nevertheless there are 
still many fundamental 
issues and questions which remain unresolved. 
For instance, the complete process of evaporation of a 
black hole\index{black hole}, 
or the inverse process of collapsing matter to form a Black Hole, 
are not completely understood in string theory. 
The counting of microstates and verification of the Hawking-Bekenstein
entropy/area law have been understood mathematically 
only in specific cases of extremal black holes,
and  probably this is the only case that can be 
studied rigorously in such a framework. 
Other issues, like the possible existence of space-time foam\index{foam},
that is microscopic singular fluctuations of the (quantum) geometry, which 
give the space time a ``foamy'', topologically non trivial and 
possibly non-continuous 
structure at Planck\index{Planck scale} scales ($10^{-35}$ m), 
still remain far from being resolved in the context of 
string\index{string} theory.

In \cite{emntheory} it was suggested that a consistent mathematical 
framework for dealing with such issues in the context of string\index{string} 
theory was the Liouville non-critical 
srting theory approach, involving strings 
propagating in non-conformal\index{non-conformal} space-time backgrounds. 
This violation of conformal symmetry, which lies 
at the cornerstone of 
critical string\index{string} theory, is 
remedied by the non-decoupling of 
the Liouville\index{Liouville} mode, which 
enters as a whole new target space 
dimension. In certain models of stringy foam, 
this extra dimension has time-like signature, and 
hence it can be identified with a target time, 
thereby giving the time coordinate 
a fundamentally irreversible nature, 
as a result of specific properties 
of the Liouville\index{Liouville} dynamics. Indeed, the latter 
acts as a local renormalization-group scale 
on the world-sheet of the string, 
and as such is irreversible. 
This fundamental irreversibility of non-critical string 
theory makes it analogous 
to non-equilibrium systems in field theory. From this 
point of view, then,  
critical strings are viewed  
as asymptotic ``equilibrium points'' in string theory space.

Alternative approaches to Quantum Gravity, 
on the other hand, such as the loop gravity\index{loop gravity} 
approach~\cite{loop}, 
which has the ambition of formulating a space-time background independent
quantum theory of Gravity, 
have only relatively recently began to deal with non-flat space times (such as 
those with cosmological constant) 
or highly curved ones (black holes {\it etc.}), 
and hence their full potential in dealing 
with the above issues is still not explored\cite{smolin}. These are very 
elegant theories from a geometrical viewpoint, which are based on the 
analogy of gravity to non Abelian gauge theories. 
Understanding the r\^ole
of matter in such gravity theories is a pressing task, 
in order to give such mdoels phenomenological relevance. 
In addition to loop gravity, non commutative\index{non commutative geometry} 
geometry\cite{lukierski,glikman}
is another mathematically elegant route that would certainly prove to 
be relevant for a dynamical quantum theory of space time at Planck scales,
where space time may be discrete. This approach, although
existing for some time, has only recently started to be paid attention
by the bulk of the theoretical physicists, with 
a plethora of applications, ranging 
from field theoretic models to string and brane theories.

A theoretical model, however, 
no matter how detailed and elegant it might be, 
does not become a {\it physical} theory unless it makes 
some form of contact
with {\it experiment}. Thus, to understand and be guided 
in our quest for quantum gravity we need experimentally testable or 
falsifiable predictions. Critical strings\index{string}, or other approaches
to quantum gravity, 
which respect all 
local symmetries of classical General Relativity, did not make any 
predictions for low-energy theories which could be testable in the
foreseeable future. 
The reason is simple: the coupling constant
of gravity, the Newton constant $G_N \propto 1/M_P^2$ (in four dimensions)  
is very small, and, on account of local Lorentz\index{Lorentz invariance} symmetry 
and general covariance, 
quantities of
possible experimental interest, such as cross sections and probabilities,
would be characterised by quantum gravitational loop corrections which 
would be 
proportional to some power of curvature tensors. The latter having 
dimensions of momentum squared, would imply that such quantities would be 
suppressed at least 
by the inverse square (and most likely by higher powers) 
of the Planck Mass\index{Planck Mass} scale. This would make the prospects for 
detection of such quantum gravity 
effects difficult, if not impossible, for the foreseeable future.  
Of course this does not necessarily mean 
that such approaches are physically incorrect, what 
it means is that, even if they represent reality,
we would have no way of testing them in the foreseeable future, 
and as such they would remain 
solely mathematically consistent
models. 

On the other hand, recently, more and more physicists contemplate the idea
that some of the fundamental symmetries or laws that govern classical
General and Special Relativity, such as 
{\it linear Lorentz symmetry}\index{Lorentz invariance}, 
or principles such as 
{\it the equivalence principle}\index{equivalence principle}, may not be valid 
in a full quantum theory of gravity. If true, then, this would probably imply 
that the above-mentioned Planck-mass strong suppression factors could be 
modified in such a way that quantum gravity effects are enhanced, thereby 
leading to 
some testable/falsifiable predictions in the near future. 
For instance, in the non-critical string approach to quantum gravity 
advocated in \cite{emntheory}, deviation from conformal invariance due to 
peculiar backgrounds in string theory, including foamy ones, imply
in some models at an effective low-energy field theory level, {\it modified 
dispersion relations}\index{dispersion relations} 
{\it for photons or at most for some electrically neutral gauge bosons}. 
Such modifications dot not occur not for charged probes 
or in general chiral matter\cite{synchro}, thereby violating a form of 
the equivalence principle\index{equivalence principle}, in 
the sense of the non-universality of gravity 
effects. In such models it is a gauge symmetry that protects the dispersion
relation of charged or chiral matter probes, which, unlike photons,   
do not interact with space time 
defects in the foam, the latter 
consisting of point-like branes in string theory\cite{horizons}.
The modification to the dispersion relations due to such quantum gravity 
effects  are suppressed only by a single power of Planck Mass\cite{aemn}. 
Such {\it minimal suppression models for photons} are not far from being 
tested, for instance by future Gamma Ray Burst\index{Gamma Ray Burst} 
astronomy\cite{emnnature,grb}.
On the other hand, models of quantum gravity foam\index{foam} 
with universal modified
dispersions linearly suppressed by the Planck Mass scale are already excluded
by means of astrophysical observations of Synchrotron radiation
from Crab Nebula\cite{crab,jacobpoland}, 
and one is not far from reaching sensitivities
quadratic to inverse Planck mass\cite{synchro}.

In this context, interesting ``bottom-up'' 
approaches to quantum gravity have been proposed and developed rigorously, 
such as the Doubly-Special Relativity\index{Doubly-Special Relativity} (DSR) theories~\cite{nlls}, 
which are at the focus of this meeting. According to such 
approaches, the conventional 
Lorentz\index{Lorentz invariance} symmetry of flat Minkowski space time 
is not valid, but instead one has a 
symmetry under non-linear extensions of the 
Lorentz\index{Lorentz transformation} transformations.
Such non-linear extensions are not unique, and this poses an interesting
theoretical challenge for these models. The basic idea behind such theories 
is that the Planck scale should be observer independent, and hence such 
non-linear models are characterised not only by the invariance under frame
changes of the dimensionless speed of light in vacuo, but also by the
frame-invariance of a dimensionful length scale, the Planck length. For this 
reason, although at present formulated in flat space times, such 
non-linear extensions of Lorentz symmetry are viewed as a prelude to
more complete models of quantum gravity, where the local group 
is not the conventional (linear) Lorentz\index{Lorentz invariance}, 
thereby violating the strong form 
of the equivalence 
principle. However it remains to be 
proven whether such models are viable as candidates 
for a complete and realistic theory of quantum gravity. 
In other lectures in this meeting
we shall hear more about 
the mathematical foundations and properties of such theories~\cite{amelino},
and their phenomenology\index{Phenomenology of quantum gravity}~\cite{grillo,jacobpoland,piran},  
where we refer the reader for details. 

In all approaches mentioned so far as candidate theories 
for quantum gravity there is a common feature, 
associated with the violation of a theorem whose validity 
characterises all consistent flat-space time relativistic 
quantum field theories known to date. 
This is the CPT theorem\index{CPT theorem}~\cite{pauli,bell,jost,wight}. 
The violation 
of this (discrete) space-time symmetry may have important phenomenological 
implications for low energy physics, and indeed one is prompted 
immediately to think
that this may be a way of 
testing or falsifying experimentally various theoretical 
models of quantum gravity entailing such a violation.

There is a number of fundamental questions, however, that one has to ask
before embarking on a study of the phenomenology of 
CPT Violation\index{CPT Violation}: 
(i) What are the theories which 
allow for CPT breaking?,  (ii) How (un)likely is it that somebody, someday 
finds CPT Violation in the Laboratory,
and why?, (iii) What formalism does one has to adopt? Indeed, 
since our current phenomenology 
of particle physics is based on CPT 
invariance, 
how can we be sure of observing CPT 
Violation and not something else? 
And finally, (iv) there does not seem to be a single 
``figure of merit'' for CPT violation. Then how should we 
compare various ``figures of merit" of CPT tests
(e.g. direct
mass measurement between matter and antimatter (e.g. $K^0$-${\overline K}^0$ 
mass difference a la CPLEAR\index{CPLEAR Experiment}), 
quantum decoherence effects, modifications to 
Einstein-Podolsky-Rosen\index{Einstein-Podolsky-Rosen} 
(EPR) states in meson factories, 
neutrino mixing, electron g-2 and 
cyclotron frequency comparison, 
neutrino spin-flavour conversion {\it etc}.)

In some of these questions I shall try to give answers
in the context of the present set of Lectures. I shall not try to present a  
complete overview of phenomenological tests of CPT Invariance, 
however, because the subject is vast, 
and already occupies a considerable part of the 
published literature. In these lectures I will place the emphasis 
on neutrino tests of CPT invariance, because 
as I will argue below, in many instances neutrinos seem to provide at present
the best bounds on possible CPT violation. However, I must stress
that, precisely because CPT violation is a highly model dependent
feature of some approaches to quantum gravity (QG), there may be 
models in which the sensitivity of other experiments on CPT violation, 
such as astrophysical experiments, is superior to that of current neutrino 
experiments. For this reason I will also give a brief outline of 
alternative tests of CPT violation.

My lectures will focus on the following three major issues:  

{\bf (a)} \underline{What is CPT Symmetry}: I will give a definition of what we mean 
by CPT invariance, and under what conditions this invariance holds.

{\bf (b)} \underline{Wny CPT Violation ?}: Currently there are various 
{\it Quantum Gravity Models} which  
may {\it violate} Lorentz\index{Lorentz Violation} symmetry and/or
quantum coherence (unitarity {\em etc}), and through this CPT symmetry: 
(i) space-time foam~\cite{foam} (local field theories~\cite{garay}, 
non-critical strings~\cite{emntheory} {\em etc.}), 
(ii) (non supersymmetric) string-inspired 
standard model extension with Lorentz Violation\index{Lorentz Violation}~\cite{kostel},  
(iii)  Loop Quantum Gravity\index{loop gravity}~\cite{loop}. 
(iv) {\em CPT violation} may also occur at a {\em global scale}, 
{\em cosmologically}~\cite{cosmonem}, 
as a result of a 
cosmological constant in the Universe, whose presence may jeopardize 
the definition of a standard scattering matrix.

{\bf (c)} \underline{How can we detect CPT Violation?} : Here is a 
current list of most sensitive 
particle physics probes for CPT tests: 
(i) {\em Neutral Mesons}\index{meson}: Kaons\index{kaons}~\cite{ehns,emn}, B-mesons, 
and their entangled 
states in $\phi$ and $B$ factories\index{meson factories}~\cite{huet,benatti1,bernabeu}. 

(ii) {\em anti-matter factories}: antihydrogen\index{antihydrogen}~\cite{antihydro} 
(precision spectroscopic tests on free and trapped 
molecules~\cite{kostel,bluhm,mavroyoko}), 

(iii) Low energy atomic physics experiments~\cite{bluhm}, 
including ultra cold neutron
experiments in the gravitational field of the Earth.

(iv) Astrophysical Tests  
(especially  Lorentz-Invariance
violation\index{Lorentz Violation}  tests, via  modified  dispersion\index{dispersion relations} relations of matter 
probes {\it etc.})~\cite{emnnature,grb}

(iv) Neutrino Physics, on which we shall mainly concentrate 
in these lectures~\cite{mavrovenice}.

I shall be brief in my description due to space restrictions. For 
more details I refer the interested reader to the relevant literature.
I will present some elementary proofs of theorems that will be essential
for the formalism of CPT Violation and its phenomenology. 
I will not be complete 
in reviewing the phenomenology of CPT violation; in my lectures
I will place emphasis on a specific type of violation,
that through  quantum decoherence, which 
I believe to be the most likely one to charactrise space-time foam theories
of quantum gravity; this belief is based on the fact that 
decoherence may be compatible with fundamental local symmetries
of space time, such as Lorentz invariance\index{Lorentz invariance}~\cite{mill,discr}. 
For completeness, however, I will also 
give a brief exposition of alternative ways of CPT
violation, and 
refer the reader to some key references, where more detailed information
is provided on those topics. 
Needless to say that I am fully aware 
of the vastness of the topic of CPT Violation, 
which grew enormously in recent years, 
and I realize that I might not have done a perfect job here; I 
should therefore apologize beforehand 
for possible omissions in references,
and topics,  
but this was not 
intentional. I do hope, however, that I give here a rather 
satisfactory representation
of the current situation regarding this important research topic.

\section{Theoretical Motivation for CPT Violation and Formalism}

\subsection{The CPT theorem and how it may be evaded} 

The CPT theorem\index{CPT theorem} refers to quantum field theoretic models of particle physics,
and ensures their invariance under  the successive operation (in any order)
of the following {\it discrete } transformations: 
{\bf  C}(harge), {\bf P}(arity=reflection)\index{Parity}, and {\bf T}(ime reversal).
The {\em invariance} of the Lagrangian density ${\cal L}(x)$ 
of the field theory under the combined action of {\em CPT}  
is a property of any quantum field theory  in a {\em Flat} space time  
which respects:
{\em  (i) Locality}\index{locality}, {\it (ii) Unitarity}\index{unitarity} 
and {\it (iii) Lorentz Symmetry}\index{Lorentz invariance}.
\begin{eqnarray} 
\Theta {\cal L}(x)\Theta^\dagger = {\cal L}(-x)~,~\Theta =CPT~,~{\cal L}={\cal L}^\dagger 
\label{lagrcpt}
\end{eqnarray}

The theorem has been suggested first by L\"uders and 
Pauli\index{Pauli}~\cite{pauli}, and 
also by John Bell~\cite{bell},
and has been put on an axiomatic form, using Wightman axiomatic approach to 
relativistic (Lorentz invariant) field theory, by Jost~\cite{jost}. 
Recently the Lorentz covariance\index{Lorentz covariance} of the Wightmann (correlation) functions
of field theories~\cite{wight} as an essential requirement
for a proof of CPT\index{CPT} has been re-emphasized in \cite{greenb}, in 
a concise simplified exposition of the work of Jost.  
The important point to notice in that proof is the use of 
flat-space Lorentz covariance\index{Lorentz covariance}, which allows the passage onto a momentum 
(Fourier) formalism. Basically, the Fourier formalism employs 
appropriately superimposed plane 
wave solutions for fields, with four-momentum $p_\mu$. 
The proof of CPT, then, follows by the Lorentz covariance\index{Lorentz covariance}
transformation properties of the Wightman functions, and the unitarity 
of the Lorentz transformations\index{Lorentz transformation} 
of the various fields.

In curved space times, especially highly curved ones
with space-time boundaries, such as 
space-times in the (exterior)
vicinity of black holes\index{black hole}, where the boundary is provided
by the black hole horizons\index{horizons}, 
or space-time foamy 
situations, in which one has vacuum creation of 
microscopic (of Planckian size $\ell _P = 10^{-35}$ m) black-hole 
horizons~\cite{foam}, such an approach is invalid, and Lorentz
invariance\index{Lorentz invariance}, and possibly unitarity\index{unitarity}, are lost.  Hence, such models of quantum 
gravity
violate requirements (ii) \& (iii) of the CPT theorem, 
and hence one should expect its {\em
violation}.  

It is worthy of discussing briefly the basic mechanism by which 
unitarity may be lost in 
space-time foamy situations in quantum gravity. This is the lecturer's favorite
route for possible quantum-gravity induced CPT Violation\index{CPT Violation}, 
which may hold 
independently of possible Lorentz invariance\index{Lorentz invariance} 
violations. It is at the core of 
the induced decoherence\index{decoherence} by quantum gravity\index{quantum gravity}~\cite{ehns,emn}.


The important point to notice is that, in general, 
space-time may be {\em discrete} and {\it topologically non-trivial} 
at Planck scales\index{Planck scale}  $10^{-35}~m$, 
which might (but this is not necessary\cite{mill,discr}), 
imply  Lorentz symmetry Violation\index{Lorentz Violation} (LV), 
and hence CPT Violation\index{CPT Violation} (CPTV). 
Phenomenologically, at a macroscopic level, 
such LV may lead to 
extensions of the standard model which violate both Lorentz and 
CPT invariance~\cite{kostel}.

\begin{figure}[ht]
\centering
\includegraphics[height=6cm]{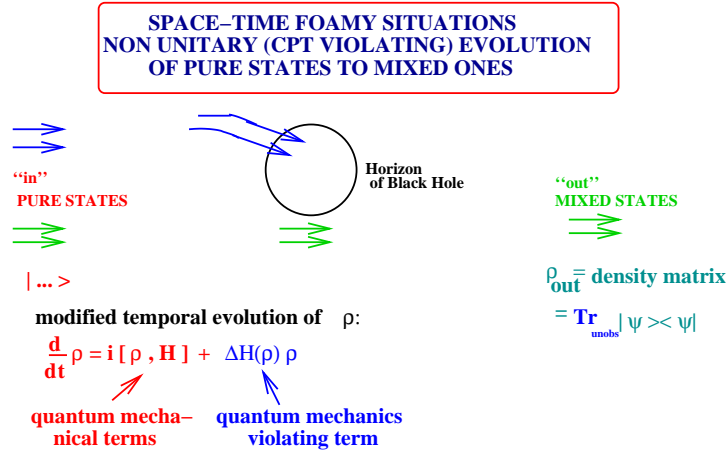} 
\caption{A basic mechanism for loss of information in a space time foamy
situation.}
\label{loss}
\end{figure}

In addition,  
there may be an {\it environment of gravitational} 
degrees of freedom (d.o.f.) {\it inaccessible} to 
low-energy experiments (for example non-propagating d.o.f., for which
ordinary 
scattering is not well defined~\cite{emn}). This will lead in general
to an 
{\it apparent information loss}\index{information loss} for low-energy observers,
who by definition  can measure only propagating low-energy
d.o.f. by means of scattering experiments. 
As a consequence, an apparent lack of unitarity and hence 
CPTV may arise, which is in principle independent of any LV effects. 
The loss of information may be understood simply 
by the mechanism illustrated in fig. \ref{loss}. In a 
foamy space time there is an ongoing creation and annihilation 
of Quantum Gravity {\em singular} fluctuations  (e.g. microscopic
(Planck size) black holes etc), which indeed implies that 
the observable space time is an open system\index{open system}. When matter particles
pass by such fluctuations (whose life time is Planckian, of order
$10^{-43}$ s),  part of 
the particle's quantum numbers ``fall into'' the horizons, and are captured
by them as the microscopic horizon disappears into the foamy vacuum\index{foamy vacuum}.   
This may imply  the exchange of information between 
the observable world and the gravitational ``environment'' consisting 
of degrees of freedom inaccessible to low energy scattering experiments,
such as back reaction of the absorbed matter onto the space time, recoil of the microscopic black hole {\it etc.}. 
In turn, such a loss of information\index{information loss} 
will imply evolution of 
initially pure quantum-mechanical states 
to mixed ones for an asymptotic observer.

As a result, the asymptotic observer will have to use 
density matrices\index{density matrix}
instead of pure states: 
$\rho_{out}={\rm Tr}_{\rm unobs}|out><out|=$ \$ $\rho_{in} 
 \$ \ne S S^\dagger$, with $S=e^{iHt}$ the ordinary 
scattering materix\index{scattering matrix}. Hence, in a foamy situation the concept of the 
scattering matrix is replaced by that of 
the {\it superscattering matrix}\index{superscattering matrix},
\$, introduced by Hawking\index{Hawking}~\cite{foam}, which   
is a {\it linear}, but {\it non-invertible} 
map between ``in'' and ``out'' density matrices; in this 
way, it quantifies the unitarity\index{unitarity} 
loss in the effective low-energy theory. 
The latter violates CPT due to a mathematical theorem by R.Wald\index{Wald}, 
which we describe in the next subsection~\cite{wald}.  

Notice that this is an effective violation,
and indeed the complete theory of quantum gravity (which though is still 
unknown) may respect some form of CPT invariance. 
However, from a phenomenological point of view, this 
effective low-energy violation of CPT is the kind of violation 
we are interested in here. A word of caution is necessary at this point.
Some theorists believe that quantum gravity does not entail 
an evolution of a pure quantum state\index{pure state} 
to a mixed\index{mixed state} one, but, 
as is the case in some quantum 
mechanical decoherence\index{decoherence}
models of open systems, to be discussed below, 
the purity of states
is maintained during the quantum-gravity 
induced decoherent evolution. If this is the case, 
then CPT may be conserved in such models, 
provided, of course, Lorentz invariance\index{Lorentz invariance} and locality\index{locality} 
of interactions are respected. 

\subsection{\$  matrix and strong CPT Violation (CPTV)}
 
The theorem of R. Wald states the following~\cite{wald}: 
{\em if \$ $\ne S~S^\dagger $, then CPT is violated, at least in its
strong form, in the sense that the CPT}\index{CPT} {\it operator 
is not well defined.}

For instructive purposes we shall give here an elementary proof.
Suppose that CPT\index{CPT} is conserved, then there exists a unitary, invertible 
CPT operator
$\Theta$: $\Theta {\overline \rho}_{in} = \rho_{out}.$
Since the density matrix\index{density matrix} 
acts on a tensor product space between ket and 
bra vectors by definition, $\rho = \psi \otimes {\overline \psi}$, the 
action of $\Theta $ is defined schematically as: $\Theta = \theta \theta^\dagger$, with $\theta$ acting on state vectors $\psi$, and being anti-unitary, i.e. 
$\theta^\dagger = -\theta^{-1}$. 

Asuming that such a $\Theta$ exists, we have: $ \rho_{out} =$ \$ $\rho_{in} \to \Theta {\overline \rho}_{in} 
=$\$ $\Theta^{-1} {\overline \rho}_{out} \to $ 
${\overline \rho}_{in} =\Theta^{-1} $\$ $\Theta^{-1} {\overline \rho}_{out} $.

But ${\overline \rho}_{out} =$\$${\overline \rho}_{in} $, 
hence : 
\begin{equation}
{\overline \rho}_{in} = \Theta^{-1} \$ \Theta^{-1}  \$ 
{\overline \rho}_{in}.
\end{equation}

The last relation implies
that \$  has an {\it inverse} 
\begin{equation}
\$^{-1} = \Theta^{-1} \$ \Theta^{-1}, 
\label{inversedollar}
\end{equation}
which, however, as 
we explain now is {\it impossible}, due to the information loss\index{information loss}
in case a pure state evolves into a mixed one.  

To prove\cite{wald} this last statement 
formally we first notice that 
from (\ref{inversedollar}) 
one also obtains the relation:
\begin{equation}
\Theta  = \$ \Theta^{-1}  \$.
\label{thetarel}
\end{equation}
Consider now a pure state 
density matrix $\rho_{in} = |IN \rangle\langle IN |$,
which evolves to the density matrix (mixed state)\index{mixed state} \$$\rho_{in}$.
As a result of (\ref{thetarel}), the mixed state 
$\Theta^{-1}$\$$\rho_{in}$ must evolve to the pure state $\Theta \rho_{in}$.
However, suppose we have an out state  $\psi$, which we obtain 
by the action of \$ on an IN density matrix\index{density matrix} $\sigma$, that is: 
\begin{equation}
\$ \sigma = \psi \otimes {\overline \psi} 
\label{proof1}
\end{equation}
where, as mentioned above, 
$\otimes $ denotes the appropriate tensor product of Hilbert spaces
spanned by ket and bra vectors. 
One may expand $\sigma$ in terms of its eigenvectors $\phi_i$, corresponding
physically to a weighted superposition of states that comprise 
the mixed\index{mixed state} state
$\sigma$: 
\begin{equation}
\sigma = \sum_{i} p_i \phi_i \otimes {\overline \phi}_i,
\label{proof2}
\end{equation}
with $p_i$ positive, and $\sum_{i}p_i = 1$.  Since by definition 
\$ is a linear map, we have: 
\begin{equation}
\sum_{i}p_i \$ (\phi_i \otimes {\overline \phi}_i) =  \psi \otimes {\overline \psi}
\label{proof3}
\end{equation}
Consider now an OUT state vector $\chi$ orthogonal to $\psi$. Taking the expectation value of (\ref{proof3}) in the state $\chi$ we obtain:
\begin{equation}
\sum_{i} p_i \langle \chi |\$(\phi_i \otimes {\overline \phi}_i)|\chi\rangle 
= 0
\label{proof4}
\end{equation}
Each term in (\ref{proof4}) is non negative, due to the positive-definiteness
of $p_i$ and the positivity\index{positivity} of 
the density matrix\index{density matrix} (by definition)
$\$(\phi_i \otimes {\overline \phi}_i)$. Therefore, (\ref{proof4})
implies 
\begin{equation} 
\langle \chi |\$(\phi_i \otimes {\overline \phi}_i)|\chi\rangle = 0
\label{proof5}
\end{equation}
for all $i$ and all $\chi$ orthogonal to $\psi$. This implies
\begin{equation}
\$ (\phi_i \otimes {\overline \phi}_i) = \psi \otimes {\overline \psi}
\label{proof6}
\end{equation}
for all $i$, i.e. each initial pure state $\phi_i$ must evolve to 
the {\it same} final pure state $\psi$. In that case, $\theta^{-1}\psi$ 
must evolve to the final state $\theta \phi_i$ for all $i$. 
This is {\it impossible} if there is more than one $\phi_i$, i.e. 
if the density matrix $\sigma$ represents a {\it mixed} state.

Hence, in case where decoherence\index{decoherence} 
implies the evolution\index{evolution} of a 
pure state to a mixed one, CPT\index{CPT} 
{\em must be violated}, at least in its strong form, in the
sense of $\Theta$ 
not being a well-defined operator, and the non existence of the 
inverse of \$, as discussed previoulsy. The non invertibility 
of \$ should not be considered as a surprise in that case, as a result
of the involved loss of information\index{information loss} in the problem.
CPT symmetry, and also by the same arguments 
microscopic time reversal\index{microscopic time reversal}\cite{wald}, 
fail in a dramatic way
in such a case: microscopic time-reversed dynamics does not merely fail to 
be the same as time evolution forward in time, which would simply
mean the non commutativity of the corresponding operators/generators
of the symmetry with the hamiltonian of the system under consideration, 
but {\it does not exist} at all.

As I remarked before, this is 
my preferred way of CPTV by Quantum Gravity\index{quantum gravity}, 
given that it may occur
in general independently of LV and thus preferred frame approaches
to quantum gravity. Indeed, I should stress at this point that 
the above-mentioned gravitational-environment induced decoherence
may be Lorentz invariant\index{Lorentz Invariant Decoherence}~\cite{mill}, 
the appropriate Lorentz transformations
being slightly modified to account, for instance, for the discreteness
of space time at Planck\index{Planck scale} length~\cite{discr}. 
This is an interesting 
topic for research, and it is by no means complete. 
Although the lack of an invertible scattering matrix
in most of these cases implies a strong violation of CPT, 
nevertheless, 
it is interesting to
demonstrate explicitly whether some form of CPT invariance holds 
in such cases~\cite{waldron}. This also includes cases with non-linear
modifications of Lorentz symmetry~\cite{nlls}, 
discussed in this School, 
which arise from the requirement 
of viewing the Planck\index{Planck scale} length 
as an invariant (observer-independent)
proper length in space time. 

It should be stressed at this stage that, if the CPT operator 
is not well defined,
then this may lead to a whole new perspective of dealing with precision 
tests in meson\index{meson} factories. 
In the usual LV case of CPTV~\cite{kostel}, the 
CPT breaking is due to the fact that the CPT operator, which is well-defined
as a quantum mechanical operator in this case, does not commute with the 
effective low-energy Hamiltonian of the matter system. This leads to 
mass differences between particles and antiparticles\index{antiparticle}. If, however, the 
CPT operator is not well defined, as is the case of the quantum-gravity
induced decoherence~\cite{ehns,emn}, 
then, the concept of the `antiparticle' gets 
modified~\cite{bernabeu}. In particular, the antiparticle space is 
viewed as an independent subspace of the state space of the system,
implying that, in the case of neutral mesons, for instance, the 
anti-neutral meson should not be treated as an 
identical particle\index{identical particles}
with the corresponding meson. This leads to the possibility of novel 
effects associated with CPTV as regards 
entangled states of 
Einstein-Podolsky-Rosen\index{Einstein-Podolsky-Rosen} (EPR) type, 
which may be testable
at meson factories\index{meson factories}~\cite{bernabeu}. We shall discuss this 
in some detail later on.

\subsection{CPT Symmetry without CPT Symmetry?}

An important issue which arises at this point is whether the 
above violation of CPT\index{CPT} symmetry 
is actually detectable experimentally.
This issue has been examined in \cite{wald}, where it was 
proposed that despite the strong CPT 
violation in cases where decoherence leads to an evolution 
of a pure state to a mixed one, 
there is the possibility for a softer (weaker) form of 
CPT invariance in such cases, compatible with the  
non-invertibility of \$. 

The main idea behind such a weak form of 
CPT invariance\index{weak form of CPT invariance} is that,
although in the full theory CPT is violated in the above sense,
nevertheless one can still define asymptotic {\it pure} scattering IN and OUT 
states as the CPT inverse of each other. In formal terms, 
although in the full theory $\Theta$ is not well defined, however 
one can define pure states $\psi \in {\cal H}_{IN}$, and 
$\phi \in {\cal H}_{OUT}$ in the respective Hilbert spaces ${\cal H}$ 
of IN and OUT states, such 
that the following equality between probabilities ${\cal P}$ holds:
\begin{equation}
{\cal P}(\psi \to \phi )={\cal P}(\theta^{-1}\phi \to \theta \psi)
\label{probscptweak}
\end{equation}
If only pobabilities are measured experimentally, which is certainly 
our experience so far, 
then the equality (\ref{probscptweak}) would imply that 
the strong form of CPT invariance would be undetectable experimentally. 

From the point of view of the superscattering matrix\index{superscattering matrix} \$, the equality
(\ref{probscptweak}) implies the following relation\cite{wald}:
\begin{equation}
\langle \phi |\$(\psi \otimes {\overline \psi})|\phi\rangle =
\langle \theta \psi |\$(\theta^{-1}\phi \otimes {\overline 
\theta^{-1}\phi})|\theta \psi\rangle 
\end{equation}
or, equivalently :
\begin{equation}
\$^{\dagger} = \Theta^{-1} \$ \Theta^{-1}
\label{dollardagger}
\end{equation}
when the action is considered on pure asymptotic states.
Relation (\ref{dollardagger}) is compatible with the non-existence of an 
inverse of \$, unless the full CPT\index{CPT} invariance 
holds, which would imply
unitarity of \$, i.e. \$$^\dagger$ = \$$^{-1}$.
Wald has argued in favour of this conclusion by considering a 
simple case of finite-dimension ($n$) Hilbert\index{Hilbert Space} 
spaces of IN and OUT states,
and assuming that every pure IN state evolves to the density matrix 
$1/n \delta ^a_b$ in the OUT Hilbert space. It is clear that in this example
\$$^{-1}$ does not exist, but for all $\psi$ and $\phi$ the relation
(\ref{probscptweak}) holds, since both sides equal $1/n$.

\subsection{Decoherence and Purity of States under Evolution}

Since the above result of weak CPT invariance requires 
the purity\index{purity} of asymptotic scattering states, 
a natural question to ask 
is whether there exist concrete models of decoherence\index{decoherence} 
where the purity of an initial state vector remains, 
while time irreversibility\index{time irreversibility} holds. 

A physically acceptable framework for discussing 
decohering evolution\index{decohering evolution} 
of
an open quantum mechanical system is that of Lindblad
or the so-called dynamical semigroup\index{semigroup} approach\cite{lindblad}, 
which ensures the 
{\it complete positivity}\index{complete positivity} 
of the density matrix\index{density matrix} 
$\rho(t)$ at any time moment $t$ during the 
evolution\index{evolution}, and the 
{\it conservation 
of probability} ${\rm Tr}\rho = 1$.
The Lindblad\index{Lindblad} evolution of 
open systems\index{open system}\cite{lindblad}, 
with Hamiltonian $H$, interacting with an environment\index{environment} 
through operators
$D_j, D^\dagger_j$, is described as a {\it linear} 
evolution in the density matrix $\rho$:
\begin{equation}
{\dot \rho} = i[\rho, H] + {\cal D}[\rho]; \qquad 
{\cal D}[\rho] = \sum_{j}\left(\{\rho, D_j^\dagger D_j\}-2D_j\rho D_j^\dagger\right)
\label{lindblad}
\end{equation}
where $\{.,.\}$ denotes an anticommutator.
The Hamiltonian $H$ in (\ref{lindblad}) 
may contain terms from the environmental entanglement\index{entanglement} 
which can be expressed
as commutators with $\rho$, and hence it should be understood as an 
effective Hamiltonian of the system. The decoherence\index{decoherence}
term ${\cal D}[\rho]$, on the other hand, 
cannot be expressed as such a commutator.

To ensure energy conservation\index{energy conservation} 
on the average, and monotonic increase  
of the von-Neumann entropy\index{entropy} 
$S=-{\rm Tr}\left(\rho{\rm ln}\rho\right)$,  
one has to impose self-adjointness\index{self-adjointness} 
of the Lindblad\index{Lindblad} environmental operators 
\begin{equation}
D_j^\dagger = D_j
\label{selfadjoint}
\end{equation}
and also require that these operators commute with the Hamiltonian
\begin{equation}
[D_j, H]\qquad {\rm for}\quad {\rm all}\quad j
\label{commutator}
\end{equation}
This leads to a double commutator structure of the decoherence terms in (\ref{lindblad}):
\begin{equation}
{\cal D}[\rho] = \sum_{j}[D_j,[D_j, \rho]]
\label{doublecomm}
\end{equation}
In general, in this type of decoherence\index{decoherence}
one has the evolution of a pure state into a mixed\index{mixed state} 
one. However, 
there exist subclasses of Lindblad evolution, in particular 
energy-driven simple decoherence models\cite{houghston,adler}, 
where the purity\index{purity} of state vectors\index{state vector} 
is preserved. 
A mathematical criterion for this feature is that 
\begin{equation}
\rho^2 = \rho, \qquad {\rm Tr}\rho = 1,
\label{purity}
\end{equation}
during the evolution\index{evolution}.

In such models, $D_j = \lambda_j H$, with $\lambda_j$ c-number 
constants. Without loss 
of generality, 
we can substitute  in such a case 
the sums in (\ref{doublecomm}) by a single environmental operator 
\begin{equation}
D = \lambda H, \qquad \lambda^2 = \sum_{j} \lambda_j^2 
\label{deqh}
\end{equation}
This simplifies the situation and will suffice for our purposes
in this work. 

In this type of decoherence\index{decoherence}, 
the density matrix\index{density matrix} evolution preserves the purity of states, and  
can be written in terms of stochastic\index{stochastic} 
Ito\index{Ito process} differential equations 
for the state vectors $|\psi \rangle $ (or equivalently the pure state density matrix
$\rho =  |\psi \rangle \langle \psi |$):
\begin{equation}
d\rho = -i[H, \rho]dt -\frac{1}{8}[D,[D,\rho]]dt + \frac{1}{2}[\rho,[\rho,D]]dW_t
\label{ito}
\end{equation}
where $t$ is the time, and $dW_t$ is an Ito stochastic differential 
obeying 
\begin{equation}
dW_t^2 = dt, \qquad dtdW_t = 0 
\label{stochcond}
\end{equation}
which are 
the equivalent of white noise\index{white noise} conditions.
Needelss to say that one can generalise the above equation to the case where sums of $D_j$ operators are involved, but as we mentioned above this will not
be necessary for our purposes here. 

We remark that, in terms of state vectors $|\psi\rangle$, 
the first term in (\ref{ito}) is nothing but 
the Schr\"odinger Hamiltonian term $-iH|\psi\rangle$, while the 
second term resembles Fokker-Planck\index{Fokker-Planck} 
stochastic diffusion\index{diffusion} terms. Unlike the 
Schr\"odinger term, the diffusion term 
is not invariant under the time reversal operation $t \to -t$ and
$i \to -i$, and hence {\it time irreversibility}\index{time 
irreversibility} occurs in the problem, despite the purity of states.

Upon using (\ref{deqh}) in (\ref{ito}), one obtains a 
stochastic\index{stochastic} equation for this energy-driven decoherence:
\begin{equation}
d\rho = -i[H, \rho] -\frac{\lambda^2}{8}[H,[H,\rho]]dt + 
\frac{\lambda}{2}[\rho,[\rho,H]]dW_t
\label{ito2}
\end{equation}
The double commutator of the Hamiltonian, together with the purity-of-states
condition (\ref{purity}), leads to the following order of the 
decoherence\index{decoherence}
term in such models, obtained by considering the vacuum expectation value
of the double commutator term in (\ref{ito2}): 
$\gamma \equiv \langle\langle {\cal D}[\rho]\rangle\rangle = 
{\rm Tr}\left(\rho\frac{\lambda^2}{8}[H,[H,\rho]] \right)$.
Using as a complete orthonormal basis of states energy eigenstates 
$|m\rangle$, then, it is straightforward to see that the above estimate 
leads to the square of the energy variance\index{energy variance} 
\begin{equation}
\gamma = \frac{\lambda^2}{8}\left(\Delta H\right)^2 = 
\langle\langle H^2\rangle\rangle 
- \left(\langle\langle H\rangle\rangle\right)^2
\label{decohvariance}
\end{equation}
for this model of decoherence. 

In quantum-gravity driven models of decoherence it is natural to assume 
that $\lambda^2 \propto 1/M_P$, where $M_P \sim 10^{19}$ GeV is the 
Planck scale, which is expected to be the characteristic scale of 
quantum gravity\index{quantum gravity}. In such models then one obtains the following 
estimate for the 
decoherence coefficient $\gamma$\cite{adler}
\begin{equation}
\gamma \sim  \left(\Delta H\right)^2/M_P
\label{estimate1}
\end{equation}
We shall come back to physical applications of this case later on, 
when we discuss sensitive probes of 
quantum mechanics, such as neutral mesons\index{meson} and 
neutrinos\index{neutrino}. 

Before closing this subsection we should remark that other types of decoherence
models, which are not energy driven, but correspond 
to spontaneous localisation\index{spontaneous localisation} 
in space, also exist. 
One such model is the one presented in \cite{ghirardi},
in which the operator $D$ is taken to be proportional to the spatial
coordinate operator $q$, thereby leading to spatial localisation.
In such a case again the decoherence coefficient $\gamma$ (\ref{estimate1})
is found to be 
proportional to the square of the position operator variance\index{position variance} 
$\gamma \propto \left(\Delta q\right)^2$, expressing, e.g. 
spatial separation between centres of wavepackets\index{wavepackets}, resulting for instance from
the mass difference. 

\subsection{More General Case: Dynamical Semi-Group Approach 
to Decoherence, and Evolution of Pure States to Mixed }

In the previous subsection we examined special cases of Lindblad 
decohering evolution, which preserved the purity of quantum states.
The Lindblad approach to decoherence, however, in general has the feature of 
implying an evolution of a pure state to a mixed one, 
in the sense of ${\rm Tr}\rho(t)^2 \ne {\rm Tr}\rho(t)$, 
thereby leading to 
a violation of the strong form of CPT, according to the theorem of \cite{wald}.
The general Lindblad\index{Lindblad} evolution can be formulated in such a way that 
no detailed knowledge of the underlying microscopic dynamics of 
the decohering environment\index{environment} 
is necessary in order 
to arrive at certain conclusions of phenomenological interest. 
This is achieved by means of the so-called {\it dynamical semigroups
approach to decoherence}~\cite{lindblad}\index{dynamical semigroup}, 
which is a generic formalism
to describe a decohering evolution obeying some basic properties.
The time irreversibility\index{time irreversibility} 
in this approach is linked to the lack of an  
inverse of an element in an appropriate semigroup\index{semigroup}. 

Consider the generic case of a decohering (of Lindblad\index{Lindblad}, 
or even more general, type) evolution 
for an $N$-level system, that is a system whose Hamiltonian
(energy) eigenstates span an $N$-dimensional state vector space.
The decohering operators, 
assumed bounded for our purposes here, can be 
represented by $N \times N$ matrices generated by a basis 
$F_\mu,~\mu=0,1, \dots N^2-1 $, endowed by the scalar product $(F_\mu, F_\nu)=\frac{1}{2}\delta_{\mu\nu}$. For the purposes in this work we shall be dealing
explicitly with $N=2,3$ level systems, in which cases the basis $\{F_\mu \}$
consists of: (i) the three $2\times 2$ 
Pauli\index{Pauli} matrices plus the $2\times 2$ identity matrix $I_2$
for the $N=2$ case,
and (ii) the $3\times 3$ Gell-Mann\index{Gell-Mann} matrices $\Lambda_i$,$i=1,\dots 8$ 
plus the $3\times 3$ 
identity matrix $I_3$ for the $N=3$ case. 

Generically the matrices $F_\mu$ satisfy the following commutation relations:
\begin{equation}
[F_i, F_j]=i\sum_{k}f_{ijk}F_k, \qquad 1 \le i,j,k \le 8
\label{comm}
\end{equation}
where $f_{ijk}$ are the structure constants\index{structure constants} 
of the $SU(N)$ group, and 
we follow the notation that Latin indices run from 1, to $N^2-1$, while
Greek indices run from $0,1, \dots N^2-1$.

Expanding the environmental operators, as well as the (effective)  
Hamiltonian and the density matrix\index{density matrix} in (\ref{lindblad}) 
in terms of the basis $\{ F_\mu \}$:
\begin{equation}
H = \sum_{\mu}h_\mu F_\mu, \qquad \rho = \sum_{\mu}\rho_\mu F_\mu, \qquad
D_j = \sum_{mu} d^{(j)}_\mu F_\mu 
\label{expansion}
\end{equation}
and imposing the hermiticity of $D$, 
which ensures the monotonic increase
of the von Neumann entropy\index{entropy} 
$S=-{\rm Tr}\rho{\rm ln}\rho$, we can 
write the decoherence term ${\cal D}[\rho]$ in (\ref{lindblad}) as:
\begin{equation} 
{\cal D}[\rho]_{\rm Lindblad} = \sum_{\mu,\nu}L_{\mu\nu}\rho_\mu F_\nu,
\label{expandedlind}
\end{equation}
where the matrix $L_{\mu\nu}$ is real and symmetric, 
with the properties:
\begin{equation}
L_{\mu 0} =L_{0\mu}=0, \qquad L_{ij}=\frac{1}{2}\sum_{k,l,m}
({\vec d}_m \cdot {\vec d}_k)f_{iml}f_{ikj},
\label{condlind}
\end{equation}
whereby ${\vec d}_\mu = (d_\mu^{(1)}, \dots d_\mu^{(N^2-1)})$.

The vanishing of the first row and column is due to entropy increase. 
Notice that if we do not impose the requirement of 
energy conservation\index{energy conservation} 
on the average,
then it is not necessary to assume the commutativity of 
the operators with the Hamiltonian, so in general  
$[D_j, H] \ne 0$. In fact below 
we shall examine some examples 
where energy may be violated due to foam interactions.

The evolution\index{evolution} equation (\ref{lindblad}), then, reads: 
\begin{equation}
{\dot \rho} = \sum_{i,j} h_i\rho_jf_{ij\mu} + \sum_{\nu}L_{\mu\nu}\rho_\nu,
\qquad \mu,\nu = 0, \dots N^2-1.
\label{finalevollind}
\end{equation}
where the overdot denotes derivative with respect to time $t$.

Probability conservation ${\rm Tr}\rho(t) =1$ at any time moment $t$ 
implies that the differential equation for the $\rho_0$ component
decouples, yielding 
\begin{equation}
\rho_0 (t) = {\rm const}
\label{probcons}
\end{equation}
The remaining differential equations (\ref{finalevollind}) 
then can be written in the form:
\begin{equation}
{\dot \rho}_k = \sum_{j}\left(\sum_{i}h_if_{ijk} + L_{kj}\right)\rho_j 
= \sum_{k}{\cal M}_{kj}\rho_j
\label{matrixform}
\end{equation}
Representing by ${\cal A}$ the matrix that diagonalises the matrix ${\cal M}$, 
and letting $\{\lambda_1, \dots \lambda_{N^2-1}\}$ be the set of eigenavalues
of ${\cal M}$, and $\{v_1, \dots v_{N^2-1}\}$ be the corresponding set of its 
eigenvectors, we have for the $ij$ elements of ${\cal A}$: 
${\cal A}_{ij}=(v_i)_j$. The solution of (\ref{matrixform}), then, can be 
written as:
\begin{equation} 
\rho_i(t) = \sum_{k,j}e^{\lambda_k t}{\cal A}_{ik}{\cal A}^{-1}_{kj}\rho(0)_j
\label{solutionmatrixform}
\end{equation}
Thus, in the dynamical semigroup\index{semigroup} approach, we have seen that 
the imposition of generic properties, such as monotonic entropy increase,
probability conservation {\it etc.}, allows for an apparently complicated
decoherence/entanglement problem to be transformed into an algebraic 
problem of determining the eigenvalues\index{eigenvalues} and 
eigenvectors\index{eigenvectors} of finite-dimensional
matrices. In general, for $N$-level systems, with $N \ge 3$, the general form 
of the decoherence matrix is too complicated to allow for 
clear physical meaning of all its entries. As we shall discuss below,
however, in the context of specific examples, one can make physically
meaningful simplicifcations, which allow for physical predictions 
to be made from such a formalism.

\subsection{State Vector Reduction (``Wavefunction Collapse'') 
in Lindblad Decoherence}

Decoherence in general is expected to 
lead to a decay with time $t$ of the off-diagonal elements
of the reduced density matrix\index{density matrix} of an 
open system\index{open system}, 
which are in general of the form\cite{mohanty,qm}. 
\begin{equation}
\rho (x,x',t) \sim {\rm exp}\left(-ND(x-x')^2t\right), 
\label{svr}
\end{equation}
where $x,x'$ denote the spatial locations of the centre of mass of  
a system of $N$ particles,
and $D$ is a generic decoherence\index{decoherence} parameter. 
Notice the dependence of the exponent on the square of the 
distance $|x-x'|^2$, and on the number of particles $N$, 
which implies that the larger the $N|x-x'|^2$ the faster the decoherence.
Hence, macroscopic bodies 
(containing, say, at least an Avogadro number\index{Avogadro number} 
of particles)
will in general decohere very fast. 
Such considerations
are general, and can also be extended to 
decoherence\index{decoherence} models that 
may have relevance to quantum gravity\index{quantum gravity}, 
such as the wormhole\index{wormhole}-induced 
decoherence\cite{mohanty}. 

I should stress  at this point that in general,  
decoherence\index{decoherence} does not necessarily solves
the problem of measurement, because it cannot explain which one of the 
diagonal entries of the density matrix is picked up during a ``measurement'',
that is an interaction of the subsystem with a macroscopic environment.

In some models of decoherence, though, especially the ones where the purity of 
states is preserved during the evolution, like the ones examined above,
it is possible to establish a mathematical criterion for the state vector 
reduction, that is the localisation of the state vector 
in a given ``measurement''\index{measurement} channel in state space. 
It is the point of this subsection to discuss briefly this issue.

First of all we note that the temporal evolution 
(\ref{lindblad}) for these specific Lindblad\index{Lindblad} systems  
can be written in terms of 
the corresponding state vectors $|\psi \rangle$ 
via the \index{stochastic} Ito\index{Ito process} form~\cite{gisin}: 
\begin{eqnarray} 
&& |d\psi \rangle = -i H|d\psi \rangle dt + \sum_{j} 
\left( \langle D_j^\dagger \rangle_\psi D_j - \frac{1}{2}D_j^\dagger D_j 
- \right. \nonumber 
\\ && \left. \frac{1}{2} \langle D_j^\dagger \rangle_\psi \langle D_j \rangle_\psi 
\right)|\psi \rangle dt + \sum_j \left(D_j - \langle D_j \rangle_\psi\right)|\psi \rangle dW_{j,t}
\label{ito2b}
\end{eqnarray}
where $dW_{j,t}$ are the stochastic\index{stochastic} differential 
random variables satisfying (\ref{stochcond}). 

The state vector reduction, or equivalently ``collapse''\index{collapse} 
of the wavefunction\index{wavefunction}
that characterises this formalism can be proven as follows~\cite{gisin}: one 
makes the assumption that the Hamiltonian of the system $H$ can be cast in a 
{\it block-diagonal} form in terms of state-space ``channels'' $\{ k\}$, which 
exist
independently of any ``measurement''\index{measurement} (i.e. interaction with 
a macroscopic measurement apparatus). 
This means that, if ${\cal P}_k$ denotes the 
projection operator\index{projection operator} on channel $k$, then
\begin{equation}
[H, {\cal P}_k ] = 0 
\label{channels}
\end{equation}
The state vector\index{state vector} reduction is 
then proven by demonstrating the 
localisation of $|\psi \rangle$ on a state-space channel $k$ due to 
the environmental entanglement in (\ref{ito2b}). A mathematical measure of 
this localisation\index{localisation} is the so-called 
{\it Quantum Dispersion Entropy}\index{Quantum Dispersion Entropy} 
${\cal K}$ defined as\cite{gisin}: 
\begin{equation} 
{\cal K} = - \sum_{k}\langle 
{\cal P}_k \rangle _\psi {\rm ln}\langle {\cal P}_k \rangle _\psi 
\label{qde}
\end{equation} 
which, if one uses (\ref{ito2b}), and the above assumptions, can be shown to 
have 
the following monotonic decrease properties:
\begin{equation}
\frac{d}{dt}\left(M{\cal K}\right) = -\sum_{k} \frac{1 - \langle 
{\cal P}_k \rangle_\psi}{\langle {\cal P}_k \rangle_\psi}\sum_j 
|\langle {\cal P}_k D_j {\cal P}_k \rangle_\psi |^2 \le 0 
\label{decrease}
\end{equation}
where $M$ denotes an average over an ensemble of theories. 
The monotonic decrease (\ref{decrease}) implies localisation of the state 
vector in state space, 
in a time which depends on the details of the environmental 
entaglement, and specifically on the so-called effective interaction
rates 
$R_k \equiv \sum_{j} |\langle {\cal P}_k D_j {\cal P}_k \rangle_\psi |^2$,
which are positive semi-definite quantities, characteristic of the 
system.
This localisation seems therefore a rather generic feature of the 
Lindblad stochastic decoherence (\ref{ito2}). 
We remark, however, that in some specific cases 
of environmental entaglement, such a localisation may not be complete,
and one may obtain pointer states\index{pointer states} 
(i.e. minimum uncertainty coherent states\index{coherent states})
from decoherence\cite{zurek}. This 
is an important topic, which however we shall not dwell upon in these lectures.

\subsection{Non-Critical String\index{Non-Critical String} Decoherence: 
a link between Decoherent Quantum Mechanics and Gravity?}

There is an interesting connection
between the above-mentioned models of decoherence 
with 
non-critical string\index{string} theory. 
The latter is viewed as a {\it non-equilibrium}
version of string theory, the equilibrium\index{equilibrium} `points' 
corresponding to the 
critical strings. In these lectures we shall not describe in detail the
corresponding formalism, 
but we shall rather give a comprehensive
outline of the approach, and concentrate 
on those aspects of the framework that 
are relevant for our purposes here. For details we refer the interested 
reader to the literature~\cite{emntheory,mavro}.

The basic idea~\cite{emntheory,kogan} 
is the identification of the target time
in non-critical strings with a world-sheet renormalization group
scale, the Liouville\index{Liouville} field zero mode. 
Non-critical strings are described, in a first-quantised framework,
by world-sheet 
sigma models with non-conformal\index{non-conformal} background fields $\{ g^i \}$.
The corresponding two-dimensional world-sheet\index{world sheet} action is then
given schematically by:
\begin{equation}
S_\sigma = S^* + \int_\Sigma g^i V_i 
\label{sigmamodel}
\end{equation} 
where $S^*$ is a two-dimensional 
conformal world-sheet action, corresponding to 
a critical string theory, and the second term 
on the right hand side of (\ref{sigmamodel})
represents deformations from this ``conformal point''\index{conformal}. 
The operators
$V_i$ are the vertex operators, which describe the 
string excitations
corresponding to the background fields $g^i$, 
over which the 
string propagates in target space time. This set 
may contain gravitons, dilatons, gauge fields, {\it etc.}, 
$\{g^i \} =\{ G_{\mu\nu}, \Phi, A_\mu \dots \}$.
The important thing 
to notice is that the background space-time fields $g^i$
appear as couplings of the two-dimensional $\sigma$-model theory.

The non conformal nature of the backgrounds implies that the 
world-sheet renormalization 
group (RG) $\beta$-functions $\beta^i =dg^i/d{\rm ln}\mu$,
where $\mu$ is a two-dimensional RG scale, 
are non zero. For a critical string
$\beta^i = 0$, which determines the ``consistent'' target space 
backgrounds over which the string propagates. These are the
equilibrium ``points'' in the (infinite dimensional) space
of string theories, spanned by the ``coordinates'' $\{ g^i \}$. 

For consistency of the world-sheet theory, such non conformal
backgrounds require dressing with the Liouville mode, an
extra $\sigma$-model field, playing the r\^ole 
of a target-space coordinate. This field restores 
conformal invariance\index{conformal invariance}, at the cost of enlarging the target
space by one extra dimension~\cite{mavro}, 
whose r\^ole is played by the world sheet zero mode of the 
Liouville field. 
Depending on the kind of 
deformation,  the Liouville mode could be space-like or
time-like in target space. 
In these lectures we shall be interested in the time-like Liouville
mode\index{time-like Liouville mode} case. 
The Liouville zero mode then can be identified with the 
target time in a consistent way~\cite{emntheory,mavro}, which in 
some cases is forced upon us dynamically, due to 
minimization, upon such an identification, 
of the effective potential of the low-energy field 
theory\cite{gravanis}. In this way, the 
low-energy theory does not have two times. 

Since the Liouville mode may be viewed as a world-sheet 
RG scale ${\rm ln}\mu$, we have a situation in which 
a target time variable is identified with a $\sigma$-model 
RG scale. The irreversibility of the 
latter has been proven for unitary theories
by means of the Zamolodchikov's c-theorem\cite{zamo}, but is expected 
to hold also for non-unitary ones, 
due to the presence of a cutoff scale on the world-sheet, which is 
associated with ``loss'' of information due to modes with two-dimensional 
momenta
beyond the cutoff\cite{kutasov}. This  
guarantees a {\it microscopic time irreversibility}\index{time irreversibility}, 
in a non trivial way.

\begin{figure}[ht]
  \centering
\includegraphics[height=3cm]{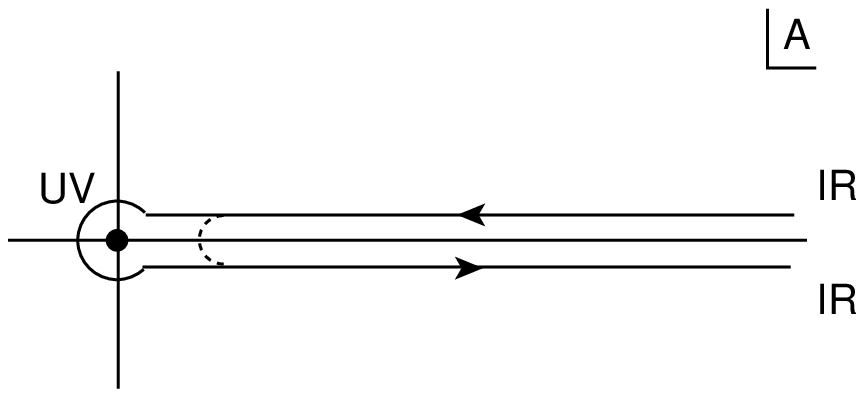}
\hfill 
\includegraphics[height=3cm]{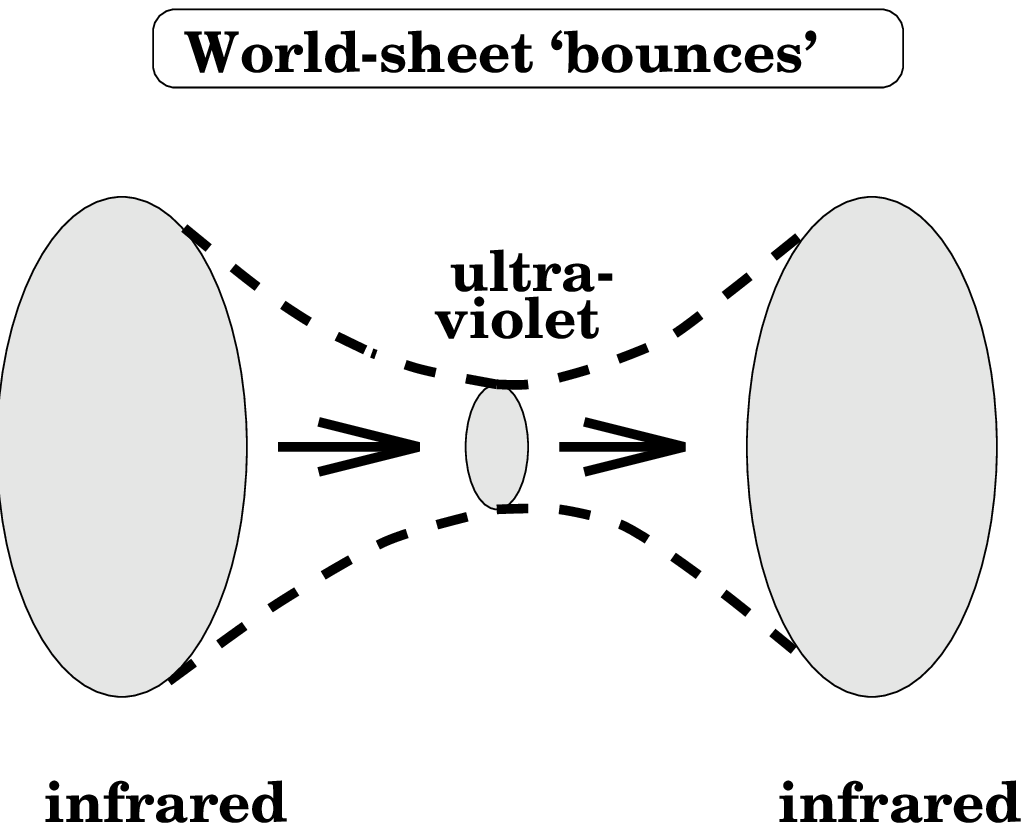}
\caption{\underline{Left Picture}: Steepest-descent curve 
for Liouville zero mode path integration, in the 
complex plane obtained after complexifying the world-sheet area $A$.
Upon the identification of this mode with target time, such curves resemble
closed time paths of non equilibrium field theories, 
in agreement 
with the non-equilibrium nature of the Liouville 
string. \underline{Right Picture}:
The ``breathing world-sheet'', 
as a result of the path on the left. The target-space 
irreversibility arises
from a ``bounce'' interpretation of this process.}
\label{liouvpath}
\end{figure}

Formally, in Liouville strings, the world-sheet correlators
of vertex operators are identified with well-defined 
\$-matrix elements
rather than scattering amplitudes. 
The non-factorisability\index{non-factorisability} 
of the \$-matrix into proper S-matrix amplitudes, 
\$$\ne SS^\dagger$, 
is obtained by noting 
that in Liouville\index{Liouville} strings, which by definition propagate
on non-conformal backgrounds\index{non-conformal}, 
one may define the Liouville zero mode 
world-sheet path integral
in a steepest-descent fashion by means of the curve indicated in figure
\ref{liouvpath}\cite{kogan,emntheory,mavro}. Upon the identification of the 
Liouville zero mode with time, such a curve resembles closed-time-paths\index{Closed Time Paths}
in non-equilibrium\index{non-equilibrium} 
field theories. It is the short-distance world-sheet 
singularities (UV) near the origin of the curve of fig. \ref{liouvpath}
that cause the aforementioned non factorizability of the \$ matrix. 
One may link the breathing world sheet, arising from the 
steepest-descent path of the Liouville mode, to a ``bounce'' 
on the infrared  (IR, large world-sheet area) limit\cite{kogan}, 
implying an irreversible
RG flow from the ultraviolate to infrared fixed points of the 
world-sheet system. 
Details are given in the literature\cite{emntheory,mavro}, where we refer
the interested reader for details. For our purposes we only 
mention that
this property links the time irreversibility\index{time irreversibility}  
of the Liouville mode,
stemming from world-sheet RG properties, to 
fundamental properties of space-time \$-matrix elements, 
in a similar fashion to the analysis in \cite{wald}.

The theory space ``coordinates''/backgrounds fields $g^i$ become 
quantum operators upon summing up world-sheet 
genera~\cite{emntheory}; decoherence in this theory space
is induced precisely by the non vanishing $\beta$-functions,
that is the departure from the conformal point~\cite{emntheory}. 
To see this one invokes the principle of world-sheet 
renormalization group\index{renormalization group} 
invariance of target-space quantities 
with physical significance 
for the string propagating in
such non-conformal\index{non-conformal} backgrounds. One such quantity is the 
density matrix of this string matter  $\rho_s$.
The RG invariance implies that $\frac{d}{dt}\rho_s =0$, 
where $t \equiv {\rm ln}\mu$ is the world-sheet RG scale. 

In the quantum theory this equation reads~\cite{emntheory,kogan}:
\begin{equation}
{\dot \rho_s} = i[\rho_s, H] + :\beta^j{\cal G}_{ji}[g^i,\rho_s]:
\label{liouveq}
\end{equation}
where the overdot denotes partial derivative with respect to $t$, 
$H$ is the effective low-energy string Hamlitonian, 
and ${\cal G}_{ij} =z^2{\overline z}^2\langle V_i(z)V_j(0)\rangle$
is the Zamolodchikov's metric in ``theory space''~\cite{mavro}. 
The notation $: \dots :$ denotes appropriate ordering of the 
quantum operators.

Equation (\ref{liouveq}) has similar form to that of a 
`decoherent
evolution' in the parameter $t$. Clearly, for 
critical backgrounds\index{critical backgrounds}
$\beta^i=0$, and hence the evolution in RG space does not imply
any such ``decoherence''. 
However, this decoherence would acquire physical significance
only if the identification
of the scale $t$ with the real target time variable in string 
theory holds~\cite{emntheory}. 
This is not a trivial issue, and in fact it can be shown that 
it does not hold for any non conformal deformation. However,
as already mentioned, 
there are physically interesting cases, among which 
strings in de Sitter space times~\cite{cosmonem}, 
to be discussed 
separately in the next subsection, 
or colliding brane 
cosmologies\index{colliding brane cosmologies}~\cite{gravanis},
which are non conformal backgrounds in string theory,
and in which the above-mentioned identification of time
with the world-sheet RG scale, that is the Liouville 
zero mode, occurs due to dynamical reasons, leading to 
minimization of energy. 

Under such an identification, the RG evolution (\ref{liouveq})
becomes a real temporal evolution for the reduced density
matrix of a string interacting with the non conformal background,
which leads to the presence of decoherence\index{decoherence} terms proportional to 
the RG $\beta^i \ne 0$. 
Using (\ref{liouveq}) 
it can be shown\cite{emntheory} 
that such Liouville-string decoherence
has the following properties: 

(i) {\it Conservation of Probability}, 

(ii) {\it Von-Neumann entropy monotonic increase}: one calculates the 
relevant rate as:
\begin{equation}\label{entropyliouv} 
\frac{\partial}{\partial t }\left({\rm Tr}\rho{\rm ln}\rho \right)= 
\beta^i {\cal G}_{ij}\beta^j \left({\rm Tr}\rho{\rm ln}\rho \right)
\end{equation}
which is positive semi-definite, since $\beta^i {\cal G}_{ij}\beta^j \ge 0$ 
due to Zamolodchikov's c-theorem for unitary\index{unitary} theories 
or its extension for non-unitary ones\cite{kutasov}. 

(iii) {\it Energy conservation}\index{energy conservation} 
on the average, since 
\begin{equation} 
\frac{\partial}{\partial t }\langle\langle H \rangle\rangle =
\frac{\partial}{\partial t} 
\left(\frac{\partial \beta^i}{\partial g^i}\right) = 0
\end{equation}
due to the fact that there is no explicit RG scale $t$ dependence on the 
$\beta^i$ function, due to renormalizability of the $\sigma$-model.
However, a word of caution should be placed here. In some cases, 
in particular 
logarithmic conformal 
field theories\index{logarithmic conformal field theories}, 
such as D-particle recoil\cite{kogan},
where the short-distance limit of two deformation operators 
contain explicit logarithms 
$V_i (z) V_j (z´) \sim {\rm ln}z c_{ijk}V_k/|z-z´|^2$, 
there is explicit $t$ dependence in the Operator Product Expansion 
coefficients appearing in the 
perturbative expansion of 
the $\beta$-function in powers of 
coupling constants $g^i$\cite{szabo2}. For instance, in the recoil 
problem, the anomalous dimension coefficients are $t$-dependent\cite{kogan}. 
In such cases,
the energy conservation on the average may be spoiled.

This type of Liouville-string decoherence\index{decoherence} leads to 
``localisation''\index{localisation} in 
theory space $g^i$~\cite{emntheory}, which can be seen 
as follows:  the RG $\beta$-functions are expressed 
as a power series in the coordinates/background fields $g^i$, 
$\beta^i =C^i_{i_1\dots i_n}g^{i_1} \dots g^{i_n}$. The linear term 
is the anomalous dimension\index{anomalous dimension} term. In a weak field expansion, i.e. 
when $g^i$ are assumed sufficiently weak so that perturbation theory
holds, one may assume to a good approximation
$\beta^i \simeq y_i g^i $,
with $y_i$ the anomalous scaling dimension of the 
$\sigma$-model coupling/background field $g^i$. 
Note also that this is an exact result 
(in terms of a $g^i$ expansion) in 
some non conformal cosmological backgrounds 
of string theory, such as de Sitter\index{de Sitter space} space, i.e. a 
space time with a non zero cosmological constant\index{cosmological constant} $\Lambda > 0$.
In such a case, 
the graviton $\beta$-function, to order
$\alpha ' =M_s^{-2}$, with $M_s$ the string mass scale,
is given by the Ricci tensor:  
$\beta^{\mu\nu} = \alpha ' R_{\mu\nu} = \Lambda g_{\mu\nu}$, 
and thus is linear
in the graviton background. We shall examine this case in some detail 
in the next subsection.

In such linearised cases, one may choose the antisymmetric 
quantum ordering prescription which leads to a double commutator
structure in the theory space coordinate operators  $g^i$, so that 
(\ref{liouveq}) reads: 
\begin{equation}
{\dot \rho_s} = i[\rho_s, H] + \frac{1}{2}y_i [g_i, [g^i,\rho_s]],
\label{liouveq2}
\end{equation}
where we have used the fact that, 
to leading order in $g^i$, the Zamolodchikov ``metric in theory space'' 
${\cal G}_{ij} \simeq \delta_{ij} + {\cal O}(g^2)$, 
which can always be arranged by an appropriate choice of 
a Renormalization scheme~\cite{mavro}.

Comparing (\ref{liouveq2}) with (\ref{lindblad}),(\ref{doublecomm})
we observe that we are encountering here exactly an analogous 
situation, but instead of energy driven or position 
localisation\index{localisation} 
decoherence models, we have a 
non-critical string theory induced 
decoherence. Since $g^i$ are generalised ``position vectors''
in theory space, the same arguments leading to localisation
of the state vector 
in those models will lead here to ``localisation in
theory space $g^i$''. From a physical viewpoint this 
would imply the emergence of the equilibrium\index{equilibrium} target space
of string theories in a dynamical way, due to evolution 
of a non critical string theory to those equilibrium points. 
Moreover, the double commutator structure in (\ref{liouveq2}) 
will also lead to variances $(\Delta g^i)^2$ for the background fields
$g^i$, expressing the back reaction of string matter on those 
backgrounds. 
In the next subsection we shall examine a concrete and physically
interesting example of such a situation, 
that of a de Sitter space\index{de Sitter space} 
time background. As we shall discuss below, in 
such a case one also 
obtains an interesting case of CPT Violation of unconventional form,
which may be related to some energy-driven decoherence models
mentioned above\cite{houghston,adler}.

\subsection{Cosmological CPTV?}

\begin{figure}[ht]
  \centering
\includegraphics[height=6cm]{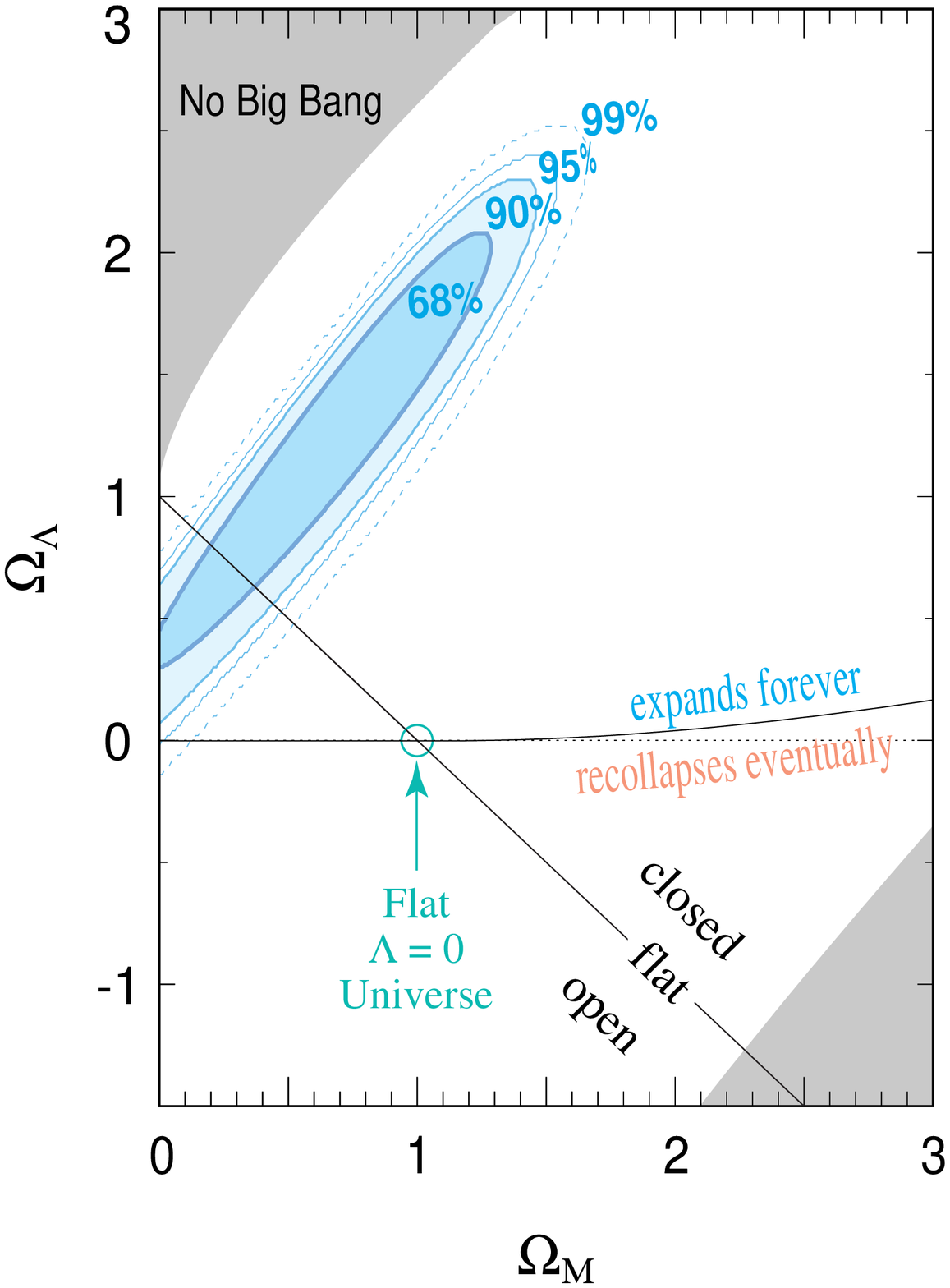}
\hfill 
\includegraphics[height=6cm]{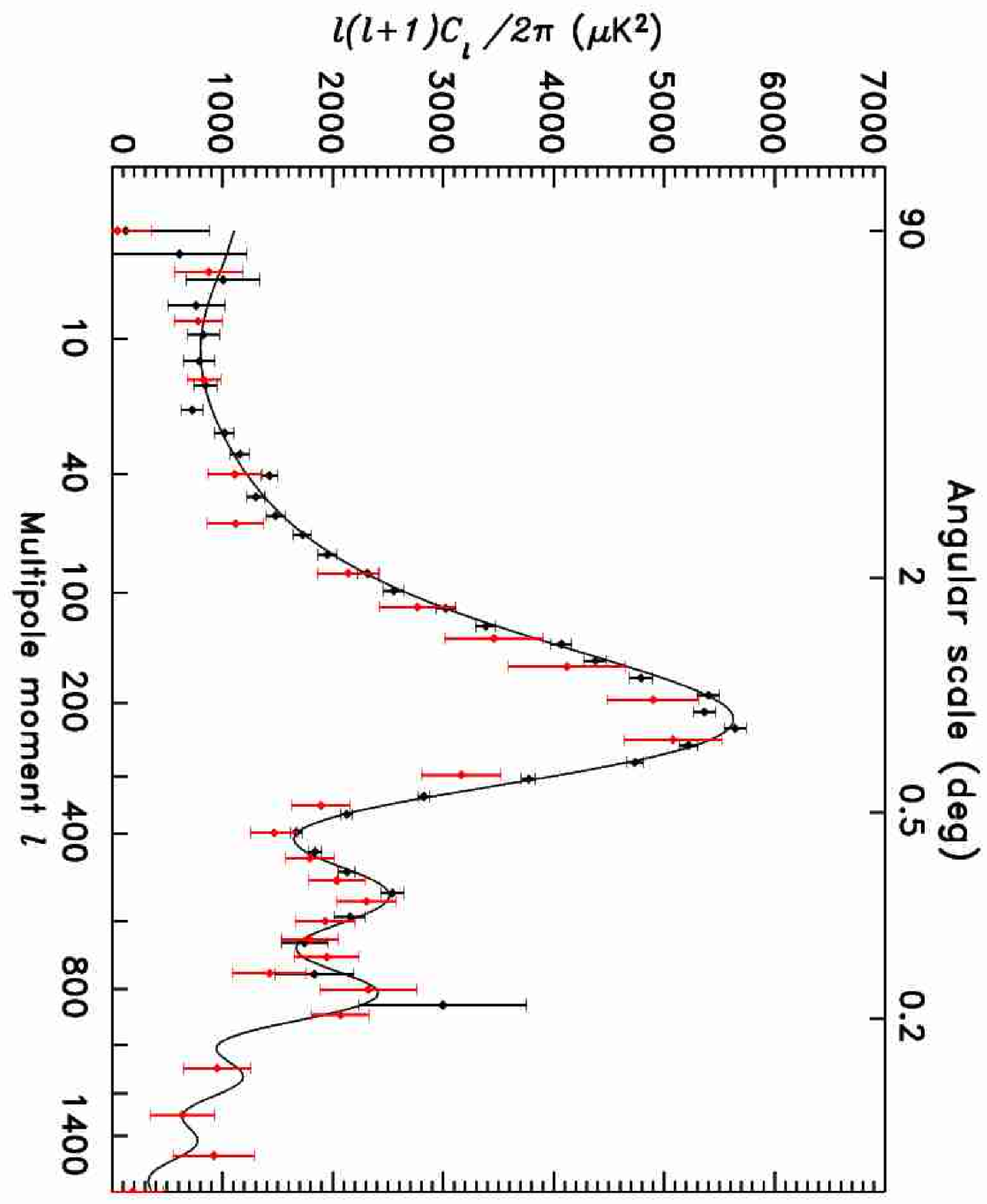}
\hfill 
  \includegraphics[height=4cm]{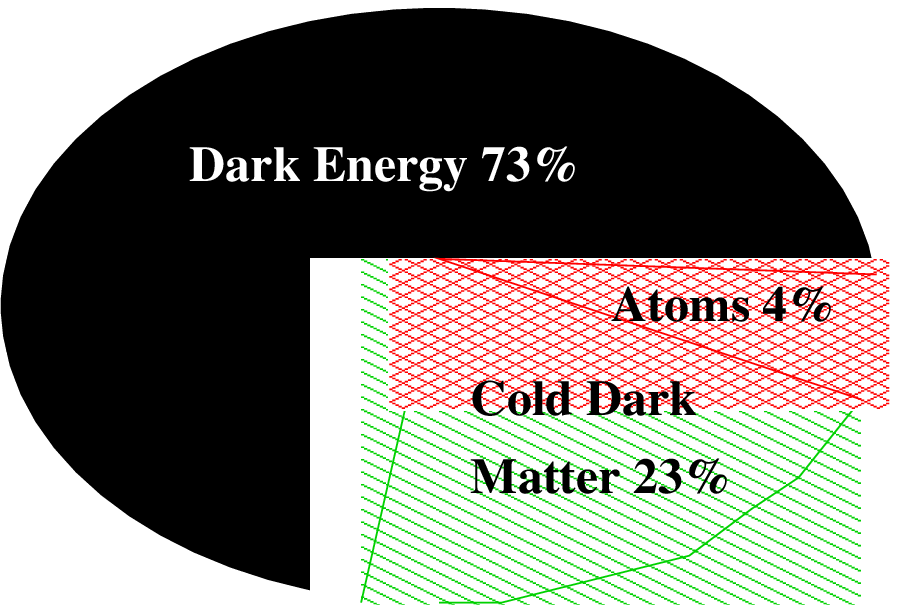} 
\caption{Recent observational evidence for Dark Energy of the Universe 
({\em upper left figure}: evidence from SnIa (Ref. [54]), {\em upper right figure}: 
evidence from CMB measurements (Ref. [55])) 
and a pie graph ({\em lower central figure}) of the energy budget of our world 
according to these observations.}
\end{figure}

One of the reasons that make me prefer the Violation of CPT 
via the \$ matrix decoherence
approach over other approaches to CPT Violation,
concerns a novel type 
of CPT Violation\index{CPT Violation}{\it at a global scale}, 
which may characterize
our Universe. This has been proposed in ref.~\cite{cosmonem}, and 
was given the name 
{\it cosmological CPT Violation}\index{cosmological CPT Violation}.
This type of CPTV is prompted by recent 
astrophysical Evidence for the existence of a Dark Energy\index{Dark Energy} 
component of the Universe. For instance, there is direct evidence 
for a current era acceleration of the Universe, based on measurements
of distant supernovae SnIa~\cite{snIa}, which is supported also by 
complementary observations on Cosmic Microwave Background (CMB) 
anisotropies (most spectacularly by the recent data of WMAP\index{WMAP} 
satellite 
experiment)~\cite{wmap}. 

Best fit models of the Universe from such combined data 
appear at present consistent with 
a non-zero, positive 
{\em cosmological constant}\index{cosmological constant} $\Lambda \ne 0$. 
Such a $\Lambda$-universe will eternally accelerate, as it will enter 
eventually an inflationary\index{inflationary phase} (de Sitter) phase again, in which the scale 
factor will diverge exponentially   
$a(t) \sim e^{\sqrt{\Lambda/3}t}$,
$t \to \infty$. 
This implies that there 
exists a {\it cosmological horizon}. 

The existence of such horizons\index{horizons} implies incompatibility with 
a S-matrix: 
no proper definition of asymptotic state vectors is possible, 
 and there is always an 
environment of d.o.f. crossing the horizon.  
This situation  may be considered as dual to 
that of a black hole\index{black hole}, 
depicted in fig. \ref{loss}: in that case the 
asymptotic observer was in the exterior of the black hole horizon,  
in the cosmological case the observer is inside the horizon.
However, both situations are characterized 
by the lack 
of an invertible scattering matrix\index{scattering matrix}, 
hence the above-described 
theorem by Wald\index{Wald}\cite{wald} 
on \$-matrix and CPTV applies~\cite{cosmonem}, 
and thus CPT is violated, at a global scale,
due to a cosmological constant\index{cosmological constant} $\Lambda >0 $.

It has been argued in 
\cite{cosmonem} that such a violation is described effectively by a 
modified temporal evolution of matter in such a $\Lambda$-universe, 
which is given by 
\begin{equation}  
\partial_t \rho = [\rho, H] +  {\cal O}(\Lambda M_s)\rho 
\label{densityevol}
\end{equation} 
where $\Lambda$ is a dimensionless cosmological constant in four dimensions,
and $M_s$ is the quantum gravity scale (which may 
be different from the four-dimensional Planck scale, see discussion below).

This form has been derived in the above-described context 
of non critical strings.
Indeed, a de Sitter space\index{de Sitter space} 
time constitutes a non conformal string background,
and according to the ideas presented in the previous subsection
the temporal evolution of string matter in such a space time is
described by the decohering evolution 
(\ref{liouveq}). Since, as mentioned
previously, the main source of departure from conformal\index{conformal} 
symmetry comes from
the graviton $g_{\mu\nu}$ 
background, whose $\beta^{\mu\nu}=\Lambda g_{\mu\nu}$, one actually has
the evolution (\ref{liouveq2}), with 
the double commutaror structure for the background $g_{\mu\nu}$. 
The order of the decoherence parameter $\gamma$, then, in such a case
is: 
\begin{equation}
\gamma \sim \Lambda M_s (\Delta g_{\mu\nu})^2,
\label{cosmodecohest}
\end{equation}
where $M_s$ is the string scale, and $\Lambda$ is a dimensionless
cosmological constant in the $d$-dimensional space time the string
propagates on. One may use the modern view point 
that our four dimensional world 
is actually a string membrane (D-brane\index{brane}), 
embedded in a ten-dimensional target (bulk) space. 
The Standard Model matter is localised on such brane worlds.
In the bulk,  
only fields belonging to the gravitational multiplet of the string spectrum 
are allowed to 
propagate. From this view point, then, the string scale $M_s$ may be 
different from the four dimensional Planck scale\index{Planck scale}. 
However, since string matter
is confined on the brane world, it essentially
interacts effectively only with the four-dimensional graviton fields,
that lie on, or cross, our brane world, 
and hence one arrives at the estimate (\ref{densityevol}), with 
$\Lambda$ the effective four-dimensional cosmological 
constant\index{cosmological constant} 
on the brane. 

An important 
issue concerns the order of the variance\index{variance} of the 
metric fluctuations\index{fluctuations} $(\Delta g_{\mu\nu})^2$. 
To arrive at the estimate
(\ref{densityevol}) one has to assume that such variances are of order one.
However, there are models of space time 
foam in string/membrane theory\cite{horizons,synchro}, 
where the foam is represented as a gas of D-particle
(point-like) defects on three branes, 
which recoil\index{recoil} upon interaction with matter strings.
As a result of recoil, there are induced space-time distortions,
of the form $g_{0i} \sim u_i $, where $i$ is a spatial three-brane index, 
and $u_i$ is the recoil velocity 
of the D-particle. By momentum conservation\index{momentum conservation}, 
$u_i \sim g_s \Delta k_i/M_s$,
where $\Delta k_i$ is the momentum transfer, which is of order of the 
incident momentum, $k$,   
$g_s$ is the (weak) string coupling, and $M_s/g_s$ is the mass of 
the D-particle. Upon summing world-sheet genera\index{world-sheet genera}, 
$u_i$ becomes an operator,
which acts on energy eigenstates\index{energy eigenstates}, 
yielding appropriate eigenvalues of order $g_s \Delta k_i/M_s$.

Considering the case of a two state system, say neutrinos\index{neutrino} oscillating 
between two energy states, with the 
corresponding energy difference arising from a mass-squared difference
$\Delta m^2$ in the neutrino Hamiltonian $H \simeq p + m^2/p + \dots $, 
one has for the model of \cite{horizons}:
\begin{equation}
(\Delta g_{0i})^2 \sim (g_s\Delta E/M_s)^{2} = g_s^2 (\Delta m^2)^2/E^2M_s^2
\label{gravuncert}
\end{equation}
where $E$ is 
the energy of the low-energy neutrino 
interacting with the foam\index{foam}. 

From (\ref{gravuncert}) and (\ref{liouveq}), then, we obtain
the order of the decoherence\index{decoherence}  coefficient
for this case:
\begin{equation}
\gamma \sim \Lambda g_s^2(\Delta m^2)^2/E^2M_s
\label{equiv}
\end{equation}
Comparing with (\ref{estimate1}) we observe that it is of the same form as 
in the energy-driven decoherence model of \cite{houghston,adler},
provided the decoherence coupling with the environment 
is of order $\lambda^2 \sim g_s^2\Lambda /M_s$. In fact, one gets exactly
the result (\ref{estimate1}), if one identifies $M_s/g_s = M_P$, and 
assumes a $\Lambda \sim 1/g_s$, which could be induced by quantum 
string loop effects (but, of course, this is too big for 
a realistic cosmological constant\index{cosmological constant}). 
The equivalence with energy 
driven decoherence of the D-particle 
foam model should not have come as a surprise, given that the space-time
distortion due to the recoil\index{recoil} 
of the D-particle is driven by the energy content 
of the matter probe, 
on account of energy conservation\index{energy conservation}. 

For realistic values of $\Lambda \sim 10^{-122}$ in Planck units, 
the above effects are undetectable in any oscillation experiment. 
Although the order of the
cosmological CPTV effects in this scenario is tiny, 
if we accept that the Planck scale\index{Planck scale} 
is the ordinary four-dimensional one 
$M_P \sim 10^{-19}$ GeV, and hence undetectable in direct particle
physics interactions, 
however, such cosmological-constant induced CPTV may have already 
been  
detected indirectly through the (claimed) observational 
evidence for a current-era acceleration of the Universe
\index{acceleration of the Universe}! Of course, 
the existence of a cosmological constant brings up 
other interesting challenges, such as the possibility of  
a proper quantization of de Sitter space as an open system, 
which are still unsolved.

At this point I should mention that 
time Relaxation\index{Relaxation} models 
for Dark Energy\index{Dark Energy}, e.g. 
quintessence model\index{quintessence models}, where
eventually the vacuum energy asymptotes (in cosmological time) 
an equilibrium zero value, are still currently 
compatible with the data~\cite{relax}. In such cases it might be possible that 
there is no cosmological CPTV, since a proper S-matrix can be defined,
due to lack of cosmological horizons. 

From the point of view of string theory the impossibility of defining a 
S-matrix in de Sitter space\index{de Sitter space} times
is very problematic, because critical strings by their very definition
depend crucially on such a concept. However, this is not the case of 
non-critical string theory, which can accommodate
in their formalism $\Lambda$ universes~\cite{cosmonem}.
It is worthy of mentioning briefly that such non-critical (non-equilibrium) 
string theory cases are capable of accommodating models with large 
extra dimensions, in which the string
gravitational scale\index{string mass scale}  $M_s$ is not 
necessarily the same as the Planck scale\index{Planck scale} 
$M_P$, but it could be much 
smaller, e.g. in the range of a few TeV.
In such cases, the CPTV effects in (\ref{densityevol}) may be much larger,
since they would be suppressed by $M_s$ rather than $M_P$. 

It would be interesting to study further the cosmology\index{cosmology} of such models
and see whether the global type of CPTV proposed in \cite{cosmonem}, 
which also entails primordial\index{primordial} CP violation of similar order, 
distinct from the ordinary (observed) CP violation which 
occurs at a later stage in the evolution of the Early Universe\index{Early Universe}, 
may provide a realistic explanation of the initial 
matter-antimatter\index{cosmic matter-antimatter asymmetry} asymmetry 
in the Universe, and the fact that antimatter is 
highly suppressed today. In the standard CPT invariant 
approach this asymmetry is supposed to be 
due to ordinary CP Violation\index{CP Violation}. 
In this respect, I mention that 
speculations about the possibility that 
a primordial CPTV space-time foam is responsible
for the observed matter-antimatter asymmetry in the Universe 
have also been put forward in \cite{ahluwal} but from a different
perspective than the one I am suggesting here. In ref. \cite{ahluwal}  
it was suggested that a novel CPTV foam\index{foam}-induced phase difference 
between a space-time spinor and its antiparticle 
may be responsible for the required asymmetry. 
Similar properties of spinors\index{spinors} may also characterize space times 
with deformed 
Poincare symmetries~\cite{agostini}, which may also be viewed as 
candidate models of quantum gravity\index{quantum gravity}.
In addition, other attempts to discuss the origin of such an asymmetry 
in the Universe 
have been made within the 
loop gravity approach to quantum gravity~\cite{singh} 
exploring Lorentz Violating 
modified dispersion relations\index{dispersion relations} for matter probes, 
especially neutrinos\index{neutrino}, 
which we shall discuss below.

\section{Phenomenology of CPT Violation} 

\subsection{Order of Magnitude Estimates of CPTV} 

Before embarking on a detailed 
phenomenology\index{Phenomenology of quantum gravity} of CPTV it is worth asking whether such a task is really sensible, in other words how feasible it is to detect 
such effects in
the foreseeable future. 
To answer this question
we should present some estimates of the expected effects in some 
models of quantum gravity. 

The order of magnitude of the CPTV effects 
is a highly model dependent issue, and it depends 
crucially on the specific way CPT is violated in a
model. As we have seen  
cosmological (global) CPTV effects are tiny, 
on the other hand, quantum Gravity   
(local) space-time effects (e.g. space time foam) may be much larger.

Naively, Quantum Gravity (QG) has a dimensionful constant:
$G_N \sim 1/M_P^2$, $M_P =10^{19}$ GeV. Hence, CPT violating 
and decohering 
effects may be expected to be suppressed
by  $E^3/M_P^2 $ , where $E$ is a typical energy scale of the low-energy 
probe. This would be practically undetectable in neutral 
mesons\index{neutral mesons},
but some neutrino 
flavour-oscillation experiments (in models where flavour symmetry\index{flavour symmetry}
is broken by quantum gravity), or some 
cosmic neutrino future observations 
might be sensitive to this order: 
for instance, in models with LV, one expects 
modified dispersion relations (m.d.r.) which could yield significant effects
for ultrahigh energy ($10^{19}$ eV)
$\nu$ from Gamma Ray Bursts\index{Gamma Ray Burst} (GRB)~\cite{volkov}, 
that could be close to observation. 
Also in some astrophysical cases, e.g. 
observations of synchrotron radiation
\index{synchrotron radiation} from 
Crab Nebula or Vela Pulsar, one is able to  
constraint electron m.d.r. almost near this 
(quadratic) order~\cite{synchro}. 

However, resummation and other effects 
in some theoretical 
models may result in much larger CPTV effects of order: 
 $\frac{E^2}{M_P}$. This happens, e.g., in some loop gravity 
models~\cite{loop}, or in 
some (non-critical) stringy models of quantum gravity involving 
open string excitations~\cite{emn}.
Such large effects may  already be accessible in current experiments,
and most of them are excluded by current observations. 
Indeed, 
the Crab nebula synchrotron constraint~\cite{crab} for instance already 
excludes such effects
for electrons. Nevertheless, similar effects for photons are still escaping 
exclusion at present, and in 
view of possible 
violations of the equivalence principle\index{equivalence principle}, 
which might occur
in some theoretical models of foam\cite{synchro}, according to which only
photons are susceptible to such QG-induced m.d.r.,
the last word on minimal suppression QG effects has not been spoken yet.

On the other hand, as we discussed previously, 
in some models of decoherence~\cite{adler} one may have single Planck 
mass suppression, $1/M_P$, however the decoherence
parameters $\gamma$ depend on the energy variance, rather than 
the average energy of the probe, $\Delta E =E_2 - E_1$
between, say, the two energy eigenstates\index{energy eigenstates} 
of a two-state 
system, such as neutral kaon, or two-generation neutrino 
oscillations\index{neutrino oscillations} in hierarchical neutrino models, $\gamma \sim (\Delta E)^2/M_P$.
This will also be undetectable in oscillation experiments in the 
foreseeable future, despite the minimal Planck scale suppression of the 
effect in this case. 

From the above discussion 
it is therefore clear that we are in need 
of guidance by experiment in our quest for 
the order of decoherence or other non-trivial 
quantum gravity effects, since theoretically 
the situation is far from being resolved. 
Since very little is known 
about such models, 
it is important to obtain as much experimental information on bounds 
of the relevant parameters as possible. 
Hopefully, this will help us focusing 
our future research in the phenomenology of quantum gravity
on the right track.

\subsection{Mnemonic Cubes for CPTV Phenomenology}

\begin{figure}[ht]
\centering
  \includegraphics[height=6cm]{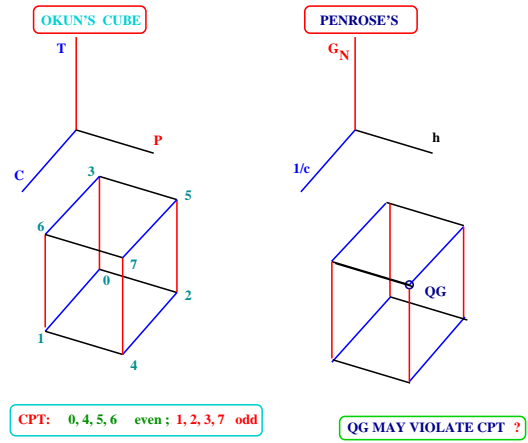}
\caption{Mnemonic cubes for CPT Violation: \underline{Left} its phenomenology.
\underline{Right}: its possible theoretical origin.}
\label{mnemonic}
\end{figure}

When CPT is violated there are many possibilities, due to the fact that 
C,P and T may be violated individually, with their violation 
independent from one another.
This was emphasized by Okun~\cite{okun} some years ago, who presented a set of 
mnemonic rules for CPTV phenomenology\index{Phenomenology of quantum gravity}, which are summarized in 
fig.~\ref{mnemonic}.
In this figure I also draw a kind of 
Penrose cube, indicating where the violations 
of CPT may come from. 
The diagram has to be interpreted as follows: CPTV may come 
from violations of special relativity (axis $1/c$), 
where the speed of light does not have its value, 
exhibiting some sort of refractive index {\it in vacuo}\index{refractive index in vacuo}, or  
from departure from quantum mechanics (axis $h$), 
or 
from gravity considerations,
where the gravitational constant departs from its value (axis $G_N$), 
or finally (and most likely)  
from quantum gravity\index{quantum gravity} considerations,  where all such effects may 
coexist.

\subsection{Lorentz Violation and CPT: The Standard Model Extension (SME)} 

We start our discussion on phenomenology of CPT violation\index{CPT Violation} 
by 
considering CPTV models 
in which requirement (iii) of the CPT theorem\index{CPT theorem} 
is violated, that 
is Lorentz invariance. As mentioned previously, 
such a violation may be a consequence
of quantum gravity fluctuations. 
In this case Lorentz\index{Lorentz Violation} symmetry is violated (LV) and hence CPT,
but there is no necessarily quantum decoherence or unitarity loss. 
Phenomenologically, at low energies, 
such a LV will manifest itself as an extension of the standard 
model in (effectively) flat space times, whereby LV terms will be introduced
by hand in the relevant lagrangian, with coefficients whose magnitude
will be bounded by experiment~\cite{kostel}. 

Such SME\index{Standard Model Extension} 
lagrangians may be viewed as the low energy limit of some 
string theory vacua, in which  
some tensorial fields acquire non-trivial vacuum expectation values 
$<A_\mu>  \ne 0~, <T_{\mu_1 \dots \mu_n}>  \ne 0.$
This implies a 
{\it spontaneous}\index{Spontaneous Breaking} 
breaking of Lorentz symmetry by  
these (exotic) string vacua~\cite{kostel}.

The simplest phenomenology of CPTV in the context of SME 
is done by studying the physical consequences of a modified Dirac\index{Dirac} 
equation 
for charged fermion fields in SME. 
This is relevant for phenomenology using data from the recently 
produced antihydrogen\index{antihydrogen} factories~\cite{antihydro,mavroyoko}. 

In these lectures I will not cover this part in detail, 
as I will concentrate mainly 
on neutrinos within the SME context. It suffices to mention that 
for free hydrogen  
$H$  (anti-hydrogen ${\overline H}$) one may consider 
the spinor $\psi$  representing electron  
(positron)  with charge $q=-|e| (q=|e|)$ 
around a proton (antiproton\index{antiproton})   
of charge $-q$, which obeys the modified Dirac equation (MDE):   
\begin{eqnarray}
&& ( i\gamma^\mu D^\mu - M -  
a_\mu \gamma^\mu - b_\mu \gamma_5 \gamma^\mu - \nonumber \\
&& -\frac{1}{2}H_{\mu\nu}\sigma ^{\mu\nu}
+  ic_{\mu\nu}\gamma^\mu D^\nu + id_{\mu\nu}\gamma_5\gamma^\mu D^\nu  )\psi =0 \label{esem}
\end{eqnarray} 
where $D_\mu = \partial_\mu - q A_\mu$, and $A_\mu = (-q/4\pi r, 0)$ is the 
Coulomb  potential. CPT \& Lorentz violation is described by terms
with parameters $a_\mu~, b_\mu~,$ while 
Lorentz violation only is described by the terms
with coefficients $c_{\mu\nu}~, d_{\mu\nu}~,  H_{\mu\nu} $.  

One can perform spectroscopic\index{spectroscopic measurements} 
tests on free and magnetically trapped molecules,
looking essentially for transitions that were forbidden if CPTV and SME/MDE 
were not taking place. 
The basic conclusion is that for sensitive tests of CPT in antimatter 
factories frequency resolution in spectroscopic measurements has to 
be improved down to a range of 
a 1 mHz, which at present is far from being achieved~\cite{mavroyoko}.

\begin{figure}[ht]
\centering
\includegraphics[height=10cm]{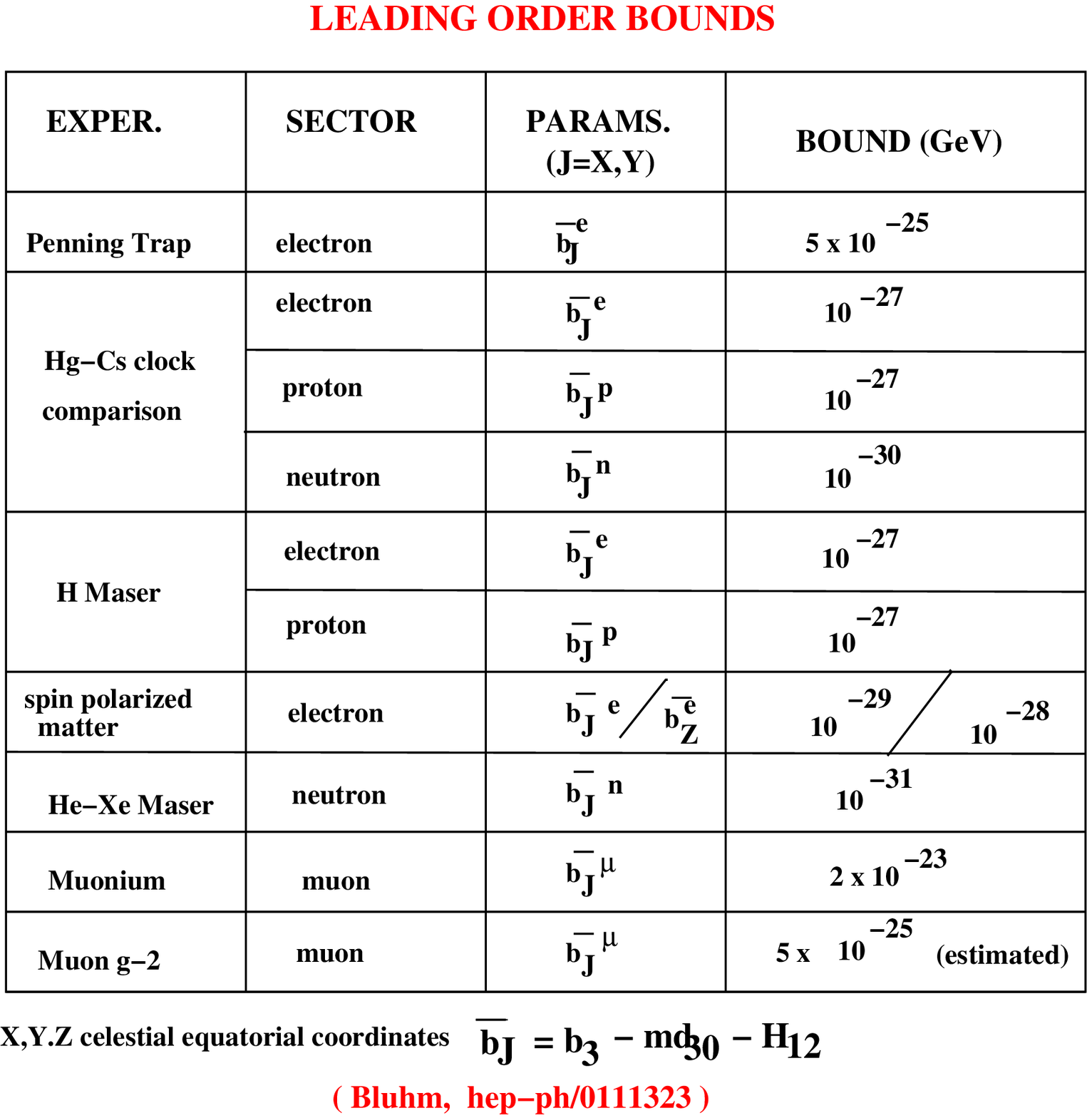}
\caption{Sensitivities of CPTV and LV parameters appearing in 
SME/Modified Dirac equation for charged probes,  
from various atomic and nuclear physics experiments.}
\label{bluhm}
\end{figure}

Since the presence of LV interactions in the SME affects dispersion relations
of matter probes, other interesting precision tests of such extensions
can be made in atomic and nuclear physics experiments, exploiting the fact
of the existence of a preferred frame\index{preferred frame} where observations take place. 
The results and the respective sensitivities of the various parameters
appearing in SME are summarized in the table of figure \ref{bluhm},
taken from ref. \cite{bluhm}. As we see, 
the frame dependence of such LV effects 
leads to very stringent bounds of the values of the LV parameters 
in some cases,
which are far more superior than the corresponding bounds 
obtained at present in antihydrogen factories.

\subsection{Direct SME Tests and  Modified Dispersion relations (MDR)}

Many LV Models of Quantum Gravity (QG) predict modified dispersion 
relations\index{dispersion relations} 
relations (MDR) for matter probes, including  
neutrinos $\nu$~\cite{emnnature,volkov,grb}.
This leads to one important class of experimental tests using $\nu$: 
each mass eigenstate of $\nu$ 
has QG deformed dispersion relations, which 
may, or may not, be the same for all flavours: 
\begin{eqnarray} 
E^2 = {\vec k}^2 + m_i^2 + f_i(E,M_{qg},{\vec k})~, \qquad 
{\rm e.g.}~ 
f_i = \sum_{\alpha =1,2 ...} C_{\alpha} {\vec k}^2 \left(\frac{|{\vec k}|}{M_{qg}}\right)^\alpha
\label{mdrnu}
\end{eqnarray}
There are stringent bounds on $f_i$ from oscillation experiments, as 
we shall discuss below. 

It must be stressed that such MDR also characterize SME, although the 
origin of MDR in the 
approach of \cite{emnnature,volkov,grb} is due to an induced 
non-trivial microscopic curvature of space time as a result of a 
back reaction of matter interacting with a stringy space time foam vacuum.
This is to be contrasted with 
the SME approach~\cite{kostel}, 
where the analysis is done 
exclusively on flat Minkowski space times, at a
phenomenological 
level. 

In general there are various experimental tests that can set bounds
on MDR parameters, which can be summarized as follows: 

(i) astrophysics tests - arrival time fluctuations for photons 
(model independent analysis of astrophysical GRB data~\cite{grb}

(ii) Nuclear/Atomic\index{atomic} Physics precision measurements (clock comparison, 
spectroscopic\index{spectroscopic measurements} tests on free and trapped 
molecules, quadrupole moments {\it etc})~\cite{bluhm}.

(iii)antihydrogen\index{antihydrogen} factories (precision spectroscopic tests 
on free and trapped 
molecules: e.g. $1S \to 2S$ forbidden transitions)~\cite{mavroyoko}, 

(iv) Neutrino mixing and spin-flavour conversion\index{spin-flavour conversion}, 
a brief discussion of which we now turn to.

\subsection{Neutrinos and SME }

The SME formalism naturally includes the neutrino sector. 
Recently a 
SME-LV+CPTV\index{Standard Model Extension} 
phenomenological model for neutrinos has been given 
in \cite{mewes}. The pertinent lagrangian terms are given by: 
\begin{eqnarray} 
&&{\cal L}^\nu_{SME} \ni 
\frac{1}{2}i{\overline \psi}_{a,L} \gamma^\mu D_\mu \psi_{a,L}  
- (a_L)_{\mu ab}  
{\overline \psi}_{a,L} \gamma^\mu \psi_{b,L}  
+ \nonumber \\&& \frac{1}{2}i  (c_L)_{\mu\nu ab}
 {\overline \psi}_{a,L} \gamma^\mu D^\nu \psi_{b,L}
\label{smenulagr}
\end{eqnarray} 
where $a,b$ are flavour indices. The model has (for simplicity)  
no $\nu$-mass differences. 
Notice that the presence of LV induces directional dependence
(sidereal effects)!

To analyze the physical consequences of the model, one passes to an 
Effective Hamiltonian~\cite{mewes} 
\begin{eqnarray} 
(H_{\rm eff})_{ab} = |{\vec p}|\delta_{ab} + \frac{1}{|{\vec p}|}
( (a_L)^\mu  p_\mu  -  (c_L)^{\mu\nu}
p_{\mu}p_{\nu})_{ab} 
\label{effham} 
\end{eqnarray} 
Notice that $\nu$ oscillations are now controlled by
the (dimensionless) quantities $a_L L$ \& $c_L L E$ 
where L is the oscillation length\index{oscillation length}. 
This is to be contrasted 
with the conventional case, where the relevant parameter
is associated necessarily with a $\nu$-mass difference $\Delta m$: 
$\Delta m^2 L/E$. 

There is an important feature of the SME/$\nu$: 
despite CPTV, the oscillation \index{neutrino oscillations} 
probabilities are the same between 
$\nu$ and their antiparticles\index{antiparticle}, if there are no mass differences
between $\nu$ and ${\bar \nu}$:   
$P_{\nu_x \to \nu_y} = P_{{\bar \nu_x} \to {\bar \nu_y}}$. 

Experimentally, it is possible to 
bound LV+CPTV SME parameters in the neutrino sector 
with high sensitivity, if we use data 
from high energy long baseline experiments~\cite{mewes}. Indeed, 
from the fact that 
there is   
no evidence for $\nu_{e,\mu} \to \nu_{\tau}$
oscillations, for instance, 
at $E \sim 100$ GeV ,  $L \sim 10^{-18}$ 
GeV$^{-1}$  we conclude that 
$a_L < 10^{-18}$ GeV, $c_L < 10^{-20}$. 

Similarly for an explanation of the LSND\index{LSND Experiment} anomaly~\cite{lsnd}, 
claiming evidence for oscillations 
between \index{antineutrino} (${\bar \nu_\mu}-{\bar \nu_e}$) 
but not for the corresponding neutrinos, a mass-squared difference of order 
 $\Delta m^2 = 10^{-19} $ GeV$^2=10^{-1}$ eV$^2$
is required, which implies that  
$a_L \sim 10^{-18}$ GeV,
$c_L \sim 10^{-17}$. This would affect other experiments, and in fact 
one can easily come to the conclusion that SME/$\nu$ does not offer a 
good explanation for LSND, if we accept the result of that experiment
as correct, which is not clear at present.   

\begin{figure}[ht]
\centering
  \includegraphics[height=4cm]{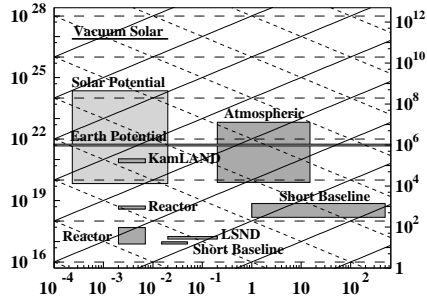}
\caption{Approximate experimental sensitivities for SME for neutrinos.
Lines of constant $L/E$ (solid),
$L$ (dashed), and $LE$ (dotted) are shown,
which give sensitivities to $\Delta m^2$, $a_L$, and $c_L$,
respectively.}
\label{tablenu}
\end{figure}

A summary of the Experimental Sensitivities for $\nu$'s SME 
parameters are given in the 
table of figure \ref{tablenu}, taken from \cite{mewes}.

\subsection{Lorentz non-invariance, MDR and $\nu$-oscillations}

Models of quantum gravity predicting MDR of the type (\ref{mdrnu})
for neutrinos~\cite{volkov,alfaro}, with a leading order $E^2/M_{qg}$
modification,  
can be 
severely constrained by a study of the induced oscillations 
between neutrino flavours, as a result of the departure from 
the standard dispersion\index{dispersion relations} relations provided that 
the quantum-gravity foam responsible for the MDR breaks
flavour symmetry, which however is not always the case~\cite{emnnu}.

\begin{figure}
\centering
  \includegraphics[height=4cm]{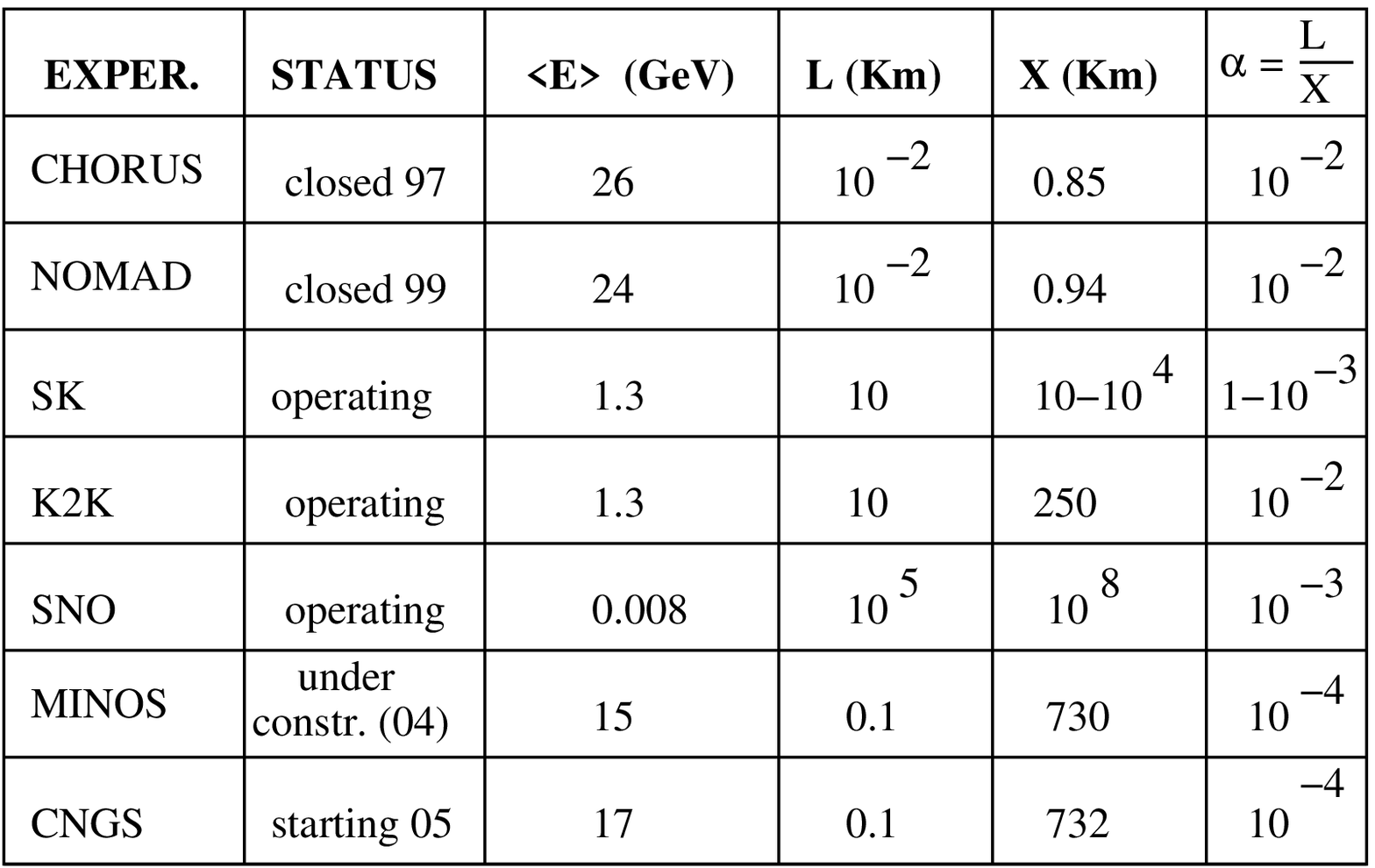}
\caption{Shown for each experiment are: (i) operation status, 
(ii) mean value of observed $\nu$ energy, (iii)  
the oscillation length $L$, (iv) typical $\nu$-flight distance $X$, 
and the ratio $\alpha = L/X$, which, in models where the foam 
induces $\nu$ flavour oscillations, coincides 
with the phenomenological parameter that controls the size of MDR effects.}
\label{brustein}
\end{figure}

This approach has been 
followed in \cite{eichler}, where it was shown that if flavour\index{flavour} symmetry
is not protected in such MDR models, then the 
extra terms in (\ref{mdrnu}), proportional to $E^2/M_{qg}$ 
will  
induce an oscillation 
length $L \sim  2\pi M_{qg}/(\alpha E^2)$, 
where $\alpha$ is a phenomenological 
parameter that controls the size of the effect. 
This should be contrasted 
to the Lorentz Invariant case where $L_{\rm LI} \sim 4\pi E/\Delta m^2$,
with $\Delta m^2 $ the square mass difference between neutrino flavours.
From a field theoretic view point, terms in MDR proportional to
some positive integer power of $E^2/M_{qg} $ may behave as 
non-renormalizable\index{non-renormalizable} 
operators, for instance, 
dimension five~\cite{myers} in the case of leading order 
QG effects suppressed only by a single power of $M_{qg}$. 

The sensitivity  
of the various neutrino oscillation experiments to the parameter
$\alpha$ is shown in figure \ref{brustein}~\cite{eichler}.  
The conclusion from such analyses, therefore, is that, 
if the flavour number symmetry 
is not protected in such MDR foam models with 
minimal $1/M_P$ suppression in the correction terms, then 
neutrino observatories and long base-line experiments should have 
already observed such oscillations. As remarked above, however,
not all foam models that lead to such MDR predict such 
oscillations~\cite{emnnu}, 
and hence such constraints are highly foam-model dependent.

\subsection{Lorentz Non Invariance, MDR and $\nu$ spin-flavor conversion}

An interesting consequence of MDR in LV quantum gravity theories is 
associated with modifications to the well-known phenomenon of 
spin-flavour conversion\index{spin-flavour conversion} in $\nu$ interactions~\cite{lambiase}. 
To be specific, we shall consider an example of a MDR for $\nu$ 
provided by a Loop Gravity\index{loop gravity} approach to quantum gravity.
According to such an approach, the dispersion relations for neutrinos 
are modified to~\cite{alfaro}:
\begin{eqnarray}  
E^2_{\pm} =  A_p^2  p^2 +  \eta  p^4 \pm 2\lambda  p + m^2 
\label{lgmdrnu}
\end{eqnarray} 
where $A_p = 1 + \kappa_1\frac{\ell_P}{{\cal L}}, ~\eta = \kappa_3\ell^2_P, ~\lambda =  \kappa_5 \frac{\ell_P}{2{\cal L}^2}$, 
and ${\cal L}$ is a characteristic scale 
of the problem, which can be either (i) $ {\cal L} \sim E^{-1} $, or (ii) ${\cal L}$=constant.

It has been noted in \cite{lambiase} that such a modification in the 
dispersion relation will affect the form of the spin-flavour conversion
mechanism. Indeed,  
it is well known through the Mikheyev-Smirnov-Wolfenstein (MSW) 
effect~\cite{msw} that 
Weak interaction Effects of $\nu$ propagating in a medium 
result in an 
energy shift $\sqrt{2}G_F(2n_e - n_n)$, where $n_e (n_n)$'s denote 
electron (neutron) densities. In addition to such effects, 
one should also take into account 
the interaction of $\nu$ with external magnetic 
fields, $B$,
via a radiatively induced magnetic moment\index{neutrino magnetic moment} $\mu$, corresponding to 
a term in the effective lagrangian: 
${\cal L}_{\rm int} = \mu {\overline \psi}\sigma^{\mu\nu}F_{\mu\nu}\psi$,
with $\psi $ the neutrino fermionic field. 

According to the standard theory, the equation for evolution describing 
the {\it spin-flavour} conversion phenomenon due to 
the above-described medium and magnetic moment effects for, say, two neutrino
flavours ($\nu_e, \nu_\mu$) is given by: 
\begin{eqnarray} 
i\frac{d}{dr} \left(\begin{array}{c} \nu_{e L} 
\\\nu_{\mu L}\\\nu_{e R}\\\nu_{\mu R}\end{array}\right) = {\cal H}\left(\begin{array}{c} \nu_{e L} \\\nu_{\mu L}\\\nu_{e R}\\\nu_{\mu R}\end{array}\right)~,
\label{spinflcon} 
\end{eqnarray}
where the effective Hamiltonian  
${\cal H}$ should be corrected in the loop gravity case 
to take into account $\lambda$-effects, associated with MDR 
(\ref{lgmdrnu}): 
{\small 
\begin{equation}
{\cal H} = 
\left(\begin{array}{cccc} -\frac{\Delta m^2}{4p}{\rm cos}2\theta 
-  \lambda  + \sqrt{2}G_Fn_e 
&\frac{\Delta m^2}{4p}{\rm sin}2\theta  &\mu_{ee}B  &\mu B \\
\frac{\Delta m^2}{4p}{\rm sin}2\theta 
&\frac{\Delta m^2}{4p}{\rm cos}2\theta 
-  \lambda  + \sqrt{2}G_Fn_e  
&\mu B  &\mu_{\mu\mu}B \\
\mu_{ee} B  &\mu B  
&-\frac{\Delta m^2}{4p}{\rm cos}2\theta 
+  \lambda  
&\frac{\Delta m^2}{4p}{\rm sin}2\theta \\
\mu B  &\mu_{\mu\mu}B  &\frac{\Delta m^2}{4p}{\rm sin}2\theta 
 &\frac{\Delta m^2}{4p}{\rm cos}2\theta + \lambda 
\end{array}\right) 
\label{mswhamilt}
\end{equation}
}
where $\mu \equiv \mu_{e\mu}$, $\Delta m^2 = m_2^2 - m_1^2$, 
and $B$ is the magnetic field. 
We should notice at this stage that 
the above formalism refers to Dirac $\nu$; for Majorana\index{Majorana} $\nu$ 
one should replace: $\nu_{i L} \to \nu_i$, $\nu_{i R}
\to {\overline \nu}_i$. Details can be found in \cite{lambiase}.

For our purposes we note that 
the Resonant Conditions for Flavour-Spin-flip are~\cite{lambiase}:
\begin{eqnarray} 
&& \nu_{e L} \to \nu_{\mu R}:  \quad 
2\lambda + \frac{\Delta m^2}{2p}{\rm cos}2\theta - \sqrt{2}G_Fn_e(r_{res}) =0
\nonumber \\
&& \nu_{\mu L} \to \nu_{e R}: \quad 
2\lambda - \frac{\Delta m^2}{2p}{\rm cos}2\theta - \sqrt{2}G_Fn_e(r_{res}) =0~.
\label{resonant}
\end{eqnarray} 
One can use the above conditions to obtain bounds for $\lambda,\kappa_i$
via the oscillation probabilities for spin-flavour conversion:
\begin{eqnarray} 
P_{\nu_{eL} \to \nu_{\mu R}}=\frac{1}{2}(1 - 
{\rm cos}2{\tilde \theta}{\rm cos}2\theta)~,
\label{probsfcon}
\end{eqnarray} 
where ${\rm tan}2{\tilde \theta}(r)=\frac{4\mu B(r)E}{|\Delta m^2|{\rm cos}2\theta - 
4E\lambda  + 2\sqrt{2}G_FEn_e(r)}$.

To obtain these bounds the author of \cite{lambiase} made 
the following physically relevant assumptions: 
(a) Reasonable profiles for solar $n_e \sim n_0 e^{-10.5r/R_\odot}$, 
$n_0=85N_A {\rm cm}^{-3}$. (b) Also: $\mu \sim 10^{-11}\mu_B$.  
Then, an 
upper bound on $\lambda$ is obtained of order: 
$\lambda \le \frac{1}{2}\left(10^{-12}e^{-10.5r_{res}/R_\odot}{\rm eV} +
\frac{|\Delta m^2|}{2E}\right)$. 

To obtain bounds on $\kappa$ we need to distinguish two cases: 

{\bf (I)}  \underline{${\cal L}$=universal constant}: 
In this case, we already know from photon
dispersion tests on GRB and \index{Active 
Galactic Nuclei} (AGN)~\cite{grb,alfaro} that   
 ${\cal L} \sim 10^{-18}$ eV$^{-1}$. 
Then, from best-fit solar $\nu$-oscillations induced by MSW, 
one may use experimental values of $\Delta m^2$, 
${\rm sin}^22\theta$, and obtain the following bound on 
$\kappa_i$:  $\kappa_5 < 10^{-25}$. 
From atmospheric\index{atmospheric} 
oscillations, in particular LSND experiment\index{LSND Experiment}~\cite{lsnd},
$\nu_\mu \to \nu_e$ fits the data with: $|\Delta m^2| \sim eV^2$,
${\rm sin}^22\theta \sim (0.2 - 3)\times 10^{-3}$, $E_{\rm max} \sim
10$ MeV, then  $\kappa_5 < 10^{-17}$. 

{\bf (II)} \underline{${\cal L} \sim p^{-1}$  a mobile  scale}: 
In that case,  
from SOLAR oscillations, with $p \sim 1-10 {\rm MeV}$ 
one gets $\kappa_5 ={\cal O}(1-100)$, which is a natural  
range of values from a quantum-gravity view point. 
From atmospheric oscillations, for the 
maximum $\nu$ $E \sim 10$ MeV, and ${\cal L} \sim E^{-1}$, one obtains 
 $\kappa_5 \sim 10^4$, which is a very weak bound. 

The conclusion from these considerations, therefore, is that the 
experimental data seem to 
favour case {\bf (II)}, at least from a naturalness point of view.

\subsection{$\nu$-flavour states and modified Lorentz Invariance (MLI)}

An interesting recent idea~\cite{blama}, 
which we would like to discuss now briefly, 
arises from the observation of the  
peculiar way in which flavour $\nu$ states experience 
Lorentz Invariance. 
Indeed, neutrino flavour\index{flavour} states 
are  {\it a superposition of mass eigenstates} 
 with standard dispersion relations of {\it different 
mass}. If one computes the expectation value 
of the Hamiltonian with respect to flavour states, e.g. in a 
simplified two-flavour scenario discussed in \cite{blama},
then one finds: 
\begin{eqnarray} 
E_e &=& <\nu_e|H|\nu_e> = \omega_{k,1}{\rm cos}^2\theta + \omega_{k,2}{\rm sin}^2\theta~, \nonumber \\
E_\mu &=& <\nu_\mu|H|\nu_\mu> = \omega_{k,2}{\rm cos}^2\theta + \omega_{k,1}{\rm sin}^2\theta~, 
\label{energystates}
\end{eqnarray} 
with $\theta$ the mixing angle. 

One has: $H|\nu_i>=\omega_i |\nu_i>$, $i=1,2$, where the 
$\omega_{k,i} = \sqrt{{\vec k}^2 + m_i^2}$ is a standard dispersion relation.
However, since 
the sum of two square roots in not in general a square root, 
one concludes that flavour states do not satisfy the 
standard dispersion relations. 
In general this poses a problem, as it would naively imply 
the introduction of a preferred frame, due to an apparent 
violation of the standard
linear Lorentz symmetry.

The idea of \cite{blama}, whose validity of course remains to be seen,
but which I find rather 
intriguing, and this is why I decided to include it in these lectures, 
is to avoid using preferred frames\index{preferred frame} 
by  introducing instead 
non-linearly  
modified Lorentz transformations to account for the modified dispersion 
relations of the flavour states. The idea is formally similar, but physically
very different, to the approach of \cite{nlls}, in which, in order to 
ensure observer independence of the Planck length, viewed as an ordinary 
length in quantum gravity, and not as a universal coupling constant, 
one has to modify non linearly the Lorentz transformations.
The result is that flavour states satisfy the following MDR: 
\begin{equation}
E_i^2 f_i^2(E_i) - {\vec k}^2 g_i^2(E_i) = M_i^2 \quad i=e,\mu
\label{mdrflavour}
\end{equation}
One can 
determine~\cite{blama} the 
$f_i(E_i,\theta, m_i),g_i(E_i,\theta,m_i), M_i(m_i,\theta)$ 
by comparing with 
$E_i=E_(\omega_i,m_i)$ above ((c.f. (\ref{energystates})). 

Then, in the spirit of \cite{nlls}, one can 
identify the non-linear Lorentz transformation that leaves the MDR
(\ref{mdrflavour})   
invariant:  $U \circ (E, {\vec k}) = (Ef,{\vec k}g)$.

\begin{figure}[ht]
\centering
  \includegraphics[height=4cm]{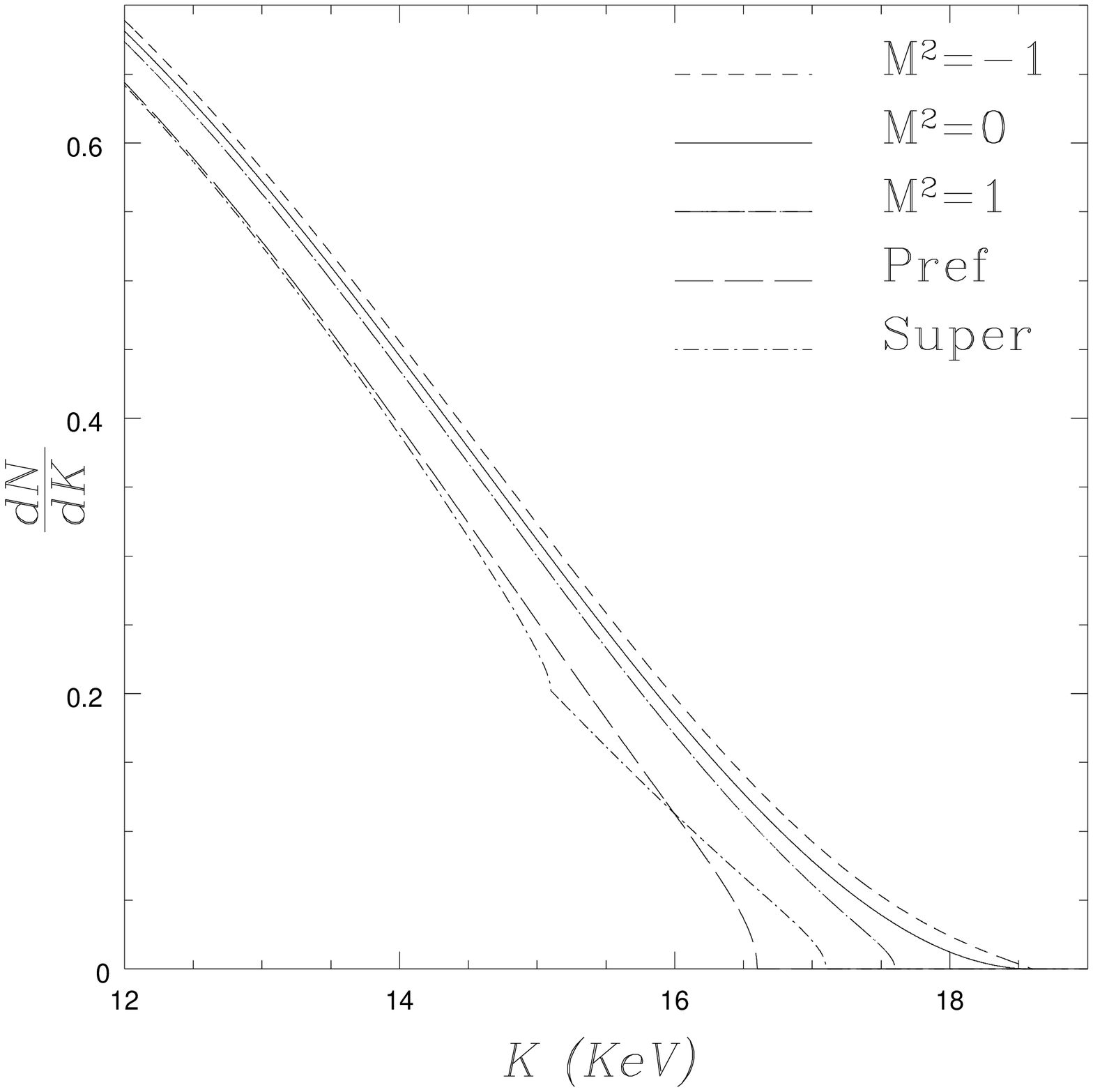}
\hfill  \includegraphics[height=4cm]{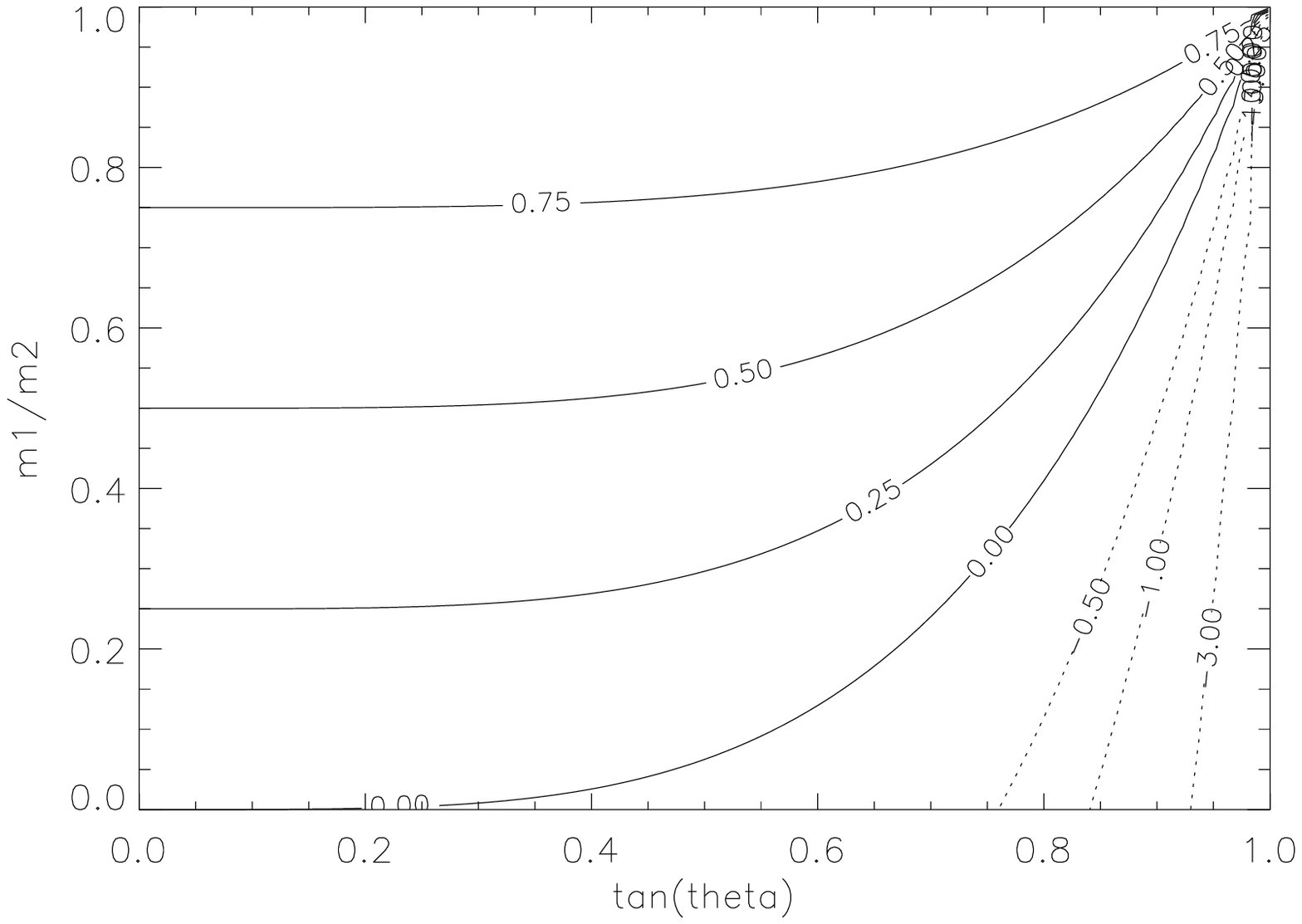}
\caption{{\underline{Left}: Tail of tritium 
$\beta$-decay spectrum, for massless $\nu$ (solid) 
and for LI flavour states (dashed and dot-long-dashed). Also plotted
is the 
preferred frame case. \underline{Right}: Likelihood Contours of $M^2$ (in units of $m_2^2$) upon which $\beta$-decay 
depends.}}
\label{bdecay}
\end{figure}

The interesting feature is that 
these ideas can be tested experimentally, e.g. in $\beta$-decay
experiments:
$N_1 \to  N_2 + e^- + {\bar \nu}_e $, where e.g. $N_1 =~^3{\rm H}$, 
$N_2=~^3{\rm He}$. 

Energy conservation in conventional $\beta$-decay implies: 
$E_{N_1} = E_{N_2} + E + E_e$, where $E$ is the energy of $e$, 
which would unavoidably introduce a preferred
frame. However, in the 
non-linear LI case for flavour\index{flavour} states, where 
the use of preferred frame is avoided, this relation 
is modified~\cite{blama}: $E_{N_1} = E_{N_2} + E + E_ef_e(E_e)$. 

These two choices are reflected in different predictions for the 
endpoint of the $\beta$-decay, that is the maximal kinetic energy 
the electron can carry (c.f. figure \ref{bdecay}). 
We refer the interested reader to \cite{blama} for further discussion
on the experimental set up to test these ideas. 

From the point of view of CPTV, which is our main topic of discussion
here, I must mention that in such non-linearly modified Lorentz symmetry
cases it is not clear what form the CPT theorem, if any, takes. 
This is currently 
under investigation~\cite{waldron}. In this sense, the link between CPTV 
and modified flavour-state dispersion\index{dispersion relations} 
relations, and therefore
the interpretation of the associated experiments from this viewpoint,
are issues which are not yet clear. 

\subsection{CPTV and Departure from Locality for Neutrinos}

\begin{figure}
\centering
  \includegraphics[height=6cm]{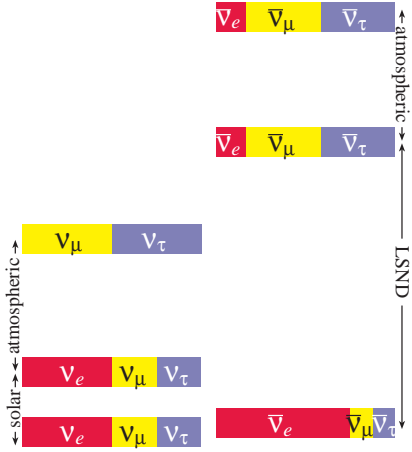}
 \caption{The CPTV neutrino spectrum 
proposed by Murayama-Yanagida to explain LSND. One needs 
$m^2_\nu - m^2_{\bar \nu} \sim 0.1~{\rm ev}^{-2} = 10^{-19}~{\rm GeV}^2$.}
\label{nuspectrum}
\end{figure}

As another way of violating CPT one can relax the requirement 
of {\it locality}\index{locality}. This 
idea has been pursued in~\cite{barenboim},
in an attempt to present a concrete model for 
CPT Violation\index{CPT Violation} for neutrinos, with CPTV Dirac masses, 
in an attempt to explain the LSND\index{LSND Experiment} anomalous 
results~\cite{lsnd}, according to which there is experimental 
evidence for oscillations in the antineutrino\index{antineutrino} sector, 
${\overline \nu}_e \to {\overline \nu}_\mu$, 
but not in the corresponding 
neutrino one.
In fact, the idea of invoking CPTV Dirac mass spectra for neutrinos
in order to account for the LSND results without invoking a 
sterile\index{sterile} 
neutrino 
is due to the authors 
of \cite{mura} (see figure \ref{nuspectrum}). However no concrete theoretical
model was presented there.

The model lagrangian of \cite{barenboim} reads: 
\begin{eqnarray} 
&& S = \int d^4x {\bar \psi} i\partial_\mu \gamma^\mu 
\psi + \frac{im}{2\pi} \int d^3x dt dt' {\bar \psi}(t) \frac{1}{t - t'}
\psi (t') 
\label{lagrangian}
\end{eqnarray} 
The on shell equations (in momentum space) for the (Dirac) spinors are: 
\begin{eqnarray} 
&&(p_\mu \gamma^\mu - m\epsilon (p_0) )u_\pm (p) =0~, 
\end{eqnarray} 
with $\epsilon (p_0)$ the sign~function, and
\begin{eqnarray} 
&&\psi_+(x) = u_+(p)e^{-ip\cdot x}, \qquad p^2 = m^2, ~p_0 > 0 \nonumber \\
&&\psi_-(x) = u_-(p)e^{-ip\cdot x}, \qquad p^2 = m^2, ~p_0 < 0 
\end{eqnarray} 
Notice that on-shell Lorentz invariance is maintained due to the presence of  
$(\epsilon (p_0)$) but \index{locality} is relaxed.  

As remarked in \cite{greenb}, however, the model of \cite{barenboim},
although respecting Lorentz symmetry on-shell\index{on-shell}, has 
correlation
functions (which are in general off-shell quantities) that 
do violate Lorentz symmetry, in the sense that they transform 
non covariantly under Lorentz transformations. Therefore, 
the CPTV
in this model is ultimately connected to LV. 

\begin{figure}
\centering
  \includegraphics[height=5cm]{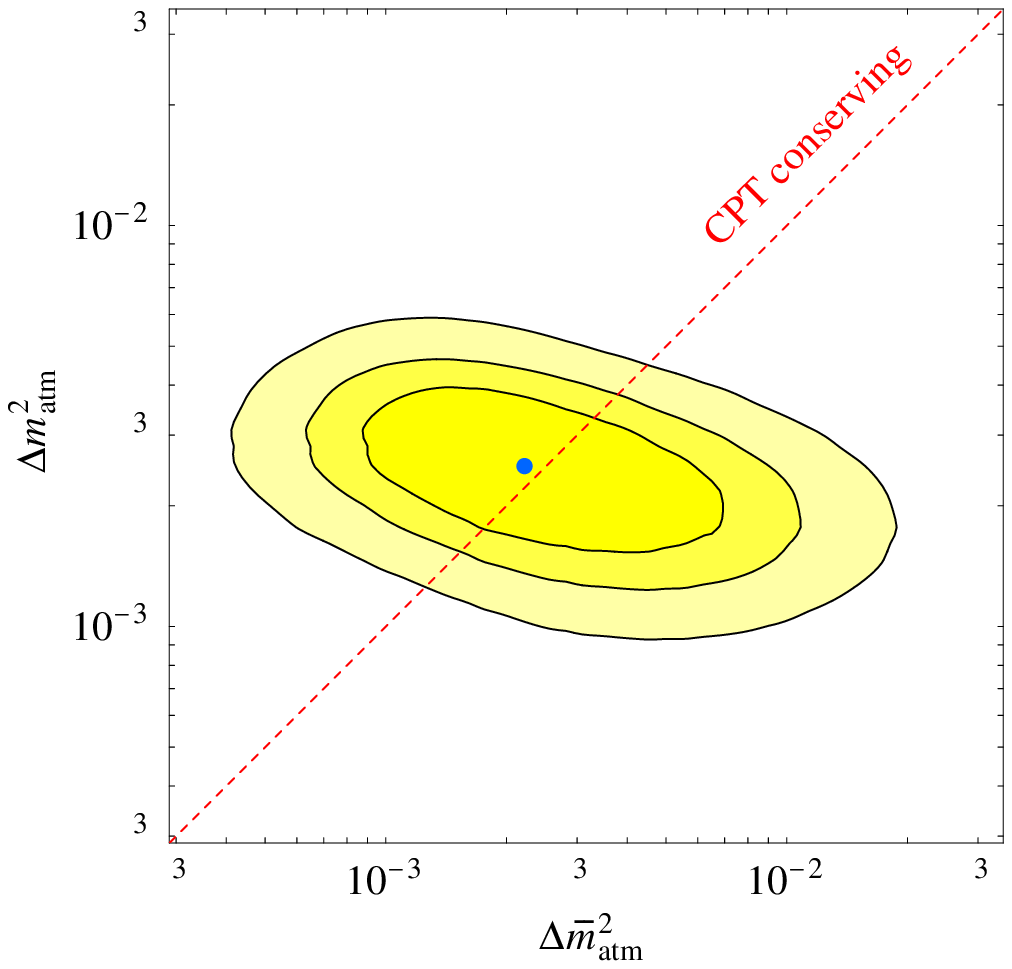}
\hfill
\includegraphics[height=5cm]{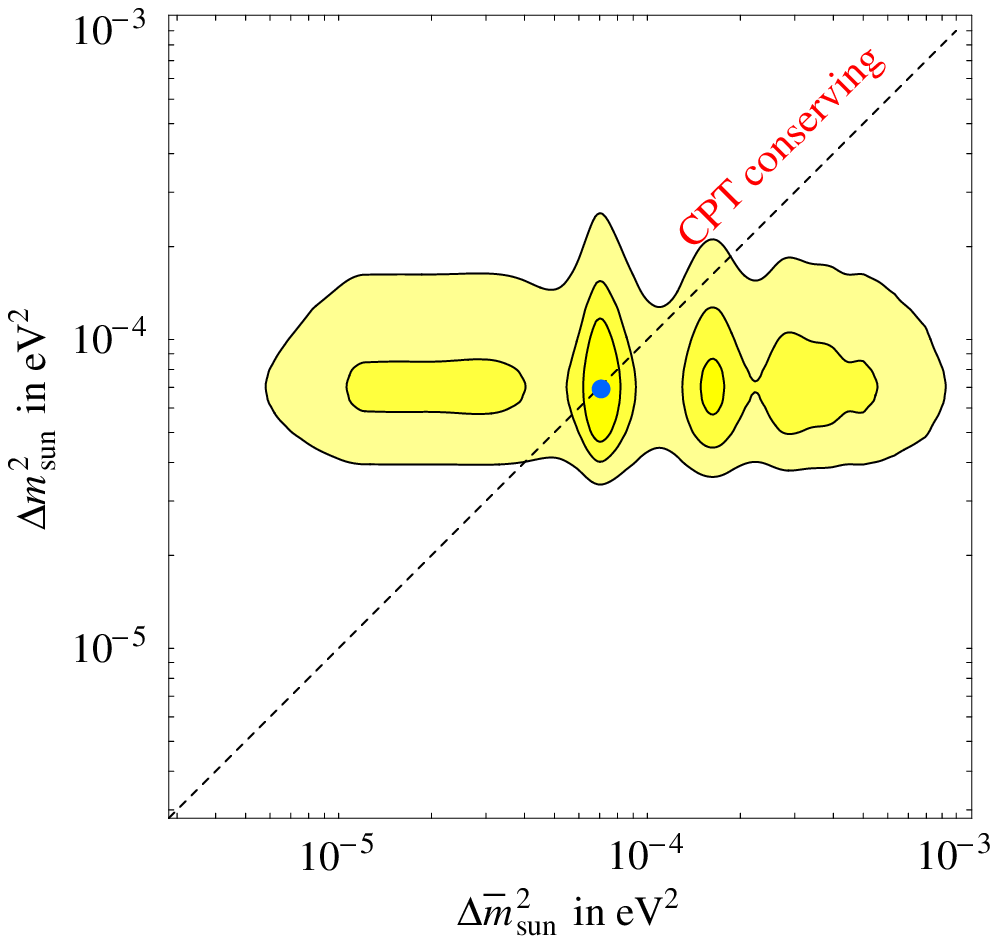}
\caption{\underline{Left:} Atmospheric $m_\nu - m_{\overline \nu}$ (68, 90, 99 \%, 2 d.o.f.). \underline{Right:} 
For solar \& reactor data (68, 90, 99 \%, 2 d.o.f.).}
\label{strumiafig} 
\end{figure}

The two-generation non-local model of \cite{barenboim} 
seems to be marginally disfavoured by the current neutrino data,
as claimed in \cite{strumia} (see figure \ref{strumiafig}).

\begin{table}\small
 \renewcommand{\arraystretch}{1.8}
$$\begin{array}{lc|cccc}
\multicolumn{2}{c|}{\hbox{model \& no.\ of free parameters}} & \Delta\chi^2& \hbox{mainly  incompatible with} &\hbox{main future test}\\ \hline
\multicolumn{2}{c|}{\hbox{ideal fit (no known model)}}& 0 &&?\\
\Delta L = 2\hbox{ decay }\bar\mu\to\bar e \bar\nu_\mu\bar\nu_e & 6 &  12 &\hbox{\sc Karmen}&\hbox{TWIST}  \\
3+1:~\Delta m^2_{\rm sterile} = \Delta m^2_{\rm LSND} & 9  & 6+9?     & \hbox{{\sc  Bugey} + cosmology?}  &\hbox{\sc MiniBoone}\\
\hbox{3 $\nu$ and CPTV~ (no $\Delta \bar m^2_{\rm sun}$)}& 10 &  15  & \hbox{KamLAND} & \hbox{\sc KamLAND}\\
\hbox{3 $\nu$ and CPTV~ (no $\Delta \bar m^2_{\rm atm}$)}& 10 &  25  & \hbox{SK atmospheric} & \hbox{$\bar\nu_\mu$ LBL?}\\
\hbox{normal 3 neutrinos}                             & 5  & 25        &\hbox{LSND}&\hbox{\sc MiniBoone} \\
2+2:~\Delta m^2_{\rm sterile} = \Delta m^2_{\rm sun}  & 9  & 30     & \hbox{SNO}& \hbox{SNO}  \\
2+2:~\Delta m^2_{\rm sterile} = \Delta m^2_{\rm atm}  & 9  & 50    & \hbox{SK atmospheric} & \hbox{$\nu_\mu$ LBL} \\
\end{array}$$ 
  \caption{Interpretations of solar, atmospheric and LSND data, ordered according to
the quality of their global fit.
A $\Delta \chi^2 = n^2$ roughly signals an incompatibility 
at $n$ standard deviations.}
\end{table}

A summary of data and interpretations of current models, including 
those which entail CPT violation is given in Table 1, taken  from
the first paper in \cite{strumia}. In that paper 
it has also been claimed that the recent WMAP\index{WMAP}~\cite{wmap} 
data on neutrinos 
seem to disfavour 3 + 1 scenaria
which conserve CPT\index{CPT} invariance. In my opinion one has to wait for 
future data from WMAP, before definite conclusions on this issue are reached,
given that the current WMAP data are rather crude in this respect. 
I will not go further into a detailed
discussion of this topic, as such summaries of neutrino data and their 
interpretations can be found in the literature, where 
I refer the interested 
reader~\cite{smirnov}.

Before closing this section, I would like to remark 
that most of the theoretical analyses 
for QG-induced CPTV 
in neutrinos have been done in simplified two-flavour oscillation models.
Including all three generations in the formalism 
may lead to differences in the corresponding 
conclusions regarding sensitivity (or conclusions about exclusion) 
of the associated CPTV effects. 
In this respect the measurements of the mixing angle\index{mixing angle} angle $\theta_{13}$
in the immediate future~\cite{minos}, as a way of detecting 
generic three-flavour effects,  
will be very interesting. In the current phenomenology, CPT invariance
is assumed for the theoretical estimates of this parameter~\cite{lindner}.

\subsection{Four-generation $\nu$ models with CPTV}

\begin{figure}
\centering
  \includegraphics[height=6cm]{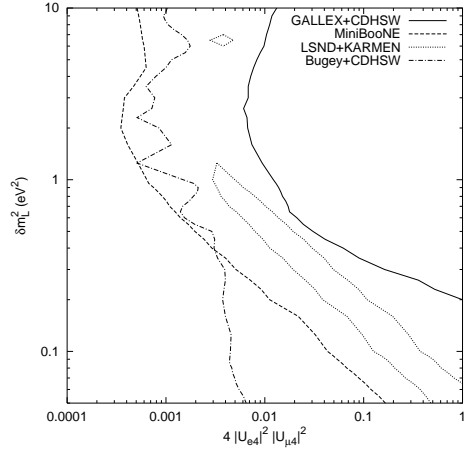}
\caption{Upper bound (solid) on the $\nu_\mu \to \nu_e$ oscillation
amplitude $4 |U_{e4}|^2 |U_{\mu4}|^2$ from the GALLEX limit on 
$|U_{e4}|$ and the CDHSW 
limit on $|U_{\mu4}|$ (90\%~C.~L.
results are used in both cases). The dot-dashed line is the 99\% C.~L. 
upper bound from Bugey and CDHSW if $CPT$ is conserved.
Also shown are the expected sensitivity (dashed) of
the MiniBooNE experiment  and, for comparison, the
allowed region (within the dotted lines) 
for $4 |\bar U_{e4}|^2 |\bar U_{\mu4}|^2$ from a
combined analysis of LSND and KARMEN data, both at 
the 90\% C.~L.}
\label{3+1}
\end{figure}

A natural question arises at this point, concerning 
($3 + 1$ or $2 + 2$)  $\nu$ scenaria which violate CPT symmetry. 
This issue has been studied recently in \cite{barger}. 
These authors postulated 
a model for CPTV with four generations\index{generation} for neutrinos 
which leads to different  
flavor mixing between $\nu$, ${\bar \nu}$:
$\nu_a = \sum_{i=1}^{4} U^*_{a i}\nu_i, \qquad 
{\bar \nu}_a = \sum_{i=1}^{4} {\bar U}_{a i}{\bar \nu}_i,$
with $U \ne {\bar U}$ due to CPTV. There are various cases
to be studied:

\begin{figure}
\centering
  \includegraphics[height=8cm]{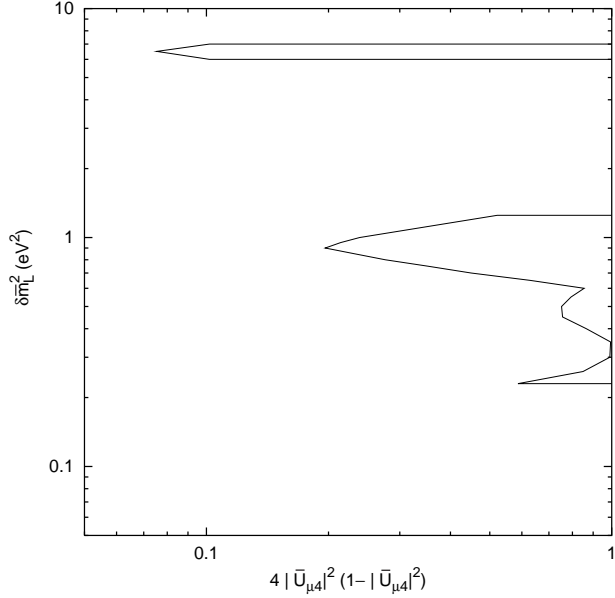}
 \caption{Lower bounds on 
$4 |\bar U_{\mu4}|^2 (1 - |\bar U_{\mu4}|^2)$
(the amplitude for atmospheric $\bar\nu_\mu$ survival at the LSND mass
scale) from the Bugey limit on $\bar\nu_e$ disappearance and the
$\bar\nu_\mu \to \bar\nu_e$ oscillation amplitude indicated by LSND and
KARMEN (90\%~C.~L. results are used in both cases).}
\label{3+1b}
\end{figure}

\begin{figure}
\centering
  \includegraphics[height=8cm]{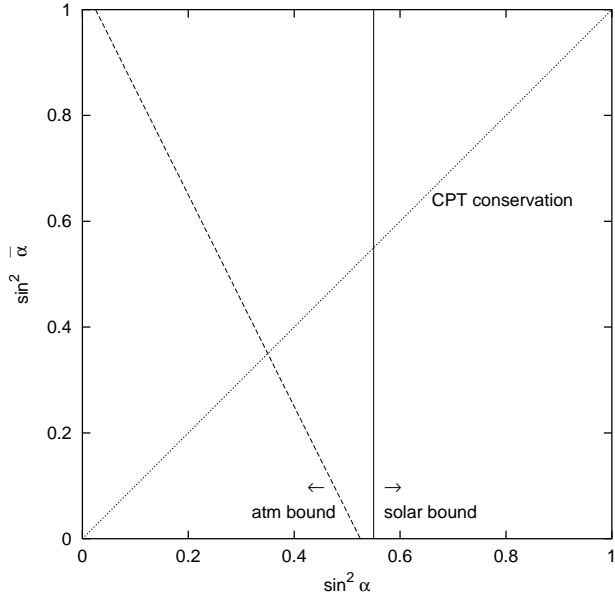}
 \caption{Constraints on sterile neutrino mixing angles $\alpha$ and
$\bar\alpha$ from \index{solar} (solid line) 
and \index{atmospheric} (dashed line) data.
The dotted line is the prediction if $CPT$ is conserved.}
\label{2+2}
\end{figure}

\begin{itemize} 
\item{} 3 + 1 models (see figs. \ref{3+1},\ref{3+1b}): one $\nu$ mass well separated from others,
sterile $\nu$ couples only to isolated state. The relevant 
Oscillation probabilities are: $P_{\nu_i \to \nu_i }(|U_{ij}|^2) \ne 
 P_{{\bar \nu}_i \to {\bar \nu}_i }(|{\bar U}_{ij}|^2)  $
  
Experimentally one may bound $|{\bar U}_{e4}|$ and $U_{\mu4}$ 
but there are no tight constraints for $|{\bar U}_{\mu 4}|$, 
$U_{e 4}$. This is to be contrasted with (3 + 1)$\nu$ CPT conserving models 
where $U = {\bar U}$. Hence (3 + 1)$\nu$ + CPTV seems still viable.

 \item{} 2 + 2 models (see fig. \ref{2+2}): sterile $\nu$ couples to solar and 
atmospheric $\nu$ oscillations.  This structure
is only permitted in ${\bar \nu}$ sector. Even with 
CPT Violation, however,  2+2 models are strongly disfavoured by data. 

\end{itemize}

Although the introduction of a fourth neutrino generation\index{generation}
with CPTV within conventional field theory
seems to be consistent with the current neutrino data,
however, there seems to be no concrete 
evidence for a forth generation from any experiment to date, 
including the most recent
astrophysical WMAP data, as we have seen above. This 
prompts one to examine alternative 
ways of explaining the current neutrino ``anomalous data'', like LSND\index{LSND Experiment},
employing 
unconventional ways of CPT Violation\index{CPT Violation} by means of quantum 
decoherence\index{decoherence}, which are in principle independent of mass differences
between particle and antiparticles. 

In the next subsection
we shall be dealing with this topic, reviewing first 
the relevant phenomenological formalism
which allows direct comparison with experiment.
I will start with the phenomenology of CPT violation 
in neutral mesons\index{meson} and neutrons\index{neutrons}, 
as a historical introduction to the general reader, 
and then I will proceed to the neutrino case.
I shall argue that minimal decoherence models with CPTV 
differences in the decoherence parameters between 
particle and antiparticle sectors, but not CPTV mass differences, 
can account for all existing neutrino data, 
including LSND results, 
without the need for enlarging the neutrino sector,
that is staying within three generation
models.

\subsection{CPTV through QG Decoherence: Neutral Mesons}

In this subsection I will  
discuss CPTV through decoherence, which is my preferred way of QG-induced
CPTV. As mentioned above, in this case the matter systems are viewed as 
open quantum mechanical or quantum-field theoretic 
systems interacting with a gravitational `environment', consisting
of degrees of freedom inaccessible to low-energy scattering experiments.
The presence of such an environment\index{environment} 
leads to modified quantum evolution\index{evolution},
which however is 
{\it not necessarily Lorentz Violating}~\cite{mill}. Thus, such an approach
to CPTV should in principle be studied separately, and indeed
it is possible for the 
CPTV decoherence effects to be disentangled experimentally 
from the LV ones, due to the
frame dependence of the latter. 

Currently, the most sensitive particle physics probes 
of such a modification from quantum 
mechanical behavior (often called `quantum mechanics violation'\index{Quantum Mechanics Violation} QMV~\cite{ehns,emn}) are: 
(i) neutral kaons\index{kaons} and B-mesons\index{B-mesons}~\cite{ehns,emn}  
and $\phi$-, B-factories\index{meson factories}~\cite{huet,benatti1,bernabeu} 
(ii) neutron interferometry\cite{ehns}, 
(iii) ultracold (slow) neutrons\index{neutrons} 
in Earth's gravitational field, 
and (iv) Neutrino\index{neutrino} 
flavour\index{flavour} mixing\index{mixing}, 
which is induced independently of 
masses and mass differences between neutrino species, as we shall 
discuss below.  In these lectures I will discuss briefly (i),(iii) and 
(iv).

Let us start with the neutral Kaon case. 
This is a typical two-state system of decoherence. One could follow the 
Lindblad\index{Lindblad} parametrization~\cite{benatti1}, in which the requirement 
of complete positivity\index{positivity} 
would imply a single decoherence\index{decoherence} parameter $\gamma$.
The requirement of energy conservation on the average in such models 
would then imply the double commutator structure (\ref{doublecomm}) 
for the decoherence term, which however would depend on the 
square of the energy variance\index{energy variance} between the two energy-eigenstates of the 
neutral Kaon system, as in (\ref{estimate1}). This would be too small to
be detected experimentally in neutral meson\index{meson} experiments and factories\index{meson factories}
in the foreseeable future. 

However, as argued in \cite{emnlarge}, 
complete positivity\index{complete positivity} may not be 
valid in generic models of quantum gravity, such as 
the non-critical string decoherence 
models (\ref{liouveq}). Indeed, in that case, the decoherence terms contain 
the Zamolodchikov metric $\langle V_i V_j \rangle $ 
and as such are non-linear in the probe 
state density matrix\index{density matrix} $\rho$, 
given that $\langle \dots \rangle = 
{\rm Tr}(\rho \dots )$  depends on it. Complete positivity for 
non-linear effective theories (e.g. 
Hartree-Fock\index{Hartree-Fock effective theories} type, 
mean field approaches)
is in general a non well defined concept~\cite{kunna}.

In fact, in the original parametrization\cite{ehns}  
of the QG-induced decoherence effects
for the neutral Kaon system  
this requirement has not been 
imposed. In such a paremetrization, which 
has also been followed in more recent, and more complete, 
phenomenological analyses of this system~\cite{emn,huet}, in addition to the 
basic principle of entropy increase, one also imposes 
the requirement of conservation of strangeness\index{strangeness} 
by the quantum gravity\index{quantum gravity} 
interactions. This follows from the so-called $\Delta S = \Delta Q$ rule
which seems to characterise the leading-order 
Kaon weak-interaction\index{weak-interaction} physics, 
which in general violates strangeness, 
but the charge transfer in the neutral Kaon physics is a much more 
subleading effect than the dominant CP violation\index{CP Violation} 
effects. This feature is assumed to be 
obeyed by the quantum gravity interactions~\cite{ehns},
which are thus assume to conserve strangeness\index{strangeness} 
to leading order.

According to our general discussion in section 2 on 
the dynamical-semigroup\index{dynamical semigroup} approach to decoherence\index{decoherence}, 
on which 
the formalism of ref.~\cite{ehns} is based, 
for the neutral-Kaon 
two-level system the non-Hamiltonian 
decoherence term in the
evolution equation for $\rho$ can be parametrized by a $4\times 4$ matrix
$\delta\H_{\alpha\beta} $, where the indices $\alpha, \beta, \dots$ enumerate
the Hermitian $\sigma$-matrices $\sigma _{0,1,2,3}$, which we represent in the
so-called $K_{1,2}$ basis, defined as 
$|K^{1,2}\rangle =\frac{1}{\sqrt{2}}
\left(|K^0\rangle \pm |\overline{K}^0\rangle\right) $. In Neutral Kaons\index{kaons},
the CP eigenstates are not energy (physical) eigenstates, thereby leading to
mixing. We refer the reader to the literature \cite{ehns,emn}
for details of this description, noting here the
following forms for the neutral kaon Hamiltonian
\begin{equation}
  H = \left( \begin{array}{cc}
  M - \coeff{i}{2}\Gamma - {\rm Re} M_{12} + \coeff{i}{2} {\rm Re} \Gamma _{12}
&  \coeff{1}{2}\delta M - \coeff{i}{4} \delta \Gamma
  -i {\rm Im} M_{12}  - \coeff{1}{2} {\rm Im} \Gamma _{12}  \\
  \coeff{1}{2}\delta M - \coeff{i}{4} \delta \Gamma
  + i {\rm Im} M_{12}  - \coeff{1}{2} {\rm Im} \Gamma _{12} &
  M - \coeff{i}{2}\Gamma + {\rm Re} M_{12} - \coeff{i}{2} {\rm Re} \Gamma _{12}
 \end{array}\right)
\label{nkham}
\end{equation}
in the $K_{1,2}$ basis, or
\begin{equation}
 H_{\alpha\beta}
 =\left( \begin{array}{cccc}  - \Gamma & -\coeff{1}{2}\delta \Gamma
& -{\rm Im} \Gamma _{12} & -{\rm Re}\Gamma _{12} \\
 - \coeff{1}{2}\delta \Gamma
  & -\Gamma & - 2{\rm Re}M_{12}&  -2{\rm Im} M_{12} \\
 - {\rm Im} \Gamma_{12} &  2{\rm Re}M_{12} & -\Gamma & -\delta M    \\
 -{\rm Re}\Gamma _{12} & -2{\rm Im} M_{12} & \delta M   & -\Gamma
\end{array}\right)
\label{hnk12}
\end{equation}
in the $\sigma$-matrix basis. Above, $M$ denotes mass parameters,
$\Gamma$ denotes widths, and $\delta(\dots)$ denotes CPTV differences
between particle and antiparticle sectors, which are due to quantum mechanical
effects, such as LV {\it etc.}.

As mentioned previously, we
assume that the dominant violations of quantum mechanics conserve strangeness,
so that $\delta\H_{1\beta }$ = 0, and that $\delta\H_{0\beta }$ = 0 so as to
conserve probability. Since $\delta\H_{\alpha\beta }$ is a symmetric
matrix, it follows that also $\delta\H_{\alpha 0}=\delta\H_{\alpha 1}=0$.
Thus, we arrive at the general parametrization
 \begin{equation}
  {\delta\H}_{\alpha\beta} =\left( \begin{array}{cccc}
 0  &  0 & 0 & 0 \\
 0  &  0 & 0 & 0 \\
 0  &  0 & -2\alpha  & -2\beta \\
 0  &  0 & -2\beta & -2\gamma \end{array}\right)
\label{nine}
\end{equation}
where, as a result of the positivity\index{positivity} 
of the hermitian density matrix $\rho$
\cite{ehns}
\begin{equation}
\alpha, \gamma  > 0,\qquad \alpha\gamma>\beta^2\ .
\label{positivity}
\end{equation}

We recall \cite{emn} that the decoherence terms violate CP, given that the  
latter transformation can be expressed as
a linear combination of $ \sigma _{2,3}$ in the $K_{1,2}$ basis :
${\rm CP} = \sigma_3 \cos\theta + \sigma_2 \sin\theta$,
for some choice of phase $\theta$. It is
apparent that none of the non-zero terms $\propto   \alpha ,  \beta ,
 \gamma $ in $\delta\H_{\alpha\beta}$  (\ref{nine})
commutes with this CP transformation. In other words, each of the three
parameters $\alpha$, $\beta$, $\gamma$ violates CP\index{CP Violation}.
Moreover, in the problem there is evolution of pure to mixed states,
as we shall discuss below, leading, according to the theorem of \cite{wald},
described above, also to a strong form of CPT Violation\index{CPT Violation}.  
Thus, the decoherent CPTV evolution in the neutral Kaon system 
leads to a
much richer phenomenology than in conventional CPT Violations 
within a quantum mechanical framework, in the absence of 
decoherence, where the CPT may be violated only through 
differences in masses $\delta M$ and widths $\delta \Gamma$ 
between particles and antiparticles\index{antiparticle}. This is
because the symmetric $\delta\H$ matrix has three parameters in its
bottom right-hand $2\times 2$ submatrix, whereas the $h$ matrix
appearing in the time evolution within quantum mechanics 
has only one complex CPT-violating parameter $\delta$,
\begin{equation}
\delta = -\coeff{1}{2}
\frac{\coeff{1}{2}\delta\Gamma+i\delta M}{\coeff{1}{2}|\Delta\Gamma|+i\Delta m}
\ ,
\label{cptdelta}
\end{equation}
where $\delta M$ and $\delta \Gamma $ violate CPT, but do not induce any mixing
in the time evolution of pure state vectors\index{state vector}\cite{emn}. 
The parameters
$\Delta m = M_L -M_S$ and $|\Delta\Gamma|=\Gamma_S-\Gamma_L$ are the
usual differences between mass and decay widths\index{widths}, 
respectively, of 
the long-lived $K_L$ and short-lived 
$K_S$ energy (physical) eigenstates. 
For more details we refer the reader to the literature
\cite{emn}. The above results imply that the experimental constraints
\cite{pdg} on CPT Violation\index{CPT Violation}
have to be rethought. As we shall discuss later on,
there are essential differences between quantum-mechanical CPT Violation and
the non-quantum-mechanical CPT violation induced by the effective parameters
$\alpha, \beta, \gamma$ \cite{ehns}.

Useful observables are associated with the decays of neutral kaons\index{kaons} to $2\pi$ or
$3\pi$ final states, or semileptonic\index{semileptonic} 
decays to $\pi l \nu$. In the
density matrix\index{density matrix} formalism introduced above, their values are given by
 expressions of the form \cite{ehns,emn}
\begin{equation}
     \VEV{O_i}= {\rm Tr}\,[O_i\rho]\ ,
\label{13and1/2}
\end{equation}
where the observables $O_i$ are represented by $2 \times 2$ hermitian
matrices.
For instructive purposes we give their
expressions in the $K_{1,2}$ basis
\begin{eqnarray}
 O_{2\pi} &=& \left( \begin{array}{cc} 0 & 0 \\
0 & 1 \end{array} \right)\ ,\qquad  O_{3\pi} \propto
\left( \begin{array}{cc} 1 & 0 \\
0 & 0 \end{array} \right)\ , \label{2pi-obs} \\
O_{\pi^-l^+\nu} &=& \left( \begin{array}{cc}
1 & 1 \\1 & 1 \end{array} \right)\ ,\qquad
O_{\pi^+l^- \bar\nu} = \left( \begin{array}{rr}
1 & -1 \\
-1 & 1 \end{array} \right)\ .
\label{semi-obs}
\end{eqnarray}
which constitute a complete hermitian set. As we discuss in detail in 
\cite{emn},
it is possible to measure the interferecne\index{quantum-mechanical 
interference} 
between $K_{1,2}$ decays into $\pi
^{+}\pi^{-}\pi^0$ final states with different CP properties, by restricting
one's attention to part of the phase space $\Omega$, \eg, final states with
$m(\pi^+\pi^0)>m(\pi^-\pi^0)$. In order to separate this interference from that
due to $K_{S,L}$ decays into final states with identical CP properties,
due to CP Violation\index{CP Violation} 
in the $K_{1,2}$ mass matrix or in decay amplitudes, we
consider the difference between final states with
$m(\pi^+\pi^0) > m(\pi^-\pi^0) $ and $m(\pi^+\pi^0) < m(\pi^-\pi^0) $.
This observable is represented by the matrix
\begin{equation}
 O_{3\pi}^{\rm int} = \left( \begin{array}{cc} 0 & {\cal K} \\
{\cal K}^* & 0 \end{array} \right)
\label{XI}
\end{equation}
where
\begin{equation}
{\cal K} \equiv
\frac{\left[\int _{m(\pi^+\pi^0) > m(\pi^-\pi^0)} d\Omega
-\int _{m(\pi^+\pi^0) < m(\pi^-\pi^0)} d\Omega\right]
A_2 (I_{3\pi}=2)A_1 (I_{3\pi}=1)}
{\int d\Omega |A_1 (I_{3\pi}=1)|^2 }
\label{Y}
\end{equation}
where ${\cal K}$ is expected to be essentially real, so that the $O_{3\pi}^{\rm
int}$ observable provides essentially the same information
as $O_{\pi^-l^+\nu}-O_{\pi^+l^-{\overline \nu}}$.

In this formalism, pure $K^0$ or ${\bar K}^0$ states, such as the ones used as
initial conditions in the CPLEAR\index{CPLEAR Experiment} 
experiment \cite{cplear},
are described by the following density matrices
\begin{equation}
\rho _{K^0} =\coeff{1}{2}\left( \begin{array}{cc}
1 &1 \\1 & 1 \end{array} \right)\ , \qquad
\rho _{{\bar K}^0} =\coeff{1}{2}\left( \begin{array}{rr}
1 & -1 \\-1 & 1 \end{array} \right)\ .
\label{rhos}
\end{equation}
We note the similarity of the above density matrices (\ref{rhos})
to the semileptonic\index{semileptonic} decay observables in (\ref{semi-obs}), which is
due to the strange quark ($s$) content of the kaon $K^0 \ni {\bar s}
\rightarrow {\bar u} l^+ {\nu} , {\bar K}^0 \ni s \rightarrow  u l^- \bar\nu$,
and our assumption of the validity of the $\Delta S = \Delta Q$ rule.

One can apply the above formalism to compute the time evolution
of certain quantities that are of relevance to 
experiment\cite{cplear}, being directly observable. 
These are asymmetries\index{asymmetries} associated with
decays of an initial $K^0$ beam as compared to corresponding decays of an
initial ${\bar K}^0$ beam
\begin{equation}
    A (t) = \frac{
    R({\bar K}^0_{t=0} \rightarrow
{\bar f} ) -
    R(K^0_{t=0} \rightarrow
f ) }
{ R({\bar K}^0_{t=0} \rightarrow
{\bar f} ) +
    R(K^0_{t=0} \rightarrow
f ) }\ ,
\label{asym}
\end{equation}
where $R(K^0\rightarrow f)\equiv \Tr[O_{f}\rho (t)]$, denotes the decay rate
into the final state $f$, given that one starts from a pure $ K^0$ at $t=0$,
whose density matrix is given in (\ref{rhos}), and
$R({\bar K}^0 \rightarrow {\bar f}) \equiv \Tr [O_{\bar f} {\bar \rho}(t)]$
denotes the decay rate into the conjugate state ${\bar f}$, given that one
starts from a pure ${\bar K}^0$ at $t=0$.
One considers the 
following set of asymmetries:
(i) {\it identical final states}: 
$f={\bar f} = 2\pi $: $A_{2\pi}~,~A_{3\pi}$,
(ii) {\it semileptonic} : $A_T$
(final states $f=\pi^+l^-\bar\nu\ \not=\ \bar f=\pi^-l^+\nu$), $A_{CPT}$ (${\overline f}=\pi^+l^-\bar\nu ,~ f=\pi^-l^+\nu$), 
$A_{\Delta m}$.  

Typically, for instance when the final states are $2\pi$,
one has  a time evolution of the decay rate $R_{2\pi}$: 
$ R_{2\pi}(t)=c_S\, e^{-\Gamma_S t}+c_L\, e^{-\Gamma_L t}
+ 2c_I\, e^{-\Gamma t}\cos(\Delta mt-\phi)$, where 
$S$=short-lived, $L$=long-lived, $I$=interference term, 
$\Delta m = m_L - m_S$, $\Gamma =\frac{1}{2}(\Gamma_S + \Gamma_L)$. 
One may define the {\it decoherence parameter} 
$\zeta=1-{c_I\over\sqrt{c_Sc_L}}$, as a measure 
of quantum decoherence\index{decoherence} induced in the system. 
For larger sensitivities one can look 
at this parameter in the presence of a 
regenerator~\cite{emn}. 
In our decoherence scenario, it can be shown\cite{emn} 
that $\zeta$ depends primarily on $\beta$,
hence the best bounds on $\beta$ can be placed by 
implementing a regenerator.

Let us illustrate the formalism by two explicit examples. We may compute the
asymmetry  for the case where there are identical final states
$f={\bar f} = 2\pi $, in which case the observable is given in (\ref{2pi-obs}).
We obtain
\begin{equation}
A_{2\pi} = \frac{
\Tr[O_{2\pi} {\bar \rho} (t)] -
\Tr[O_{2\pi} \rho (t)]}
{
\Tr[O_{2\pi} {\bar \rho} (t)] +
\Tr[O_{2\pi} \rho (t)]}
= \frac{\Tr[O_{2\pi} \Delta \rho (t)]}{\Tr[O_{2\pi} \Sigma \rho (t)]}\ ,
\label{a2pi}
\end{equation}
where we have defined: $\Delta\rho(t)\equiv\bar\rho(t)-\rho(t)$ and
$\Sigma\rho(t)\equiv\bar\rho(t)+\rho(t)$. We note that in the above formalism
we make no distinction between neutral and charged two-pion final states.
This is because we neglect, for simplicity, the effects of $\epsilon'$.
Since $|\epsilon'/\epsilon|\lsim 10^{-3}$, this implies that our analysis
of the new quantum-mechanics-violating parameters must be refined
if magnitudes $\lsim\epsilon'|\Delta\Gamma| \simeq
10^{-6}|\Delta\Gamma|$ are to be studied\cite{huet}.

In a similar spirit to the identical final state case, one can compute the
asymmetry $A_{\rm T}$ for the semileptonic decay case, where
$f=\pi^+l^-\bar\nu\ \not=\ \bar f=\pi^-l^+\nu$.
The formula for this observable is
\begin{equation}
A_{\rm T}(t)
={\Tr[O_{\pi^-l^+\nu}\bar\rho(t)]-\Tr[O_{\pi^+l^-\bar\nu}\rho(t)]
\over\Tr[O_{\pi^-l^+\nu}\bar\rho(t)]+\Tr[O_{\pi^+l^-\bar\nu}\rho(t)]}\ .
\label{asymt}
\end{equation}
Other observables are discussed in \cite{emn}, where a complete
phenomenological description of CPTV decohering effects is presented.

To determine the temporal evolution of the above observables, which is
crucial for experimental fits, 
it is necessary to know the equations of motion
for the components of $\rho$ in the $K_{1,2}$ basis. These are
\cite{emn}\footnote{Since we neglect $\epsilon'$ effects and assume the
validity of the $\Delta S=\Delta Q$ rule, in what follows we also consistently
neglect ${\rm Im}\,\Gamma_{12}$ \cite{dafne}.}
\begin{eqnarray}
\dot\rho_{11}&=&-\Gamma_L\rho_{11}+\gamma\rho_{22}
-2{\rm Re}\,[({\rm Im}M_{12}-i\beta)\rho_{12}]\,,\label{rho11}\\
\dot\rho_{12}&=&-(\Gamma+i\Delta m)\rho_{12}
-2i\alpha\I{\rho_{12}}+({\rm Im}M_{12}-i\beta)(\rho_{11}-\rho_{22})\,,
\label{rho12}\\
\dot\rho_{22}&=&-\Gamma_S\rho_{22}+\gamma\rho_{11}
+2{\rm Re}\,[({\rm Im}M_{12}-i\beta)\rho_{12}]\,,   \label{rho22}
\end{eqnarray}
where for instance $\rho$ may represent $\Delta\rho$ or $\Sigma\rho$, defined
by the initial conditions
\begin{equation}
\Delta\rho(0)=\left(\begin{array}{cc}0&-1\\-1&0\end{array}\right)\ ,\qquad
\Sigma\rho(0)=\left(\begin{array}{cc}1&0\\0&1\end{array}\right)\ .
\label{InitialConditions}
\end{equation}
In these equations $\Gamma_L=(5.17\times10^{-8}\s)^{-1}$ and
$\Gamma_S=(0.8922\times10^{-10}\s)^{-1}$ are the inverse $K_L$ and $K_S$
lifetimes, $\Gamma\equiv(\Gamma_S+\Gamma_L)/2$, $|\Delta\Gamma| \equiv
\Gamma_S-\Gamma_L =(7.364 \pm 0.016 )\times 10^{-15}\GeV$, and $\Delta
m=0.5351\times10^{10}\s^{-1}=3.522\times10^{-15}\GeV$ is the $K_L-K_S$
mass difference. Also, the CP impurity parameter $\epsilon$ is given by
\begin{equation}
  \epsilon =\frac{{\rm Im} M_{12}}{\coeff{1}{2}|\Delta\Gamma|+i\Delta m }\ ,
\label{epstext}
\end{equation}
which leads to the relations
\begin{equation}
\I{M_{12}}=\coeff{1}{2}{|\Delta\Gamma||\epsilon|\over\cos\phi},\quad
\epsilon=|\epsilon|e^{-i\phi} \quad : \quad
\tan\phi={\Delta m\over {1\over2}|\Delta\Gamma|},
\end{equation}
with $|\epsilon|\approx2.2\times10^{-3}$ and $\phi\approx45^\circ$ the
``superweak" phase\cite{dafne}.

These equations are to be compared with the corresponding quantum-mechanical
equations, which are reviewed in \cite{emn}. 
The parameters ${\delta}M$ and $\beta$ play similar
roles, although they appear with different relative signs in different places,
because
of the symmetry of $\delta\H$ as opposed to the antisymmetry of the
quantum-mechanical evolution matrix $H$. These differences are important for
the asymptotic limits of the density matrix\index{density matrix}, 
and its impurity. In our approach,
one can readily show that, at large $t$, $\rho$ decays exponentially to
\cite{emn}:
\begin{equation}
\rho _L
\approx \left( \begin{array}{cc} 1 &
(|\epsilon | + i 2 {\widehat \beta }\cos\phi )e^{i\phi}  \\
(|\epsilon | - i 2 {\widehat \beta }\cos\phi )e^{-i\phi}  &
|\epsilon |^2 + {\widehat \gamma} -
4{\widehat \beta}^2 \cos ^2\phi - 4 {\widehat \beta} |\epsilon |\sin \phi
\end{array} \right)\ ,
\label{rhoL}
\end{equation}
where we have defined the following scaled variables
\begin{equation}
\widehat
\alpha=\alpha/|\Delta\Gamma|,\quad
\widehat\beta=\beta/|\Delta\Gamma|,
\quad \widehat\gamma=\gamma/|\Delta\Gamma|.
\label{abc}
\end{equation}
Conversely, if we look in the short-time limit
for a solution of the equations (\ref{rho11}) to (\ref{rho22}) with
${\rho}_{11}\ll  {\rho}_{12} \ll  {\rho}_{22}$, we find \cite{emn}
\begin{equation}
 \rho _S
\approx \left( \begin{array}{cc}
|\epsilon |^2 + {\widehat \gamma} -
4{\widehat \beta}^2 \cos ^2\phi + 4 {\widehat \beta} |\epsilon |\sin \phi
&
(|\epsilon | + i 2 {\widehat \beta }\cos\phi )e^{-i\phi}  \\
(|\epsilon | - i 2 {\widehat \beta }\cos\phi )e^{i\phi} &
1 \end{array} \right)\ .
\label{rhoS}
\end{equation}
These results are to be contrasted with those obtained within conventional
quantum mechanics
\begin{equation}
\rho_L\approx\left(
\begin{array}{cc}1&\epsilon^*\\ \epsilon&|\epsilon|^2\end{array}\right)\ ,
\qquad
\rho_S\approx\left(
\begin{array}{cc}|\epsilon|^2&\epsilon\\ \epsilon^*&1\end{array}\right)\ ,
\end{equation}
which, as can be seen from their {\it vanishing determinant}, 
correspond to
pure $K_L$ and $K_S$ states respectively. 

This is an important difference of the decoherence approach of \cite{ehns}
from others, as it implies an evolution of pure states to mixed.
Indeed, a pure state\index{pure state} 
will remain pure as long as $\Tr\rho^2=(\Tr\rho)^2={\rm Tr}\rho =1$,
or equivalently if $\rho^2 = \rho$ as operator relations, as discussed 
in section 2 
(the normalisation ${\rm Tr}\rho =1$ expresses conservation of probability).
In the
case of $2\times2$ matrices $\Tr\rho^2=(\Tr\rho)^2-2\det\rho$, and therefore
the purity condition is equivalently expressed as $\det\rho=0$. 
In contrast, 
$\rho_L,\rho_S$ in
eqs.~(\ref{rhoL},\ref{rhoS}) describe {\it mixed states}\index{mixed state}. 
Even in the limit 
of the imposition of complete positivity\index{complete positivity},
which according to the analysis of ref.~\cite{benatti1},
would imply $\alpha = \beta =0$, $\gamma > 0$, 
there is a non vanishing determinant for the above matrices, 
indicating the difference of the decoherence model of \cite{ehns,emn}
from others in the literature
where purity\index{purity} of states has been maintained during the 
evolution~\cite{houghston,adler}.

As mentioned above, the maximum possible order of magnitude for
the decoherence 
parameters $|\alpha|, |\beta|$ or $|\gamma |$ that we could expect theoretically is ${\cal
O}( E^2/M_{Pl})\sim {\cal O}((\Lambda _{\rm QCD}~{\rm or}~m_s)^2/M_{Pl})\sim
10^{-19}\GeV$ in the neutral kaon system. The fact that the model
is different, in general, from the double commutator Lindblad model 
of decoherence (\ref{doublecomm}), is welcome from a phenomenological
view point, given that it avoids the suppression 
(\ref{estimate1})~\cite{emnlarge}. Such unsuppressed models 
may characterise, for instance, the Liouville\index{Liouville}-string 
decoherence~\cite{emntheory},
described above, which are thus subject to direct experimental 
tests in the near future.

\begin{figure}
\centering 
\includegraphics[height=8cm]{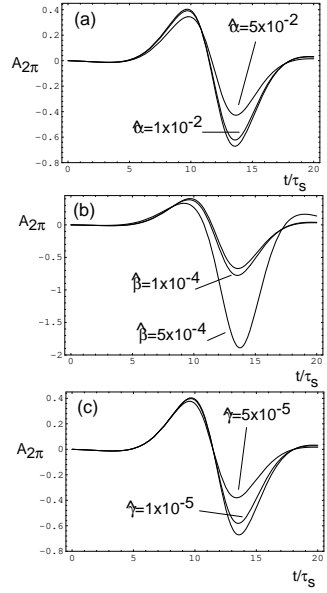}
\caption{The time-dependent asymmetry $A_{2\pi}$ for various choices of
the CPT-violating parameters: (a) dependence on $\widehat\alpha$, (b)
dependence on $\widehat\beta$, (c) dependence on $\widehat\gamma$. The
unspecified parameters are set to zero. The curve with no labels corresponds to
the standard quantum-mechanical case ($\widehat\alpha=\widehat\beta=\widehat\gamma=0$).}
\label{A2pi}
\end{figure}

\begin{figure}
\centering
\includegraphics[height=8cm]{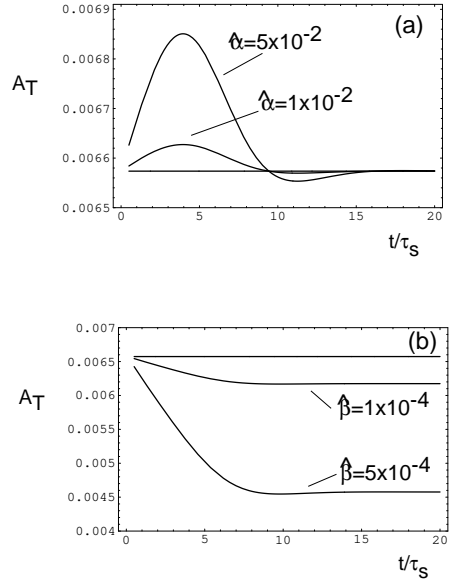}
\caption{The time-dependent asymmetry $A_{\rm T}$ for representative choices of
(a) $\widehat\alpha$ ($\widehat\beta=0$) and (b) $\widehat\beta$
($\widehat\alpha=0$). The dependence on $\widehat\gamma$ is negligible. The
flat line corresponds to the standard case.}
\label{AT}
\end{figure}

\begin{figure}
\centering 
\includegraphics[height=8cm]{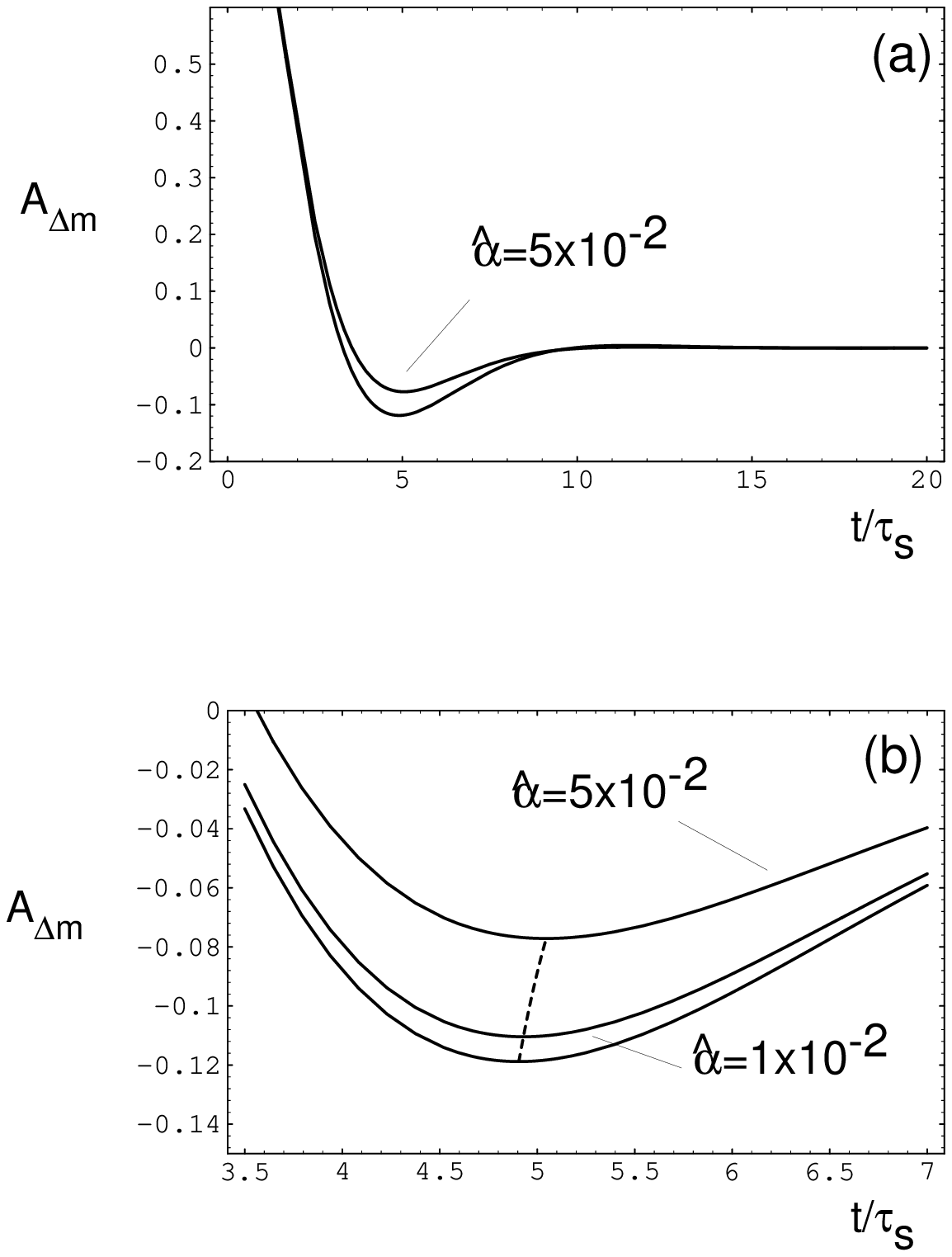}
\caption{The time-dependent asymmetry $A_{\Delta m}$ 
for representative choices
of $\widehat\alpha$ ($\widehat\beta=\widehat\gamma=0$). This asymmetry depends
most sensitively only on $\widehat\alpha$. In both panels, the bottom curve
corresponds to the standard case. In the detail (b), the dashed line indicates
the location of the minimum as $\widehat\alpha$ is varied.}
\label{ADeltam}
\end{figure}

To make a consistent 
phenomenological study of the various asymmetries\index{asymmetries} 
discussed above,
in particular to determine their time profiles and compare
them with experiment~\cite{cplear} ,
it is essential to solve the coupled system of equations (\ref{rho11})
to (\ref{rho22}) for intermediate times. 
This requires approximations in powers of the 
decoherence\index{decoherence} parameters in order
to get analytic results \cite{emn}, which we shall not describe here. 
Below we shall only outline the results briefly by demonstrating the 
time profiles of the asymmetries $A_{2\pi}$ and $A_{T}$, as well as the 
asymmetry $A_{\Delta m}$ used in the CPLEAR\index{CPLEAR Experiment} experiment~\cite{cplear}.
The relevant results are outlined in figures \ref{A2pi},\ref{AT},
\ref{ADeltam}.

The important point in such an analysis is that CPTV 
due to decoherence in neutral mesons\index{meson} can be disentangled 
from CPTV within quantum mechanics, for instance due to Lorentz 
Violation a l\'a SME~\cite{kostel}. 
The experimental tests (decay asymmetries) 
that can be performed in order to disentangle
decoherence from quantum mechanical CPT violating effects 
are summarized in table \ref{Table2}. 
Experimentally, the best available bounds to date 
for the neutral meson case come from 
CPLEAR measurements~\cite{cplear} 
$\alpha < 4.0 \times 10^{-17} ~{\rm GeV}~, ~|\beta | < 2.3. \times
10^{-19} ~{\rm GeV}~, ~\gamma < 3.7 \times 10^{-21} ~{\rm GeV} $,
which are not much different from 
theoretically expected values in some models, 
$\alpha~,\beta~,\gamma = O(\xi \frac{E^2}{M_{P}})$.

\begin{table}[ht]
\begin{center}
\begin{tabular}{lcc}
\underline{Process}&QMV&QM\\
$A_{2\pi}$&$\not=$&$\not=$\\
$A_{3\pi}$&$\not=$&$\not=$\\
$A_{\rm T}$&$\not=$&$=$\\
$A_{\rm CPT}$&$=$&$\not=$\\
$A_{\Delta m}$&$\not=$&$=$\\
$\zeta$&$\not=$&$=$
\end{tabular}
\caption{Qualitative comparison of predictions for various observables
in CPT-violating theories beyond (QMV) and within (QM) quantum mechanics.
Predictions either differ ($\not=$) or agree ($=$) with the results obtained
in conventional quantum-mechanical CP violation. Note that these frameworks can
be qualitatively distinguished via their predictions for $A_{\rm T}$, $A_{\rm
CPT}$, $A_{\Delta m}$, and $\zeta$.}
\label{Table2}
\end{center}
\hrule
\end{table}

Before closing this section it is worthy of mentioning that above we 
have considered the same set of decoherence parameters 
$\alpha,\beta,\gamma$ in both particle and antiparticle 
sectors. However, in view of the induced CPTV in the strong form\cite{wald},
it is not clear that the 
order of these two sets of parameters is the same between
particle and antiparticle sectors. 
Although we have no concrete theoretical models at present, nevertheless, 
one may envisage cases where the 
strength of the interaction with the foam 
is different
between matter and antimatter.  
An example of such a case will
be seen later on, in the context of 
neutrino physics. As we shall see there,
minimal models of QG-induced decoherence, with the latter being 
dominant only in the antiparticle\index{antiparticle}
sector, will be capable of 
explaining current neutrino\index{neutrino} anomalous data, 
such as LSND\index{LSND Experiment} reasults\cite{lsnd}, 
in a way consistent with all the other data.

\subsection{EPR Entangled Neutral Meson States and novel Decoherence-induced CPT Violating Effects}

In experiments involving multiparticle states\index{multiparticle states}, 
such as those produced in a $\phi$ or $B$ factory\index{meson factories}, 
the fact that CPT\index{CPT} may not be a well defined operation,
as a result of decoherence induced by quantum gravity~\cite{wald},
could imply novel effects~\cite{bernabeu}, 
which may affect the properties of the 
entangled\index{entangled} states, and as such are unique to such situations, and absent in single particle experiments.

In conventional 
formulations of {\it entangled} meson  
states~\cite{dunietz}
one imposes the requirement of {\it Bose statistics}\index{Bose statistics} 
for the state $K^0 {\overline K}^0$ (or $B^0 {\overline B}^0$), 
which implies that the physical neutral meson-antimeson state 
must be {\it symmetric} under the combined operation $C{\cal P}$,
with $C$ the charge conjugation and 
${\cal P}$ the operator that permutes the spatial coordinates. 
Specifically, assuming 
{\it conservation} of angular momentum, and 
a proper existence of the {\it antiparticle state} (denoted by a bar),
one observes that, 
for $K^0{\overline K}^0$ states which are $C$-conjugates with 
$C=(-1)^\ell$ (with $\ell$ the \index{angular momentum} quantum number), 
the system has to be an eigenstate of 
${\cal P}$ with eigenvalue $(-1)^\ell$. 
Hence, for $\ell =1$, we have that $C=-$, implying ${\cal P}=-$.
As a consequence 
of Bose statistics this ensures that for $\ell = 1$ 
the state of two identical bosons is forbidden~\cite{dunietz}.  
As a result, the initial entangled state 
$K^0{\overline K}^0$ produced in a $\phi $ factory 
can be written as:
\begin{equation} 
|i> = \frac{1}{\sqrt{2}}\left(|K^0({\vec k}),{\overline K}^0(-{\vec k})>
- |{\overline K}^0({\vec k}),{K}^0(-{\vec k})>\right)
\label{bbar}
\end{equation}
This is the starting point 
of all formalisms known to date, either 
in the $K$-system~\cite{dunietz} or
in the $B$-system, 
including those~\cite{huet}
where the evolution
of the entangled state is described by non-quantum 
mechanical terms, in the formalism of \cite{ehns}. 
In fact, in all these works it has been claimed  
that the expression in Eq.(\ref{bbar})
is actually independent of any assumption about CP, T or CPT symmetries. 

However, as has been alluded above, 
the assumptions leading to 
Eq.(\ref{bbar}) may not be valid if CPT\index{CPT} symmetry is violated.
In such a case 
${\overline K}^0$ cannot be considered 
as identical to ${K}^0$, and thus the requirement of $C {\cal P} = +$, imposed 
by Bose-statistics, is relaxed.
As a result, the initial entangled state (\ref{bbar}) 
can be parametrised in general as~\cite{bernabeu}:
\begin{eqnarray} 
|i> &=& \frac{1}{\sqrt{2}}\left(|K^0({\vec k}),{\overline K}^0(-{\vec k})>
- |{\overline K}^0({\vec k}),{K}^0(-{\vec k})> \right)  \nonumber \\
&+& \frac{\omega}{\sqrt{2}}\left(|K^0({\vec k}),{\overline K}^0(-{\vec k})>
 + |{\overline K}^0({\vec k}),{K}^0(-{\vec k})> \right)  
\label{bbarcptv}
\end{eqnarray}
where $\omega = |\omega| e^{i\Omega}$ 
is a {\it complex} CPTV parameter, 
associated with the non-identical particle nature 
of the neutral meson
and antimeson\index{antimeson} 
states. This parameter describes a {\it novel} phenomenon, 
not included in previous analyses.

Notice that an equation such as the one given in  
(\ref{bbarcptv}) could also be produced  as a result 
of deviations from the laws of quantum mechanics
during the initial decay of the $\phi$\index{$\phi$ particle} or
$\Upsilon$ states\index{$\Upsilon$ states}. Thus, Eq.(\ref{bbarcptv}) 
could receive contributions from two different effects,
and can be thought off as simultaneously parametrizing 
both of them.

In terms of physical (energy) eigenstates, $|K_{S,L}\rangle$,
the state (\ref{bbarcptv})
is written as (we keep linear terms in the small parameters $\omega$, 
$\delta$, i.e. in the following we ignore higher-order terms 
$\omega \delta$, $\delta^2$  {\it etc.})
{\small 
\begin{eqnarray} 
|i> &=& 
C \bigg[ \left(|K_S({\vec k}),K_L(-{\vec k})>
- |K_L({\vec k}),K_S(-{\vec k})> \right)\nonumber \\  
&+& \omega \left(|K_S({\vec k}), K_S(-{\vec k})>
- |K_L({\vec k}),K_L(-{\vec k})> \right)\bigg]  
\label{bph}
\end{eqnarray}}
with $C = \frac{\sqrt{(1 + |\epsilon_1|^2)
(1 + |\epsilon_2|^2 )}}{\sqrt{2}(1-\epsilon_1\epsilon_2)}
\simeq \frac{1 + |\epsilon^2|}{\sqrt{2}(1 - \epsilon^2)}$.
Notice again the presence of combinations $K_S K_S$ and $K_L K_L$ states,
proportional to the novel CPTV parameter $\omega$.

\begin{figure}[ht]
\centering
\includegraphics[width=0.6\textwidth]{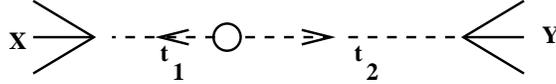}
\caption{A typical amplitude corresponding to the decay 
of, say, a $\phi$ state into final states $X,Y$; $t_i, i=1,2$ denote the 
corresponding time scales for the appearance of the final products of the 
decay.}
\label{amplitude} 
\end{figure}

Such terms become important when one considers decay channels.
Specifically, consider the decay 
amplitude $A(X,Y)$, corresponding to the appearance of a final state $X$ 
at time $t_1$ and $Y$ at time $t_2$, as illustrated in  
fig. \ref{amplitude}. 

One assumes (\ref{bph}) for the initial two-Kaon system, 
after the $\phi$ decay. The time is set $t=0$ at the moment of the decay.
Next, one integrates 
the square of the amplitude 
over all accessible times $t= t_1 + t_2$,
keeping the difference $\Delta t = t_2 - t_1$ as constant.   
This defines the ``intensity'' $I(\Delta t)$~\cite{bernabeu}:
\begin{eqnarray} 
I (\Delta t) \equiv \frac{1}{2} \int_{|\Delta t|}^\infty dt\, |A(X,Y)|^2  
\label{intensity} 
\end{eqnarray} 
In what follows we concentrate on identical final states $X=Y=\pi^+\pi^-$, 
because as we shall argue they are the most sensitive channels to probe 
the novel effects associated with the CPTV parameter $\omega$. 
Indeed~\cite{pdg}, the amplitudes 
of the 
CP violating\index{CP Violation} decays $K_L \to \pi^+\pi^-$ are 
suppressed by factors of order ${\cal O}(10^{-3})$, 
as compared to the principal
decay mode of $K_S \to \pi^+\pi^-$.
In the absence of CPTV $\omega$, (\ref{bbar}), due to the 
$K_SK_L$ mixing, such decay rates would be suppressed. This would not 
be the case, however, when the CPTV $\omega$ (\ref{bbarcptv}) parameter 
is non zero, due to the existence 
of a separate $K_SK_S$ term in that case ((\ref{bph})). 
This implies that the 
relevant parameter for CPT violation 
in the intensity is $\omega/\eta_X$, where 
$\eta_X = \langle X|K_S\rangle/\langle X|K_L\rangle$
which enhances 
the potentially observed effect. 

\begin{figure}[ht]
\includegraphics[width=4.2cm]{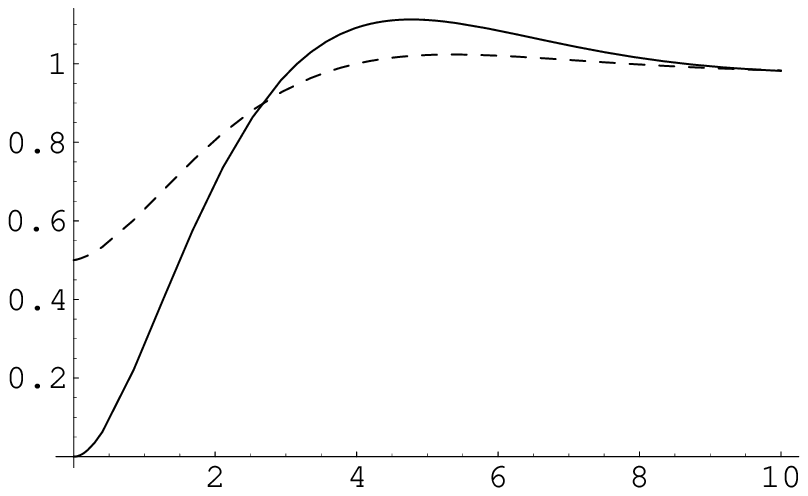}\hfill
\includegraphics[width=4.2cm]{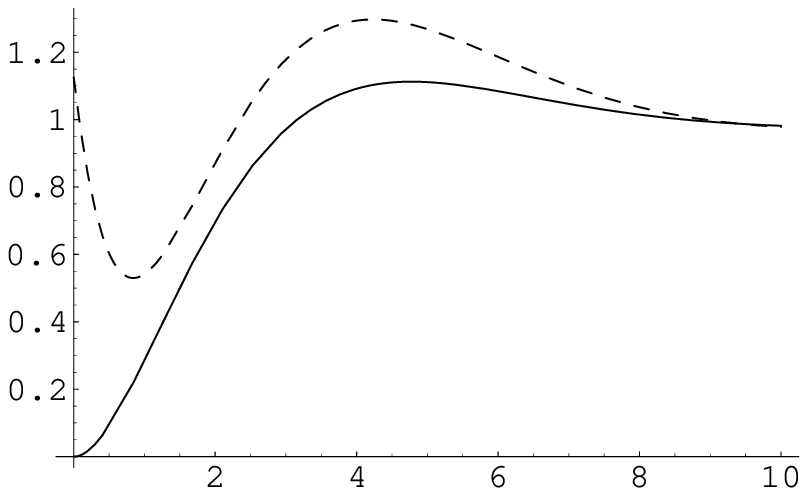}\hfill
\includegraphics[width=4.2cm]{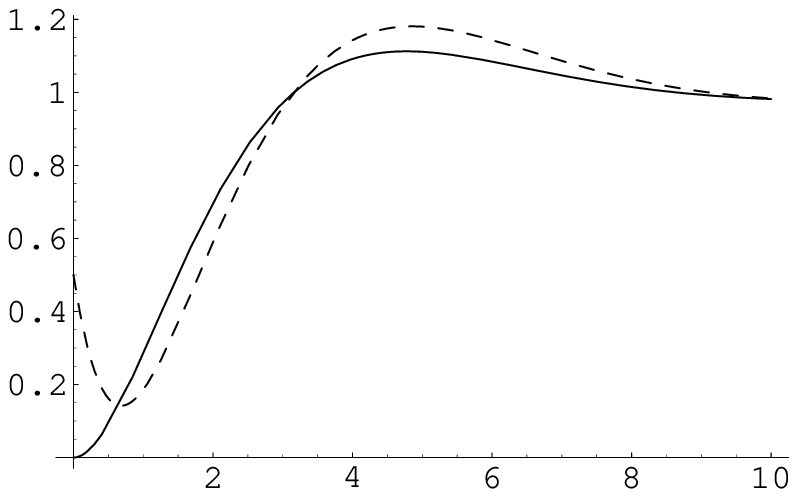}\hfill
\includegraphics[width=4.2cm]{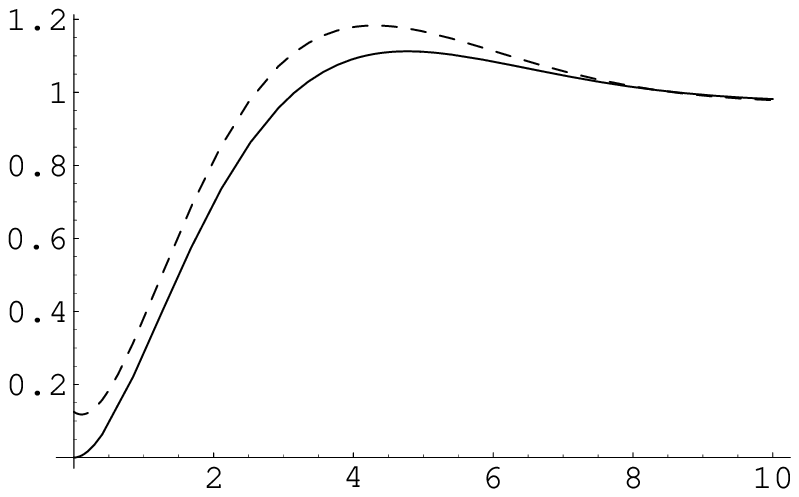}
\caption{Characteristic cases of the intensity $I(\Delta t)$, 
with $|\omega|=0$ (solid line) vs $I(\Delta t)$ 
(dashed line) with (from top left to right): (i) $|\omega|=|\eta_{+-}|$, 
$\Omega = \phi_{+-} - 0.16\pi$, (ii) $|\omega|=|\eta_{+-}|$, 
$\Omega = \phi_{+-} + 0.95\pi$, (iii) $|\omega|=0.5|\eta_{+-}|$, 
$\Omega = \phi_{+-} + 0.16\pi$, (iv) $|\omega|=1.5 |\eta_{+-}|$, 
$\Omega = \phi_{+-}$. $\Delta t$ is measured in units of $\tau_S$ (the 
mean life-time of $K_S$) and 
$I(\Delta t)$ in units of 
$|C|^2 |\eta_{+-}|^2 |\langle\pi^{+}\pi^{-}|K_S\rangle|^4 \tau_S$.} 
\label{intensityfigure} 
\end{figure}

The effects of the CPTV $\omega$ on such 
intensities $I(\Delta t)$ are indicated in figure 
\ref{intensityfigure}.
We next comment on the distinguishability 
of the $\omega$ effect\index{$\omega$ effect} 
from conventional background effects. 
Specifically, the mixing of the initial state due to 
the non-identity of the antiparticle\index{antiparticle} to  
the corresponding 
particle state 
has similar form 
to that induced by a non-resonant background 
with $C=+$~\cite{dunietz}. This latter effect is known to have a small size; 
estimates based on unitarity bounds give a size  
of many orders of magnitude smaller than the 
$C=-$ effect in the $\phi$ decays ~\cite{dunietz,dafne}. 
Terms of the type $K_S K_S$ (which dominate over $K_L K_L$) coming from
the $\phi$-resonance as a result 
of CPTV 
can be distinguished from those coming from the $C=+$ background 
because they interfere differently  
with the regular $C=-$ resonant
contribution (i.e. Eq.(\ref{bph}) with $\omega=0$). Indeed, 
in the CPTV case, the $K_L K_S$ and  $\omega K_S K_S$ terms 
have the same dependence on the center-of-mass energy $s$ 
of the colliding particles producing the 
resonance,
because both terms originate from the  $\phi$-particle. Their
interference, therefore, being proportional to the real part of the 
product of the corresponding amplitudes, still displays a peak at the 
resonance. 
On the other hand, the amplitude of the $K_S K_S$ coming 
from the $C=+$ background 
has no appreciable dependence on $s$ and 
has practically vanishing imaginary part.
Therefore, given that the real part of a Breit-Wigner 
amplitude vanishes at the 
top of the resonance\index{resonance}, this implies that the  
interference of the $C=+$ background  
with the regular $C=-$ resonant contribution 
vanishes at the top of the resonance,
with opposite signs on both sides of the latter.    
This clearly distinguishes experimentally the two cases.

We continue with a brief discussion concerning the distinguishability 
of the $\omega$ effect (\ref{bbarcptv}),(\ref{bph}) from 
non-quantum mechanical effects associated with the evolution, as 
in \cite{ehns}. 
The  $\omega$ effect 
can be distinguished from those 
of the QG-decohering evolution parameters $\alpha,\beta,\gamma$, 
when the formalism 
is applied  to the entangled\index{entangled} 
states $\phi$~\cite{huet,benatti}. 
A non-quantum mechanical evolution of the entangled Kaon state 
with $\omega =0$ has been considered in \cite{huet}.
In such a case the resulting density-matrix $\phi$ state 
$\tilde{\rho}_{\phi} ={\rm Tr}|\phi><\phi|$ can be written as
\begin{eqnarray} 
\tilde{\rho}_{\phi} &=&   
\rho_{S} \otimes \rho_{L} + 
\rho_{L} \otimes \rho_{S} 
- \rho_{I}\otimes \rho_{{\overline I}} 
- \rho_{{\overline I}}\otimes \rho_{I} 
\nonumber\\
&-& \frac{2\beta}{d} (\rho_I \otimes \rho_S + \rho_S \otimes \rho_I ) 
- \frac{2\beta}{d^*} (\rho_{{\overline I}} \otimes \rho _S + 
\rho_S \otimes \rho_{{\overline I}} )  
\nonumber\\
&+& \frac{2\beta}{d} ( \rho_{{\overline I}} \otimes \rho_L + 
\rho_L \otimes \rho_{{\overline I}}) + 
\frac{2\beta}{d^*} ( \rho_I \otimes \rho_L + \rho_L \otimes \rho_I ) 
\nonumber\\
&-&  \frac{i\alpha}{\Delta M} 
( \rho_{I}\otimes \rho_I - \rho_{{\overline I}}
\otimes  \rho_{{\overline I}})
-\frac{2\gamma}{\Delta \Gamma} 
(\rho_S \otimes \rho_S - \rho_L \otimes \rho_L)
\nonumber
\end{eqnarray}
where the standard notation $\rho_{S} = |S><S|, ~\rho_L = |L><L|,
~\rho_I = |S><L|, ~\rho_{{\overline I}} = |L><S|$ has been employed, 
$d = -\Delta M  + i \Delta \Gamma/2$,
and an overall multiplicative factor of 
$\frac{1}{2}
\frac{(1 + 2|\epsilon|^2)}
{1 - 2|\epsilon|^2{\rm cos}(2\phi_\epsilon)}$ has been suppressed.
On the other hand, the corresponding density matrix\index{density matrix} description 
of the $\phi$ state (\ref{bph}) in our case reads: 
\begin{eqnarray} 
\rho_\phi &=& 
\rho_S \otimes \rho_L + \rho_L \otimes \rho_S 
- \rho_{I}\otimes \rho_{{\overline I}} - \rho_{{\overline I}}\otimes \rho_I 
\nonumber\\
&-& \omega (\rho_I \otimes \rho_S - \rho_S \otimes \rho_I ) 
- \omega^* (\rho_{{\overline I}} \otimes \rho _S - 
\rho_S \otimes \rho_{{\overline I}} ) 
\nonumber\\
&-& 
\omega ( \rho_{{\overline I}} \otimes \rho_L - 
\rho_L \otimes \rho_{{\overline I}})  
- \omega^* ( \rho_I \otimes \rho_L - \rho_L \otimes \rho_I ) 
\nonumber\\
&-& |\omega|^2 ( \rho_{I}\otimes \rho_I + \rho_{{\overline I}}
\otimes  \rho_{{\overline I}}) + 
|\omega|^2 (\rho_S \otimes \rho_S + \rho_L \otimes \rho_L)  
\nonumber
\end{eqnarray}
with the same multiplicative factor suppressed. 
It is understood that the evolution of both $\tilde{\rho}_{\phi}$
and $\rho_\phi$ is governed by the rules given in ~\cite{ehns,emn,huet}.
As we can see by comparing the two equations, 
the terms linear in $\omega$ in our case 
are {\it antisymmetric} under the exchange of particle states 
$1 $ and $2$, {\it in contrast} to the {\it symmetry}
of the corresponding terms linear in $\beta$ in the case of \cite{huet}.
Similar differences characterize  the terms proportional 
to $|\omega|^2$, and those proportional to $\alpha$ and $\gamma$, 
which involve $\rho_I \otimes \rho_I$, 
$\rho_{\overline I} \otimes \rho_{\overline I}$,
$\rho_S\otimes \rho_S$, $\rho_L\otimes \rho_L$. 
Such differences are therefore important 
in disentangling the $\omega$ CPTV effects proposed here 
from non-quantum mechanical evolution effects~\cite{ehns,emn,huet,benatti1}.

Finally we close this subsection with a comment on the 
application of this formalism to the $B$ factories\index{meson factories}.
Although, formally, the situation is identical to the 
one discussed above, however
the sensitivity of the CPTV $\omega$ effect\index{$\omega$ effect} 
for the $B$ system is much smaller.
This is due to 
the fact that 
in $B$ factories there is no particularly ``good'' 
channel $X$ (with $X=Y$) for which 
the corresponding $\eta_X$ is small. 
The analysis in that 
case may therefore be performed in
the equal sign 
dilepton channel, where the 
branching fraction is more important, and a high statistics 
is expected.

\subsection{CPTV Decoherence and Ultra Cold Neutrons} 

\begin{figure}
\centering
\includegraphics[height=6cm]{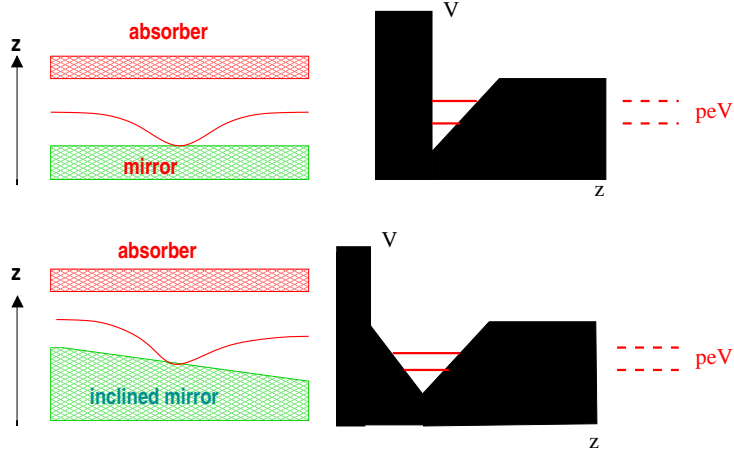} 
\caption{Inclined mirror ensures Parity invariance
of QG modifications and hence formalism similar to 
neutral kaons\index{kaons}. A few (two here) 
energy states (peV energy differences between levels) are 
inside the Earth' s potential well.} 
\label{coldneutrons}
\end{figure}

Before commencing a discussion on QG-induced decoherence in neutrinos
we would like to discuss  
briefly the application of the decoherence formalism 
of \cite{ehns} on another interesting experiment, 
which attracted some attention recently, that of ultracold neutrons 
in the gravitational field of 
Earth~\footnote{The results in this section have been 
derived in collaboration with Elias Gravanis.}. 
The arrangement of this experiment is demonstrated
in figure \ref{coldneutrons}. 

The neutrons\index{neutrons} find themselves on a quantum-mechanical
potential which is affected by the gravitational potential of Earth, 
due to their masses. A few energy states, separated by peV $\sim 10^{-15}$ 
eV energy differences, lie inside the Earth's potential\index{Earth's 
Gravitational Potential} well. 
The quantum trajectories of the neutrons are affected by this gravitational
potential in the way indicated in the figure. The neutrons are reflected
on the mirrors and are collected at the detection point. 
This has already been demonstrated experimentally,
measuring  for the first time gravitational effects
together with quantum mechanical effects~\cite{neutrons}. 

Consider for our purposes the case where two such energy states 
find themselves inside the potential well. This constitutes a two-level
system, and one may think of applying the two-state decoherence 
formalism to study QG induced effects in such a situation, 
which would modify the results concering the probabilities of 
finding the 
neutrons in one of the two available energy states at the detection point.
Like any two-state oscillation system, 
the respective probabilities should 
be equal to 1/2 in the absence of any decoherence effects, 
although in the presence of decoherence\index{decoherence} one would expect
a slight bias in the probabilities.

The quantum number which is conserved here is Parity\index{Parity}, 
which however 
is the case only if the mirror is inclined, so as to eliminate 
the effects of parity violation induced by the presence 
of the Gravitational field.
The Probability of finding the neutrons in either state 
at the detection point indicated in the figure 
can be computed following the same formalism as the two-state
parametrization of the neutral kaon system in \cite{ehns}, with the 
replacement of the strangeness\index{strangeness} 
conservation by that of parity. The results
read, to leading order in the small decoherence parameters: 
\begin{equation}
{\rm Tr}(\rho' \varrho_{1,2})=
\frac{1}{2} \pm \frac{1}{2}e^{-\frac{\alpha+\gamma}{2}t} 
\sin(\Delta E t)~, \qquad \Delta E = {\cal O}({\rm peV}) 
\label{neutronscold}
\end{equation}
where $t$ is the time. 

We next remark that, if Lorentz invariance 
is violated by Quantum Gravity, 
then the decoherence parameters $\alpha,\gamma \simeq 
\frac{E^2_{\rm kin}}{M_P}$, where $E_{\rm kin}={\cal O}({\rm peV})$ 
is the kinetic energy of the neutrons.
This is too small to be detected in this kind of experiment;
However, in case QG decoherence respects Lorentz Symmetry, which as mentioned
above is possible~\cite{mill,discr},  
then $\alpha,\gamma \simeq 
\frac{m_n^2}{M_P}$. For the duration of the experiment, which is 
or order $t \sim {\rm msec}$, then, we observe that 
the decoherence effects are 
much larger. However, 
at present, there seems to be no significant sensitivity from this 
type of experiment, as compared with other available tests of decoherence.
Nevertheless, one cannot exclude the possibility of a significant improvement
in sensitivity in similar experiments in the foreseeable future,
and this is the reason why I included this case briefly in the present 
set of lectures.

\subsection{CPTV through QG Decoherence for Neutrinos: 
the most sensitive probe to date} 

\subsubsection{Two-generation models}

We now come to discuss quantum-gravity decoherence\index{decoherence} 
in neutrinos,
whose sensitivity in this respect is far more superior
than that of neutral kaons\index{kaons}, assuming of course a universal nature of 
QG. This latter assumption, though, 
requires some second thoughts, given that, as mentioned above, 
there are theoretical models
of quantum space-time foam\index{foam}\cite{synchro}, 
in which QG effects interact differently
with various particle species.  

With this in mind we next remark that, 
QG may induce oscillations 
between neutrino flavours independently 
of $\nu$-masses~\cite{liu,lisi,benatti,klapdor}. 
We begin with the simplified case of two-neutrino generations, 
that is a two-state system, which makes the formalism very similar to the 
neutral kaon case described above. In similar spirit to the Kaon case,
the energy (physical) eigenstates of neutrinos are not flavour eigenstates,
and one has mixing. 

The basic formalism for decoherence-induced neutrino oscillations 
is described by a
QMV evolution for the density matrix of the $\nu$, which parallels
that of neutral kaon in the case of two generations\index{generation} of neutrinos: 
\begin{equation} 
\partial_t \rho = i[\rho, H] +  \delta\H \rho 
\end{equation} 
where~\cite{ehns} 
  \begin{equation}
  {\delta\H}_{\alpha\beta} =\left( \begin{array}{cccc}
 0  &  0 & 0 & 0 \\
 0  &  -2\alpha &  -2\beta  & 0 \\
 0  &  -2\beta  & -2\gamma & 0 \\
 0  &  0 & 0 & 0         \end{array}\right)
\nonumber 
\end{equation}
for energy and lepton number\index{lepton number} conservation, and 
\begin{equation}
  {\delta\H}_{\alpha\beta} =\left( \begin{array}{cccc}
 0  &  0 & 0 & 0 \\
 0  &  0 & 0 & 0 \\
 0  &  0 & -2\alpha &  -2\beta  \\
 0 & 0  & -2\beta & -2\gamma \end{array}\right)
\nonumber 
\end{equation} 
if energy and lepton number are violated,  
but flavour\index{flavour} is conserved (the latter associated formally with the 
$\sigma_1$ Pauli\index{Pauli} matrix). 

Positivity\index{positivity} of $\rho$, but not complete positivity, 
requires: $\alpha, \gamma  > 0,\qquad \alpha\gamma>\beta^2$.
The parameters $\alpha,\beta,\gamma$  violate CP, and CPT in general, 
as discussed previously.   

The relevant oscillation\index{neutrino oscillations} 
probabilities, describing the evolution 
of a neutrino\index{neutrino} 
flavour $\nu_\alpha$ , created at time $t=0$, to a neutrino
flavour $\nu_\beta$ at time $t$, are determined by means of the 
dynamical semigroup\index{dynamical semigroup} 
approach (\ref{solutionmatrixform}):  
\begin{equation}\label{genprobs}
P_{\nu_\alpha \to \nu_\beta}(t) 
= {\rm Tr}\left(\rho_\alpha (t)\rho^\beta\right)
\end{equation}
For our problem we have a two state system, 
and the computation of the eigenvalue problem is easy. 
For the two cases above, 
we obtain after some straightforward algebra~\cite{liu}:

{\bf (A)} \underline{For the flavour conserving case}: 

As a simplified example, consider the oscillation 
$\nu_e \to \nu_x$~ ($x=\mu, \tau$ or sterile):

\begin{equation} 
P_{\nu_e \to \nu_x} = \frac{1}{2} - \frac{1}{2}e^{-\gamma L}
{\rm cos}^22\theta_v  - 
\frac{1}{2}e^{-\alpha L}{\rm sin}^22\theta_v{\rm cos}
(\frac{|m_{\nu_1}^2 - m_{\nu_2}^2|}{2E_\nu}L)
\label{prob1}
\end{equation}
Here $L$ is the oscillation length and 
$\theta_v$ the mixing angle. 

In the mass basis one has:  
$|\nu_e> = {\rm cos}\theta_v |\nu_1> + {\rm sin}\theta_v|\nu_2>,$ 
$|\nu_\mu> = -{\rm sin}\theta_v |\nu_1> + {\rm cos}\theta_v|\nu_2>.$ 
Note that in this case the mixing angle $\theta_v = 0$
if and only if the neutrinos are massless.  
From the above considerations, however,  it is clear that 
there are flavour oscillations even in the massless case, due 
to a non-trivial 
QG parameter
$\gamma$, compatible with flavour conserving formalism:
$<\nu_e|\sigma_1|\nu_e>=-<\nu_\mu|\sigma_1|\nu_\mu>= 
2{\rm sin}\theta_v{\rm cos}\theta_v.$

{\bf (B)} \underline{For Energy and Lepton number conserving case}:

Again, we consider a two-flavour example: 
$\nu_e \to \nu_x$~ ($x=\mu, \tau$ or sterile). The relevant oscillation 
probability in this case is calculated to be~\cite{liu}:

\begin{equation} 
P_{\nu_e \to \nu_x} = \frac{1}{2}{\rm sin}^22\theta_v \left(1
- e^{-(\alpha + \gamma) L}
{\rm cos}(\frac{|m_{\nu_1}^2 - m_{\nu_2}^2|}{2E_\nu}L) \right)
\label{prob2}
\end{equation}
where we 
assumed for simplicity, and illustrative purposes, that 
$\alpha,\beta,\gamma \ll \frac{|m_{\nu_1}^2 - m_{\nu_2}^2|}{2E_\nu}$. 
The reader is invited to contrast this result with case {\bf (A)} above.

One can use 
the results in the cases {\bf (A)} and {\bf (B)} to bound 
experimentally $\xi \equiv \{\alpha,\beta,\gamma$\}. 
At this stage the reader is 
invited to recall that there exist 
two kinds of theoretical estimates/predictions for the order of magnitude
of the parameters $\alpha,\beta,\gamma$:
An optimistic one~\cite{emn}, according to which 
 $\xi \sim \xi_0 (\frac{E}{\rm GeV})^n, 
n=0,2,-1$, and this has a chance of being falsified in 
future experiments, if the effect is there, and a 
pessimistic one~\cite{adler}, which depends on the square of the 
neutrino mass-squared difference (\ref{estimate1}), 
$\xi \sim 
\frac{(\Delta m^2)^2}{E^2 M_{qg}}$, ($M_{qg} \sim M_P \sim 10^{19}$ GeV),
which is much smaller, and probably cannot be accessed by immediate future
neutrino oscillation experiments.

We now mention that in some models of QG-induced decoherence, 
complete positivity\index{complete positivity}  of $\rho(t)$ 
for composite systems, such as $\phi$ or $B$ mesons, may be 
imposed~\cite{benatti} (however, I must stress once more that 
the necessity of this requirement, especially in a QG context
where non-linear effects may be present~\cite{emn}, 
remains to be proven).  
This results in an ideal Markov\index{Markov} environment, 
with: $\alpha = \beta = 0, \gamma > 0$.

If this model is 
assumed for $\nu$ oscillations induced by QG decoherence~\cite{lisi}, 
then the following 
phenomenological parametrization can be made:
 $\gamma = \gamma_0 (E/{\rm GeV})^n$, $n=0,2,-1$.
with $E$ the neutrino energy.

\begin{figure}
\centering
 \includegraphics[height=4cm]{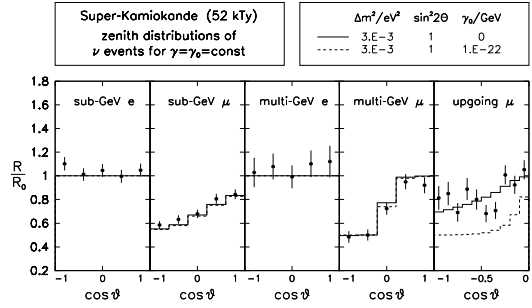}
\caption{Effects of decoherence  
($\gamma = \gamma_0 = {\rm const} \ne 0$)  on the distributions of lepton events as a function of the zenith angle 
$\vartheta$.}
\label{nudatadec}
\end{figure}

\begin{figure}
\centering
  \includegraphics[height=4cm]{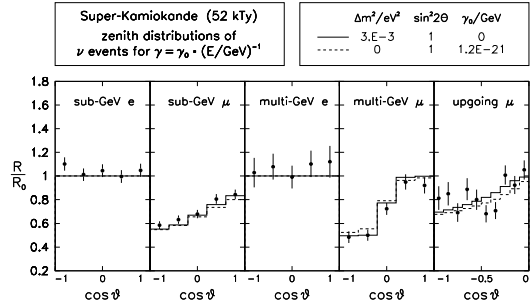}
\caption{Best-fit scenarios for pure oscillations ($\gamma =0$)
(solid line) and for pure decoherence with  $\gamma \propto 1/E$ 
(dashed line).}
\label{nudatadec2}
\end{figure}

From Atmospheric $\nu$ data one is led to the following bounds
for the QG-decoherence parameter 
$\gamma$ (c.f. figures \ref{nudatadec},\ref{nudatadec2})~\cite{lisi}:  

(a) $n=0$, $\gamma_0 < 3.5 \times 10^{-23}$ GeV. 

(b) $n=2$, $\gamma_0 < 0.9 \times 10^{-27}$ GeV. 

(c) $n=-1$, $\gamma_0 < 2 \times 10^{-21}$ GeV.  

Especially with respect to case (b) the reader is 
reminded that the 
CPLEAR bound on $\gamma$ for neutral Kaons\index{kaons} was 
$\gamma < 10^{-21}$ GeV~\cite{cplear}, i.e. the $\nu$-oscillation experiments
exhibit much higher sensitivity to QG decoherence effects than neutral meson
experiments.

Finally, I note that in \cite{klapdor} it was remarked that 
very stringent bounds on $\alpha,\beta$ and $\gamma$ 
(in the lepton number\index{lepton number} violating QG case) 
may be  
imposed by looking 
at oscillations 
of neutrinos from astrophysical sources (supernovae\index{supernovae} and AGN).
The corresponding bounds
on the $\gamma$ parameter 
from oscillation analysis of neutrinos from supernovae and AGN,
if QG induces such oscillations, are very strong: $\gamma < 10^{-40}$~GeV
from Supernova1987a, using the observed constraint~\cite{sncons} 
on the oscillation 
probability $P_{\nu_e \to \nu_\mu,\tau} < 0.2$, and  
$\gamma < 10^{-42}$~GeV from AGN, 
which exhibit sensitivity  
to order higher than $E^3/M_{qg}^2$, with $M_{qg} \sim M_P \sim 10^{19}$ GeV! 
Of course, the bounds from AGN 
do not correspond to real bounds, 
awaiting the observation of high energy neutrinos 
from such astrophysical sources. 
In \cite{klapdor} bounds have also been derived for the QG decoherence 
parameters by assuming that QG may induce neutrinoless double-beta decay. 
However, using current experimental 
constraints on neutrinoless double-beta decay\index{double-beta decay }
observables~\cite{klapdor2} one arrives at very weak bounds 
for the parameters $\alpha,\beta, \gamma$. 

One also expects stringent bounds on decoherence\index{decoherence} 
parameters, 
but also on deformed dispersion\index{dispersion relations} 
relations, if any, for neutrinos,
from future underwater neutrino telescopes\index{neutrino telescopes}, 
such as ANTARES~\cite{antares},
and NESTOR~\cite{nestor}~\footnote{As far as I understand, but I 
claim no expertise on this issue, the NESTOR experiment has an 
advantage
with respect to detection of very high energy cosmic neutrinos, which may be 
more sensitive probes of such quantum gravity effects.}.

\subsubsection{Three-Generation Models: Decoherence and the LSND Result}

As we discussed 
above, two-generation CPTV mass models for neutrinos
within quantum mechanics 
are excluded by
global fits of available data, especially solar neutrino models. 
This situation is not expected to change by the inclusion of a 
third generation, 
although I must stress that, as far as I am aware of, 
complete three-generations
analyses of this kind have not been performed as yet. 
Moreover, although four generation CPTV neutrino models 
are still consistent with experimental data, nevertheless there seems 
to be no experimental evidence 
for a forth generation,
especially after the recent WMAP astrophysical results.  
On the other hand, as we have just seen, 
two-generation neutrino analysis of decoherence
effects did not show any spectacular results, apart from 
the imposition of stringent bounds on the relevant parameters. 

This prompts one to think that the extension of the decoherence
formalism to three generations\index{generation} of neutrinos, 
which from a mathematical 
view point is a problem with considerable increase in technical 
complexity, is a futile task, with no physical importance whatsoever. 
However, there are 
the 
``anomalous'' results provided by the LSND\index{LSND Experiment} 
collaboration~\cite{lsnd}
on the evidence
for ${\overline \nu_e} \to {\overline \nu_\mu}$ oscillations, through 
${\overline \nu_e}$ disappearance, but not for the cooresponding oscillations
in the neutrino sector, 
which call for an explanation, if one, of course,  
takes them seriously into account. These effects, as we have seen, 
cannot be explained by conventional quantum field theoretic analyses,
even if CPT\index{CPT} is assumed violated. 

It is the point of this subsection to point out
that, if one extends the
decoherence analysis to three generations of neutrinos and allows for 
CPT Violation\index{CPT Violation}
among the decoherence\index{decoherence} parameters, 
it is possible\cite{bm} 
to fit all the currently availble neutrino data, including the 
LSND results, 
by simple decoherence models,
in which the dominant decoherence parameters occur in the antineutrino sector.
It is important that in such ``asymmetric'' 
decoherence models 
there is no need for 
enlarging the neutrino sector by a fourth generation, neither for 
introducing CPTV mass parameters. If the LSND results are confirmed
by future experiments, then this would be a significant result, as it 
would provide for the first time a clear experimental evidence for 
a CPTV decoherence\index{decoherence} event, 
which would be directly related to quantum gravity effects.

Let us briefly present the arguments leading to these results. 
Formally, the extension of the completely
positive decoherence scenario to the standard three-generation neutrino 
oscillations case is straightforward, and it was described in section two. 
One adopts a three-state Lindblad\index{Lindblad} problem, and, 
following the standard procedure outlined  there, 
one determines the corresponding 
eigenvectors and eigenvalues, as in the two-level case examined in the previous
subsection. It is only a considerable increase in mathematical complexity,
and obscurity in the precise physical meaning of all the non-trivial 
entries of the 
decoherence matrix that one encounters here. 

The relativistic neutrino\index{Relativistic Neutrino Hamiltonian} Hamiltonian
$H_{\rm eff} \sim p^2 + m^2/2p$, with $m$ the neutrino mass,  
is used as the effective Hamiltonian
of the subsystem in the evolution equation (\ref{lindblad}).
In terms of the generators ${\cal J}_\mu$, $\mu = 0, \dots 8$ 
of the SU(3) group, $H_{\rm eff}$ can be expanded as~\cite{gago}:
${\cal H}_{\rm eff} = \frac{1}{2p}\sqrt{2/3}\left(6p^2 + \sum_{i=1}^{3}m_i^2
\right){\cal J}_0 + \frac{1}{2p}(\Delta m_{12}^2){\cal J}_3
+ \frac{1}{2\sqrt{3} p}\left(\Delta m_{13}^2 + \Delta m_{23}^2 \right){\cal J}_8$, 
with the obvious notation $\Delta m_{ij}^2 = m_i^2 - m_j^2$, $i,j =1,2,3$.

The analysis of \cite{gago} assumed {\it ad hoc} a diagonal form 
for the $9 \times 9$ decoherence matrix ${\cal L}$ in (\ref{expandedlind}):
\begin{equation} 
[{\cal L}_{\mu\nu}]= {\rm Diag}\left(0, -\gamma_1,-\gamma_2,-\gamma_3,-\gamma_4,-\gamma_5,-\gamma_6,-\gamma_7,-\gamma_8\right)
\label{diagonal}
\end{equation}
in direct analogy with 
the two-level case of complete positivity~\cite{lisi,benatti}. 
As we have mentioned already, there is no strong physical
motivation behind such restricted forms of decoherence. This assumption,
however,
leads to  the simplest possible decoherence models, 
and, for our  
{\it phenomenological} purposes
in this work,  
we will assume 
the above form, which we will use
to fit all the available neutrino data. It must be clear to the reader though,
that such a simplification, if proven to be successful (which, as we shall 
argue below, is the case here),
just adds more in favor of decoherence models, 
given the 
restricted number of available parameters for the fit in this case. 
In fact, any other non-minimal
scenario will have it easier to accommodate data 
because it will have more degrees of freedom available for such a 
purpose.

Specifically we shall look at transition probabilities (\ref{genprobs}),
which can be computed in a straightforward manner
within the dynamical-semigroups\index{dynamical semigroup} approach outlined previously~\cite{gago}:
\begin{eqnarray} 
&& P(\nu_\alpha \to \nu_\beta) ={\rm Tr}[\rho^\alpha (t)\rho^\beta] 
= \nonumber \\
&&\frac{1}{3} + \frac{1}{2}\sum_{i,k,j}
e^{\lambda_k t}{\cal D}_{ik}{\cal D}_{kj}^{-1}\rho^\alpha_j (0)\rho_i^\beta
\label{trans}
\end{eqnarray}
where $\alpha,\beta = e, \mu, \tau$ stand for the three neutrino flavors, and 
Latin indices run over $1, \dots 8$. The quantities $\lambda_k$ 
are the eigenvalues of the matrix ${\cal M}$ appearing in the 
evolution (\ref{expandedlind}), after taking into account probability
conservation, which decouples $\rho_0(t)=\sqrt{2/3}$, leaving the remaining 
equations in the form: $\partial \rho_k /\partial t = \sum_{j} 
{\cal M}_{kj}\rho_j $. The matrices ${\cal D}_{ij}$ are the matrices
that diagonalize ${\cal M}$~\cite{lindblad}. Explicit forms of 
these matrices, the eigenvalues\index{eigenvalues} $\lambda_k$, 
and consequently the transition probabilities
(\ref{trans}), are given in \cite{gago}. 

The important point to stress is that, in generic models of oscillation plus 
decoherence, the eigenvalues $\lambda_k$ depend
on both the decoherence parameters $\gamma_i$ and the mass differences
$\Delta m^2_{ij}$. For instance, $\lambda_1 = \frac{1}{2}[-(\gamma_1 + \gamma_2)
-\sqrt{(\gamma_2 - \gamma_1)^2 -4\Delta_{12}^2}]$, with 
the notation $\Delta_{ij} \equiv \Delta m_{ij}^2/2p$, $i,j=1,2,3$.
 Note that, to leading order in the (small) squared-mass differences, 
one may replace
$p$ by the total neutrino energy $E$, and this will be understood 
in what follows.

We now note that 
it is a generic feature of the $\lambda_k$ to depend on  
the quantities 
$\Omega_{ij}$ which are given by\cite{bm,gago}
\begin{eqnarray}
\Omega_{12} &=& \sqrt{(\gamma_2 - \gamma_1)^2 -4\Delta_{12}^2} \nonumber \\
\Omega_{13} &=& \sqrt{(\gamma_5 - \gamma_4)^2 -4\Delta_{13}^2} \nonumber \\
\Omega_{23} &=& \sqrt{(\gamma_7 - \gamma_6)^2 -4\Delta_{23}^2} 
\qquad {\rm etc.}
\label{omegas}
\end{eqnarray}
{}From the above expressions  
for the eigenvalues $\lambda_k $, it becomes clear
that, when decoherence and oscillations
are present simultaneously,
one should distinguish two cases, 
according to the relative magnitudes of $\Delta_{ij}$ and 
$\Delta\gamma_{kl} \equiv \gamma_k - \gamma_l$: (i) $2|\Delta_{ij}| \ge |\Delta\gamma_{k\ell}|$, and (ii) 
$2|\Delta_{ij}| < |\Delta\gamma_{k\ell}|$.
In the former case, the probabilities (\ref{trans}) contain trigonometric
(sine and cosine) functions, whilst in the latter they exhibit  
hyperbolic sin and cosine dependence.

Assuming mixing\index{mixing} between the flavours, amounts to expressing neutrino
flavor eigenstates $|\nu_\alpha>$, $\alpha=e,\mu,\tau$ in terms of 
mass eigenstates $|\nu_i>$, $i=1,2,3$ through a (unitary) matrix $U$:
$|\nu_\alpha> = \sum_{i=1}^{3}U^*_{\alpha i}|\nu_i>$. This implies that
the density matrix of a flavor state $\rho^\alpha $ can be expressed 
in terms of mass eigenstates as: 
$\rho^\alpha=|\nu_\alpha><\nu_\alpha|=\sum_{i,j}U^*_{\alpha i}U_{\alpha j}|\nu_i><\nu_j| $. From this we can determine 
$\rho_\mu^\alpha =2{\rm Tr}(\rho^\alpha {\cal J}_\mu )$, a quantity needed to
calculate the transition probabilities (\ref{trans}).

The important comment~\cite{bm}
we would like to raise at this point is that, when considering
the above probabilities in the antineutrino\index{antineutrino}
sector, the respective decoherence\index{decoherence} parameters ${\bar \gamma}_i$ 
in general may be different from the corresponding ones in the 
neutrino sector,
as a result of the strong form of CPT violation. In fact, as we 
shall discuss next, this will be crucial for accommodating 
the LSND result without conflicting with  the rest of the available neutrino data.
This feature is totally unrelated to mass differences between flavors. 

In \cite{gago} a pessimistic conclusion was drawn on 
the ``clear incompatibility between neutrino data and theoretical
expectations", as followed by their qualitative tests for decoherence.
It is a key feature of the work of \cite{bm} 
to point out that this point of view may not be true at all. 
In fact, as we shall demonstrate below, 
if one takes into account  all the available 
neutrino data, including the final 
LSND results\index{LSND Experiment}~\cite{lsnd}, 
which the authors of \cite{gago} did not do,
and allows for the above mentioned CPT violation in the decoherence
sector, then one will arrive at exactly the opposite conclusion, namely 
that three-generation decoherence and oscillations can fit 
the data successfully!
 
As shown in \cite{bm}, compatibility of all available 
data, including CHOOZ~\cite{chooz} and LSND, can be achieved through a  
set of decoherence parameters $\gamma_j$ with energy dependences 
$\gamma^0_j E$  and $\gamma^0_j/E$,  
with $\gamma^0_j  
\sim 10^{-18}, 10^{-24}$~(GeV)$^2$, respectively, 
for some $j$'s, 
and in fact the fit ends up being  {\it significantly better} than 
the standard one (when LSND results are included)
as evidenced by an appropriate $\chi^2$ analysis. 

Some important remarks are in order. First of all, 
in the analysis of \cite{gago} 
pure decoherence is excluded
in three-generation scenaria, as in two generation ones, 
due to 
the fact that the transition probabilities in the case $\Delta m_{ij}^2=0$
(pure decoherence) are such that the survival probabilities 
in both sectors are equal, i.e. 
$P(\nu_\alpha \to \nu_\alpha ) =
P({\overline \nu}_\alpha \to {\overline \nu}_\alpha) $. 
{}From (\ref{trans}) we have in this case~\cite{gago}:
\begin{eqnarray} \label{trans2}
P_{\nu_e \to \nu_e} = P_{\nu_\mu \to \nu_\mu}
\simeq \frac{1}{3} + \frac{1}{2}e^{-\gamma_3 t}
+ \frac{1}{6}e^{-\gamma_8 t} 
\end{eqnarray}
{}From the CHOOZ experiment~\cite{chooz}, for which $L/E \sim 10^3/3$~m/MeV,
we have that $\langle P_{{\bar \nu}_e \to {\bar \nu}_e} \rangle \simeq 1$, 
while the K2K 
experiment~\cite{K2K} with $L/E \sim 250/1.3 $~ km/GeV has observed events
compatible with $\langle P_{\nu_\mu \to \nu_\mu} \rangle \simeq 0.7$,
thereby contradicting the theoretical predictions 
(\ref{trans2}) of pure decoherence. 

However, this conclusion is based on the fact that in the antineutrino sector
the decoherence matrix is the same as that in the neutrino sector. 
In general this need not be the case, in view of CPT Violation\index{CPT Violation}, which 
could imply a different interaction 
of the antiparticle with the gravitational
environment as compared with the particle. 
In fact in models where a pure state evolves 
to a mixed one, one expects a CPT Violation in the strong form, 
according to the theorem of \cite{wald}.

In our tests we took into account this possibility, but 
pure decoherence can be excluded also in this case, 
as it is clearly  incompatible
with the totality of the available data.

In order to check our model, we have performed a 
$\chi^2 $ comparison  (as opposed to a $\chi^2 $ fit which is still pending)
to SuperKamiokande sub-GeV and multi GeV data, CHOOZ
data and LSND, for a sample point in the
vast parameter space of our 
extremely simplified version of decoherence\index{decoherence} models. 
Since we have {\it not} performed as yet 
a $\chi^2$-fit, the point
we are selecting (rather visually and not by a proper $\chi^2$ analysis) 
is not optimized to give
the best fit to the existing data. Instead, it must be regarded 
as one among the many equally good members in this family of solutions,
being extremely possible to find another model that fits better 
the data, through a 
complete (and highly time consuming) scan over the whole parameter 
space.  

Cutting the long story short, and to make the analysis easier, 
we have set\cite{bm}
all the $\gamma_i$ in the neutrino sector to zero, restricting in this way
all the dominant decoherence effects 
in the antineutrino sector only. For the sake of simplicity 
we have assumed the form:
\begin{eqnarray} 
&&\bar{\gamma}_i = \bar{\gamma}_{i+1}~{\rm for}~~ i=1, 4, 6 \qquad
{\rm and} \qquad 
\bar{\gamma}_3 = \bar{\gamma}_{8} 
\label{decpars}
\end{eqnarray} 
Later on we shall 
set some of the $\gamma_i$'s to zero. Furthermore, we have also set the 
CP violating\index{CP Violation} 
phase of the NMS matrix to zero, so that all the mixing
matrix elements become real.

With these assumptions, the otherwise cumbersome 
expression (see end of section
for more detailed results)
for the transition
probability for the antineutrino\index{antineutrino} sector takes the form: 
\begin{eqnarray}
  P_{\bar\nu_{\alpha}\rightarrow \bar\nu_{\beta}} & = &\frac{1}{3}+\frac{1}{2}\left\{ 
\rho_{1}^{\alpha}\rho_{1}^{\beta}
\cos\left(\frac{|\Omega_{12}| t}{2}\right)
e^{-\bar\gamma_{1} t} \right. \nonumber \\
&+ &  \rho_{4}^{\alpha}\rho_{4}^{\beta}
\cos\left(\frac{|\Omega_{13}|t}{2}\right)  e^{-\bar\gamma_{4} t} \nonumber \\
&+& \rho_{6}^{\alpha}\rho_{6}^{\beta} 
\cos\left(\frac{|\Omega_{23}|t}{2}\right) e^{-\bar\gamma_{6} t} 
\nonumber \\
&+ &   e^{-\bar\gamma_{3}t} \left( \rho_{3}^{\alpha}\rho_{3}^{\beta}+
\rho_{8}^{\alpha}\rho_{8}^{\beta} \right) \Bigg\}. 
\label{prob}
\end{eqnarray}
where the $\Omega_{ij}$ were 
defined in the previous section 
and are the same in both sectors (due to our choice of $\gamma_i$'s) and
\begin{eqnarray}\label{mixpar}
  \nonumber  \rho^{\alpha}_{0} &=& \sqrt{\frac{2}{3}}
    \\ \nonumber \rho^{\alpha}_{1} &=& 2 Re(U_{\alpha 1}^{*}U_{\alpha 2})
    \\ \nonumber
\rho^{\alpha}_{2} &=& -2 Im(U_{\alpha 1}^{*}U_{\alpha 2})
    \\ \nonumber\rho^{\alpha}_{3} &=& |U_{\alpha 1}|^{2}-|U_{\alpha 2}|^{2}
    \\ \nonumber\rho^{\alpha}_{4} &=& 2 Re(U_{\alpha 1}^{*}U_{\alpha 3})
    \\ \nonumber\rho^{\alpha}_{5} &=& -2 Im(U_{\alpha 1}^{*}U_{\alpha 3})
    \\\nonumber \rho^{\alpha}_{6} &=& 2 Re(U_{\alpha 2}^{*}U_{\alpha 3})
    \\\nonumber \rho^{\alpha}_{7} &=& -2 Im(U_{\alpha 2}^{*}U_{\alpha 3})
    \\ \rho^{\alpha}_{8} &=&\sqrt{\frac{1}{3}} \left( |U_{\alpha 1}|^{2}
    +|U_{\alpha 2}|^{2}- 2|U_{\alpha 3}|^{2}\right)
 \end{eqnarray}
where the mixing\index{mixing} matrices are the same as in the 
neutrino sector.
For the neutrino sector, as there are no dominant 
decoherence effects, the standard
expression for the transition probability is valid.

It is obvious now that, since the neutrino sector does not suffer from
decoherence, there is no need to include the solar data into the fit.
We are guaranteed to have an excellent agreement with solar data, as long
as we keep the relevant mass difference and mixing\index{mixing} angle within
the LMA region.
As mentioned previously, CPT violation is driven by, and restricted to, the
decoherence parameters, and hence masses and mixing angles are
the same in both sectors, and selected to be 

\centerline{$\Delta m_{12}^2 = \Delta \overline{ m_{12}}^2 = 
7 \cdot 10^{-5}$~eV$^2$,}

\centerline{$\Delta m_{23}^2 = \Delta \overline{ m_{23}}^2 = 
2.5 \cdot 10^{-3}$~eV$^2$,}

\centerline{$\theta_{23} = \overline{\theta_{23}}= \pi/4$, $\theta_{12} = 
\overline{\theta_{12}}= .45$,}

\centerline{$\theta_{13} = \overline{\theta_{13}}= .05$,} 

\noindent 
as indicated by the state of the art phenomenological analysis 
in neutrino physics.

For the decoherence parameters we have chosen (c.f. (\ref{decpars})) 
\begin{equation} 
\overline{\gamma_{1}}= \overline{\gamma_{2}} = 2 \cdot 10^{-18} \cdot E ~~{\rm and}~~ 
\overline{\gamma_{3}} = \overline{\gamma_8}=
 1 \cdot 10^{-24}/E~, 
\label{decohparam}
\end{equation}
where 
$E$ is  the  neutrino energy, 
and barred quantities
refer to the antineutrinos\index{antineutrino}, 
given that decoherence\index{decoherence} takes place only 
in this 
sector in our model.
All the other parameters 
are assumed to be zero.  All in all, we have introduced only
two new parameters, two new degrees of freedom, $ \overline{\gamma_{1}}$
and $ \overline{\gamma_{3}}$,
and we shall try
to explain with them all the available experimental data.

\begin{figure}
\centering
\includegraphics[width=9cm]{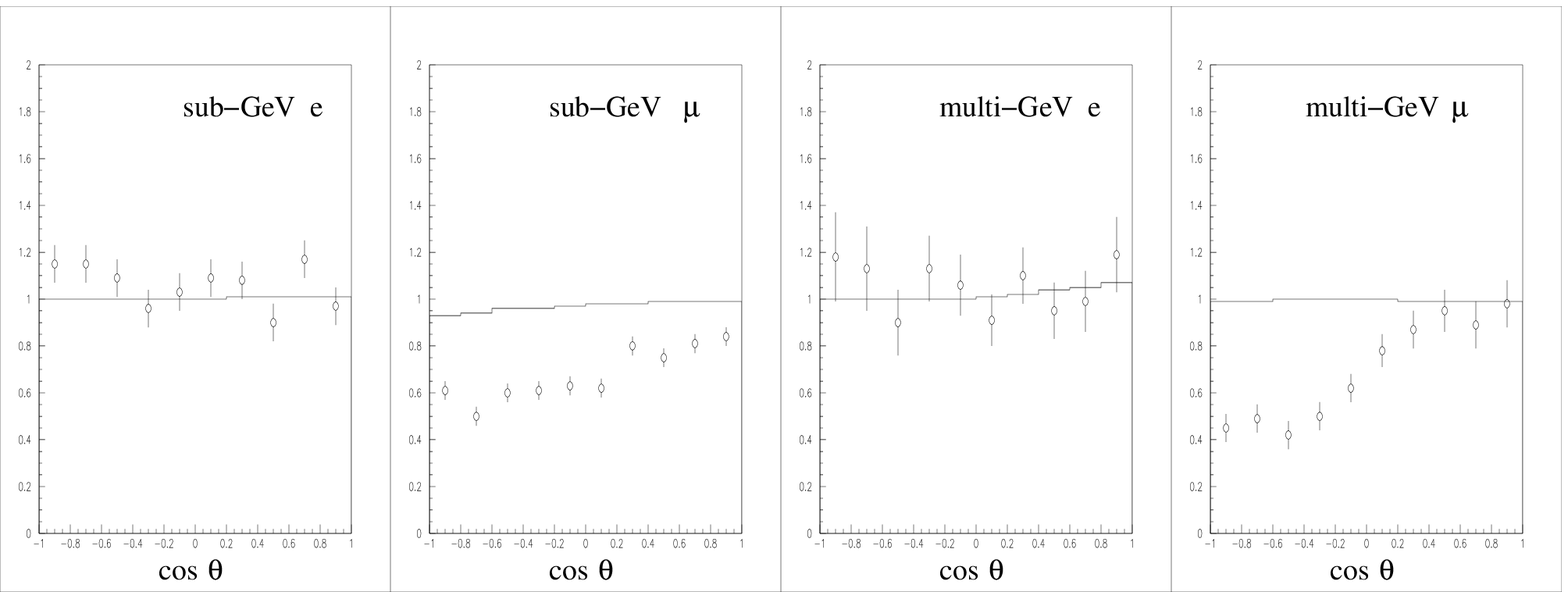} \hfill 
\includegraphics[width=9cm]{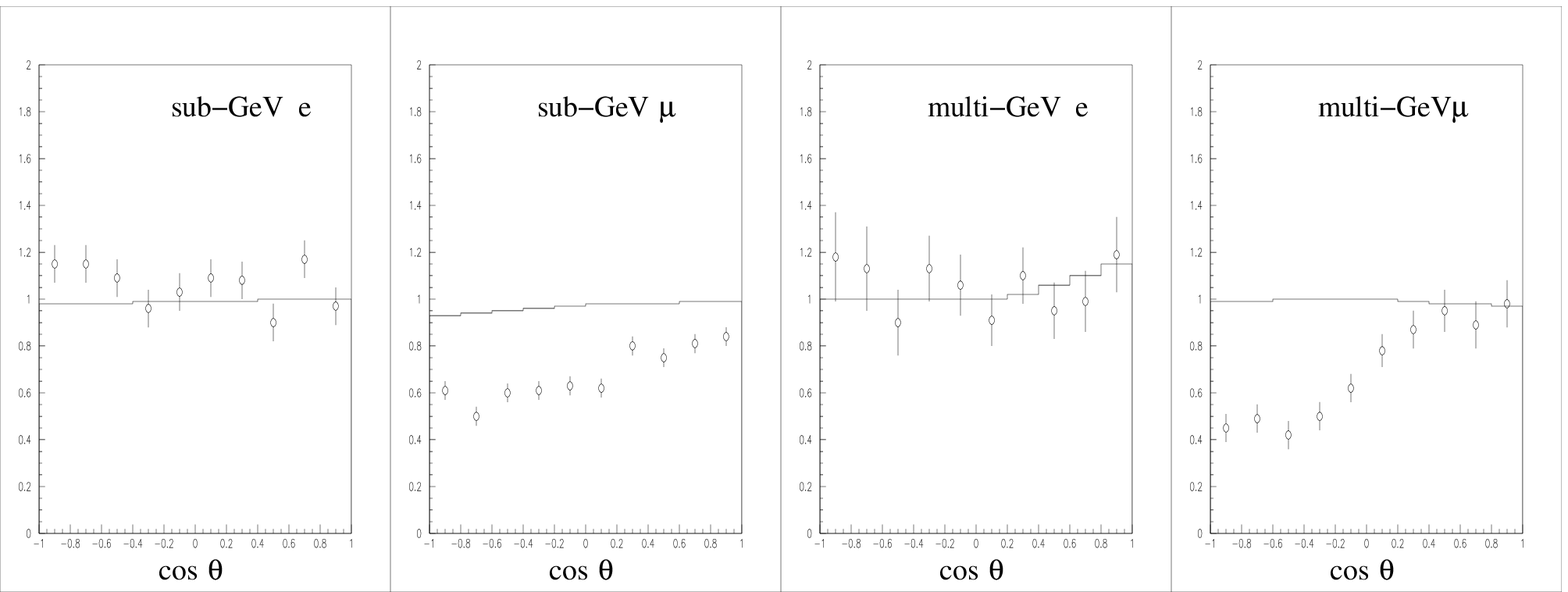} \hfill
\includegraphics[width=9cm]{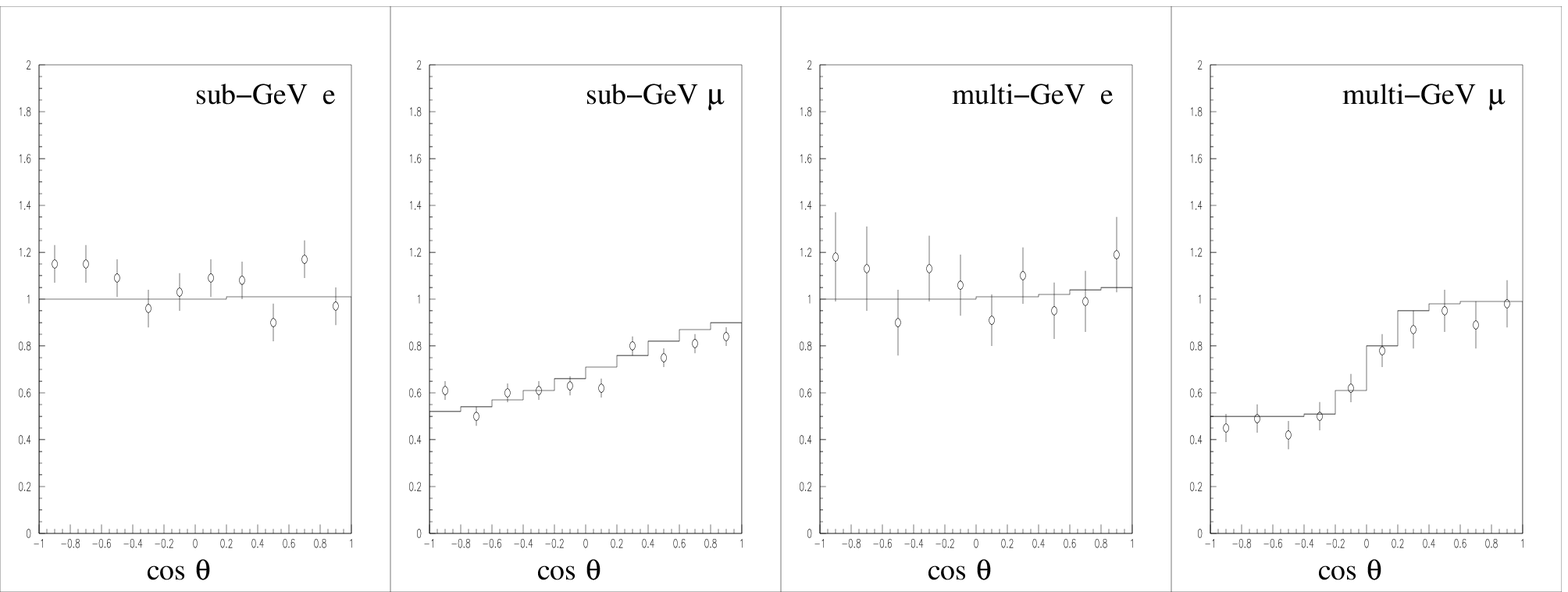}\hfill
\includegraphics[width=9cm]{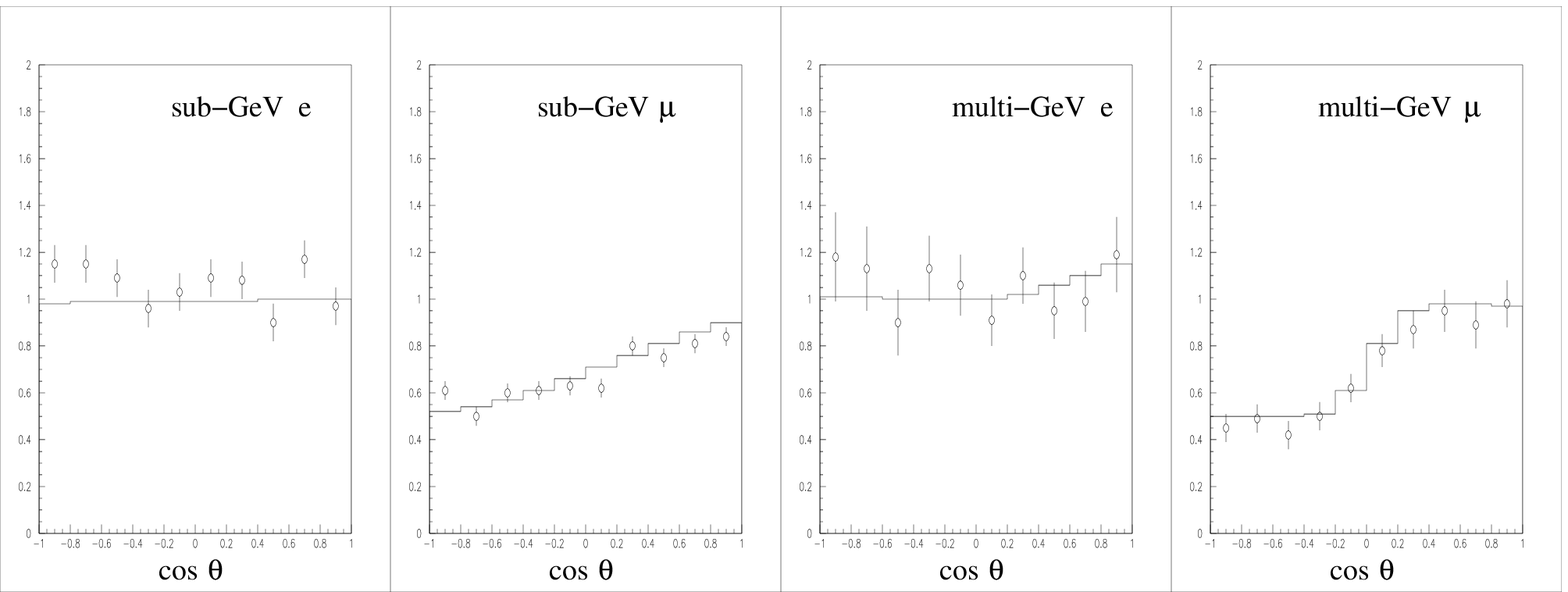}
\caption{Decoherence fits, from top to bottom: (a) pure decoherence in 
antineutrino sector, (b) pure decoherence in both sectors, (c) mixing plus
decoherence in the antineutrino sector only, (d) mixing plus decoherence
in both sectors. The dots correspond to SK data.}
\label{bestfit}
\end{figure}

In order to test our model 
with these two decoherence parameters in the antineutrino\index{antineutrino} sector,  
we have calculated the zenith angle dependence of the 
ratio ``observed/(expected in the no oscillation case)'', for muon\index{muon} and
electron atmospheric\index{atmospheric} 
neutrinos, for the sub-GeV and multi-GeV energy ranges,
when mixing is taken into account.
The results are shown in Fig. \ref{bestfit}.  
where, for the sake of comparison,
we have also included 
the experimental data.  

As bare-eye comparisons can be misleading, we have also calculated the
$\chi^2$  value for each of the cases, defining the atmospheric 
$\chi^2$  as
\begin{equation}
  \chi^2_{\rm atm}= \sum_{M, S}\sum_{\alpha=e,\mu}\sum_{i=1}^{10} 
     \frac{(R_{\alpha,i}^{\rm exp}-
     R_{\alpha,i}^{\rm th})^2}{\sigma_{\alpha i}^2} \quad .
\end{equation}
Here $\sigma_{\alpha,i}$ are the statistical errors,
the ratios $R_{\alpha,i}$ between the observed and predicted signal 
can be written as
\begin{equation}
  R_{\alpha,i}^{\rm exp}= N_{\alpha,i}^{\rm exp}/N_{\alpha,i}^{\rm MC}
\end{equation}
(with $\alpha$ indicating the lepton flavor and $i$ counting the
different bins, ten in total)
and $M,S$ stand for the multi-GeV and sub-GeV data respectively. 
For the CHOOZ experiment we used the 15 data points with their
statistical errors, where in each bin we averaged the probability
over energy and for LSND\index{LSND Experiment} one datum has been included.
The results  with which we hope  all our claims 
become crystal clear are summarized in Table \ref{table3}, were we
present the $\chi^2$ comparison for the following cases:
(a) pure decoherence in the antineutrino sector, (b) pure
decoherence in both sectors, (c) mixing plus decoherence in
the antineutrino sector, (d) mixing plus decoherence in
both sectors, and (e) mixing only - the standard scenario.

\begin{table}[h]
\centering
\begin{tabular}{|c|c|c|}
\hline\hline
model & $\chi^2$ without LSND  & $\chi^2$ including LSND  \\ [0.5 ex]
\hline\hline
(a) & 980.7 & 980.8\\ \hline
(b) & 979.8 & 980.0\\ \hline
(c) & 52.2 & 52.3\\\hline
(d) & 54.4 & 54.6\\\hline
(e) & 53.9 & 60.7\\[1ex]
\hline\hline
\end{tabular}
\caption{$\chi^2$ obtained for (a) pure decoherence in 
antineutrino sector, (b) pure decoherence in both sectors, (c) mixing plus
decoherence in the antineutrino sector only, (d) mixing plus decoherence
in both sectors, (e) standard scenario with and without the LSND result. }
\label{table3}
\end{table}

{}From the table it becomes clear that the mixing plus decoherence scenario
in the antineutrino sector can easily account for all the available 
experimental information,
including LSND\index{LSND Experiment} data. 
It is important to stress once more that our sample
point was not obtained through a scan over all the parameter space,
but  by an educated guess, and therefore plenty of room is left
for improvements. 
At this point a word of warning is in order: although superficially
it seems that scenario (d), decoherence\index{decoherence} plus mixing in both sectors,
provides an equally good fit, one should remember that including
decoherence effects in the neutrino sector can have undesirable
effects in solar neutrinos, especially due to 
the fact that decoherence
effects are weighted by the distance traveled by the neutrino, 
something that may lead to seizable (not observed!) effects in the
solar case.

One might wonder then, whether decohering effects, which affect
the antineutrino sector sufficiently to account for the LSND result, 
have any impact on the solar-neutrino 
related parameters, measured through 
antineutrinos in the KamLAND\index{KamLAND Experiment} experiment\cite{kamland}. 
In order to answer this question,
 it will be sufficient to calculate the electron survival probability
 for KamLAND in our model, which turns out to be 
$ P_{\bar\nu_{\alpha}\rightarrow \bar\nu_{\beta}} \mid_{\mbox{\tiny  KamLAND}}
\simeq .63$, in perfect agreement with observations. It is also interesting
to notice that in our model,
  the LSND effect is not given  by the phase inside
the oscillation term ( which is proportional to the solar mass difference)
 but rather by the decoherence factor multiplying the oscillation term.
 Therefore the tension between LSND\index{LSND Experiment} and
KARMEN\cite{karmen} data is naturally eliminated, because the difference
in length leads to an exponential suppression.
 
Having said that,
it is now clear that decoherence models (once neutrino mixing is taken
into account) are the best (and arguably the only) way to explain 
all the observations including 
the LSND result. This scenario , which makes dramatic predictions for
the upcoming neutrino experiments, expresses a strong observable form of 
CPT violation in the 
laboratory, and in this sense, our fit gives a clear answer to the 
question as to whether the weak form of CPT 
invariance (\ref{probscptweak}) 
is violated in Nature. It seems that, in order to 
account for the LSND results, we should invoke such a decoherence-induced 
CPT Violation\index{CPT Violation}, which however 
is independent of any mass differences
between particles and antiparticles\index{antiparticle}.

This CPT violating pattern, with equal mass spectra for
neutrinos and antineutrinos, if true, will have dramatic signatures in 
future neutrino oscillation experiments. The most striking 
consequence
will be seen in MiniBooNE\index{MiniBooNE Experiment}~\cite{miniboone},
According to our picture, MiniBooNE will be able to confirm LSND
only when running in
the antineutrino mode and not in the neutrino one, as decoherence\index{decoherence}
effects live only in the former. Smaller but experimentally 
accessible signatures
will be seen also in MINOS\index{MINOS Experiment}~\cite{minos}, 
by comparing conjugated channels (most
noticeably, the muon\index{muon} survival probability).

We next remark that fits with decoherence
parameters with energy dependences of the form (\ref{decohparam}) 
imply that the 
exponential factors $e^{\lambda_k t}$ in (\ref{trans})
due to decoherence will  modify the amplitudes of the oscillatory terms
due to mass differences, and while one term  depends on  
$L/E$ the other one is driven by $L\cdot E$, where
we have set $t=L$, with $L$ the oscillation 
length\index{oscillation length} 
(we are working with natural units where $c=1$). 

The order of the coefficients of these quantities, 
$\gamma^0_j  \sim 10^{-18}, 10^{-24}$~(GeV)$^2$, found in 
our sample point, implies that for energies of a few GeV,
which are typical of
the pertinent experiments, 
such values are not far from  
$\gamma_j^0 \sim \Delta m_{ij}^2$. If our conclusions
survive the next round of experiments, and therefore if 
MiniBOONE experiment\index{MiniBooNE Experiment}~\cite{miniboone} 
confirms previous LSND claims,
then this may be a significant result. 

Indeed, one would be tempted to 
speculate that, if the above estimate holds, 
and the decoherence coefficients are proportional
to the neutrino mass-squared differences,  
this could even indicate that
the neutrino mass differences themselves
might 
be due to quantum gravity decoherence, in the sense of 
environmental contributions to the effective neutrino 
Hamiltonian appearing 
in the decoherent evolution (\ref{lindblad}), which could
mascarade themselves as mass terms. 
Theoretically it is still 
unknown how the neutrinos acquire a mass, or what kind of mass 
(Majorana\index{Majorana}
or Dirac\index{Dirac}) 
they possess. There are scenaria in which the mass of neutrino 
may be due to some peculiar backgrounds of string theory for instance.
If the above model turns out to be right we might then have, 
for the first time in low energy physics, 
an indication of a direct detection of a quantum gravity effect, which 
disguised itself as an induced decohering neutrino mass difference. 
Notice that in our  sample point only antineutrinos have non-trivial 
decoherence parameters $\overline{\gamma_{i}}$ , for $i=1$ and 3,
while the corresponding quantities in the neutrino sector vanish.  
This implies that there is a single cause for mass differences,
the decoherence in antineutrino\index{antineutrino} sector, 
which is compatible 
with common mass differences in both  sectors. 
This would be very interesting, if true.

Finally, before closing, we would like to remark 
on extensions of the above phenomenologiocal model for decoherence
by including non-diagonal terms in the decoherence matrix $L_{\mu\nu}$. 
As mentioned above, the physical significance of such extensions
is not clear, and indeed it 
cannot be clear from simple phenomenological analyses
like the one presented here. 
One needs a detailed knoweldge of
the QG decoherence effects so as to obtain such an understanding. 

Nevertheless one may test the phenomenological efficiency of 
the simple parametrisation of \cite{gago,bm} by comparing the results
on the oscillation probabilities versus models where off diagonal terms
are included in the decoherence matrix. 
As a simple example, consider the following form of the 
decoherence
matrix~\cite{waldron}: 
{\small  \begin{eqnarray} 
{\cal M}=\left(\begin{array}{cccccccc}
  L_{11} & -\Delta_{12}+L_{12} & 0 & 0 & 0 & 0 & 0 &0 \\
  \Delta_{12}+L_{12} & L_{22} & 0 & 0 & 0 & 0 & 0 & 0 \\
  0 & 0 & L_{33} & 0 & 0 & 0 & 0 & 0 \\
  0 & 0 & 0 & L_{44} & -\Delta_{13}+L_{45} & 0 & 0 & 0 \\
  0 & 0 & 0 & \Delta_{13}+L_{45} & L_{55} & 0 & 0 & 0 \\
  0 & 0 & 0 & 0 & 0 & L_{66} & -\Delta_{23}+L_{67} & 0 \\
  0 & 0 & 0 & 0 & 0 & \Delta_{23}+L_{67} & L_{77} & 0 \\
  0 & 0 & 0 & 0 & 0 & 0 & 0 & L_{88} \nonumber \\
\end{array}\right) \nonumber \\
\label{decmatrix}
\end{eqnarray}} 
where ${\cal M}$ is the matrix appearing in the 
decoherent evolution\index{decohering evolution} 
(\ref{matrixform}). It is again  a 
straightforward but tedious exercise to determine
the matrix which diagonalises ${\cal M}$ 
and find the eigenvalues\index{eigenvalues} 
and eigenvectors\index{eigenvectors} 
of ${\cal M}$, which determine the oscillation 
probabilities (\ref{trans}). Defining $\Gamma_{ij}$ as  
 \begin{equation}
   \Gamma_{12}\equiv \sqrt{(L_{11}-L_{22})^{2}+ 4L_{12}^{2}-4\Delta_{12}^{2}}
\label{defgammas}  
\end{equation}
and similarly for the other elements,   
and taking notice of the fact that $\Gamma_{ij}$ are  similar to 
the $\Omega_{ij}$ of the diagonal 
decoherence case (\ref{omegas}),
with the only difference being 
an extra positive term ($L_{12}^2$ {\it etc.}) under the sqare root, 
we can compute the corresponding oscillation probabilities (\ref{trans}). 
For completeness we give here the relevant expression, 
which allows the interested reader to derive the diagonal case expressions
by setting the off-digonal elements of $L_{\mu\nu}$ equal to zero.
We have: 
{\small \begin{eqnarray}\nonumber 
     P_{\nu_{\alpha}\rightarrow \nu_{\beta}}(t)= \frac{1}{3}
    + \frac{1}{2}e^{\frac{(L_{11}+L_{22})t}{2}}\left\{\left(\rho_{1}^{\alpha}
    \rho_{1}^{\beta}+\rho_{2}^{\alpha}
    \rho_{2}^{\beta}\right)\left(\frac{e^{\frac{\Gamma_{12}t}{2}}+e^{-\frac{\Gamma_{12}t}{2}}}{2}\right)
    \right. \qquad \qquad \qquad \qquad \qquad \qquad
    \\ \nonumber \left. +\left[\left(\rho_{1}^{\alpha}\rho_{1}^{\beta}-\rho_{2}^{\alpha}
    \rho_{2}^{\beta}\right)\left( \frac{L_{11}-L_{22}}{\Gamma_{12}}
    \right) +\frac{2\rho_{2}^{\alpha}\rho_{1}^{\beta}(-L_{12}+\Delta_{12})
    -2\rho_{1}^{\alpha}\rho_{2}^{\beta}(L_{12}+\Delta_{12})}{\Gamma_{12}}
    \right]\left( \frac{e^{\frac{\Gamma_{12}t}{2}}-e^{-\frac{\Gamma_{12}t}{2}}}{2}     \right) \right\}
    \\  \nonumber +  e^{\frac{(L_{44}+L_{55})t}{2}}\left\{\left(\rho_{4}^{\alpha}
    \rho_{4}^{\beta}+\rho_{5}^{\alpha}
    \rho_{5}^{\beta}\right)\left(\frac{e^{\frac{\Gamma_{13}t}{2}}+e^{-\frac{\Gamma_{13}t}{2}}}{2}\right)
    \right.\qquad \qquad \qquad \qquad \qquad \qquad
    \\  \nonumber \left. +\left[\left(\rho_{4}^{\alpha}\rho_{4}^{\beta}-\rho_{5}^{\alpha}
    \rho_{5}^{\beta}\right)\left( \frac{L_{44}-L_{55}}{\Gamma_{13}}
    \right) +\frac{2\rho_{5}^{\alpha}\rho_{4}^{\beta}(-L_{45}+\Delta_{13})
    -2\rho_{4}^{\alpha}\rho_{5}^{\beta}(L_{45}+\Delta_{13})}{\Gamma_{13}}
    \right]\left( \frac{e^{\frac{\Gamma_{13}t}{2}}-e^{-\frac{\Gamma_{13}t}{2}}}{2}     \right) \right\}
    \\  \nonumber +  e^{\frac{(L_{66}+L_{77})t}{2}}\left\{\left(\rho_{6}^{\alpha}
    \rho_{6}^{\beta}+\rho_{7}^{\alpha}
    \rho_{7}^{\beta}\right)\left(\frac{e^{\frac{\Gamma_{23}t}{2}}+e^{-\frac{\Gamma_{23}t}{2}}}{2}\right)
    \right.\qquad \qquad \qquad \qquad \qquad \qquad
    \\  \nonumber \left. +\left[\left(\rho_{6}^{\alpha}\rho_{6}^{\beta}-\rho_{7}^{\alpha}
    \rho_{7}^{\beta}\right)\left( \frac{L_{66}-L_{77}}{\Gamma_{23}}
    \right) +\frac{2\rho_{7}^{\alpha}\rho_{6}^{\beta}(-L_{67}+\Delta_{23})
    -2\rho_{6}^{\alpha}\rho_{7}^{\beta}(L_{67}+\Delta_{23})}{\Gamma_{23}}
    \right]\left( \frac{e^{\frac{\Gamma_{23}t}{2}}-e^{-\frac{\Gamma_{23}t}{2}}}{2}     \right) \right\}
    \\ + e^{L_{33}t}\rho_{3}^{\alpha}\rho_{3}^{\beta}+ e^{L_{88}t}\rho_{8}^{\alpha}\rho_{8}^{\beta}\qquad \qquad \qquad \qquad \qquad \qquad\qquad \qquad \qquad \qquad \qquad \qquad
 \end{eqnarray}}
Assuming $\Gamma_{ij}$ to be imaginary, as in the diagonal case, 
taking $\sin\left( \frac{|\Gamma_{ij}|t}{2}
\right)\approx 0$, and recalling (\ref{mixpar}), we observe that,
 with real values for the elements of the mixing\index{mixing} matrix U, 
one obtains the same form for the oscillation 
probability as in \cite{bm},
provided the choice (\ref{decohparam}) is made for the diagonal elements:
 \begin{eqnarray}\label{finaleqprob}
    \nonumber P_{\nu_{\alpha}\rightarrow \nu_{\beta}}(t)= \frac{1}{3}
    + \frac{1}{2}e^{\frac{(L_{11}+L_{22})}{2}}\left(\rho_{1}^{\alpha}
    \rho_{1}^{\beta}\right)\cos \left(\frac{|\Gamma_{12}|t}{2}\right)
    \\ \nonumber +  e^{\frac{(L_{44}+L_{55})}{2}}\left(\rho_{4}^{\alpha}
    \rho_{4}^{\beta}\right)\cos \left(\frac{|\Gamma_{13}|t}{2}\right)
    \\ \nonumber +  e^{\frac{(L_{66}+L_{77})}{2}}\left(\rho_{6}^{\alpha}
    \rho_{6}^{\beta}\right)\cos \left(\frac{|\Gamma_{23}|t}{2}\right)
    \\ e^{L_{33}t}\rho_{3}^{\alpha}\rho_{3}^{\beta}+ 
    e^{L_{88}t}\rho_{8}^{\alpha}\rho_{8}^{\beta}
 \end{eqnarray}
the difference being
that $\Gamma_{ij}$, as noted earlier (\ref{defgammas}),  
is of a slightly different form 
from the respective $\Omega_{ij}$ (\ref{omegas}), 
due to the presence of the off-diagonal
elements $L_{12} \ne 0$ {\it etc.}. 
Notice from (\ref{defgammas}) that there is a tendency of 
the off diagonal elements of the decoherence 
matrix to reduce the effects of the 
neutrino mass squared difference $\Delta_{ij}^2$
Thus, this sort of extension
beyond the diagonal form of the 
decoherence\index{decoherence} matrix (\ref{diagonal}) 
will affect the magnitude of the oscillation 
length\index{oscillation length}, as compared to the diagonal case.

It is straightforward to use such parametrizations to obtain
bounds on the extra decoherence parameters by comparison with data.
We stress again, 
that, due to CPT Violation\index{CPT Violation}, the above probabilities 
may differ between particles and antiparticles\index{antiparticle} 
sectors insofar as the order of magnitude of the corresponding 
decoherence parameters is concerned.
Moreover, in view of our comments above on the possible 
contributions of a decohering environment\index{environment} 
to the Hamiltonian
terms in (\ref{lindblad}), it is also of great theoretical 
and phenomenological interest to consider the case of modified
dispersion\index{dispersion relations} relations for neutrinos\index{neutrino}
simultaneously with the above-described decoherence 
effects, and compare with current experimental limits. 
Such modifications may indeed have a common origin with the decohering
effects, the interactions with the space time foam\index{foam}.
In view of the 
effects (\ref{defgammas}) on the oscillation length, 
analyses like the one in \cite{eichler}, bounding the coefficients 
of modified dispersion
relations by means of their effects on neutrino oscillations, need therefore 
to be 
rethought. 

\section{Conclusions} 

In these lectures I discussed various theoretical ideas and 
phenomenological tests of possible CPT Violation\index{CPT Violation}
induced by 
quantum gravity\index{quantum gravity}.
From this exposition it becomes 
clear, I hope, that CPT Violation may not be an academic issue, 
and indeed it may characterize a natural theory of quantum gravity. 

There are several probes of 
CPT Violation and there is no single figure of merit for it. 
Neutrinos seem to provide 
the most stringent constraints on CPT Violation through 
quantum decoherence to date, 
which in some cases are much 
stronger  than constraints from neutral meson experiments and factories.
In this sense neutrinos 
may provide a very useful guide
in our quest for a theory of Quantum Gravity. 

Neutrino oscillation experiments provide stringent bounds
on many quantum gravity models entailing Lorentz Invariance Violation.
There are also plenty of low energy nuclear and atomic\index{atomic} physics
experiments which yield stringent bounds in models with 
Lorentz (LV) and CPT Violation\index{CPT Violation} (notice that 
the frame dependence of LV effects
is crucial for such high sensitivities).
It is my firm opinion that neutrino factories, 
when built, will undoubtedly shed light on such important
and fundamental issues and provide definitive answers to many 
questions related to LV models of quantum space time.

However, 
as I repeatedly stressed during these lectures, 
Quantum Gravity may exhibit Lorentz Invariant\index{Lorentz Invariant Decoherence} 
(and hence frame independent) CPTV Decoherence.
 Theoretically, the presence of an environment may be consistent
with Lorentz Invariance.
This scenario 
is still compatible with all the existing $\nu$ data, 
including LSND\index{LSND Experiment} ``anomalous'' results, within three generation\index{generation} 
models,
and without the need for introducing matter-antimatter 
mass differences.
Of course the order of the decoherence\index{decoherence} 
parameters of such models is highly model dependent,
and, hence, at present it is the experiment that may guide 
the theory insofar
as properties and estimates of QG decoherence effects are concerned. 
It is interesting to remark 
that, in cases where quantum gravity induces neutrino oscillations
between flavours or violates lepton number, the sensitivity of experiments
looking for astrophysical neutrinos from extragalactic sources 
may exceed the order of $1/M_P^2$ in the respective figures of merit, 
and thus
is far more superior than the sensitivities of meson factories 
and nuclear and 
atomic\index{atomic} physics experiments as probes of quantum 
mechanics. 

However, as I remarked previously, 
the reader should be alert to the fact that there is no single figure of 
merit for CPT Violation\index{CPT Violation}; thus, as we have seen, 
there may be novel CPTV effects unrelated, in principle, 
to LV\index{Lorentz Violation} and locality\index{locality} 
violations, which are 
associated with modifications of 
EPR correlations. Such effects may be inapplicable to neutrinos, and thus 
testable only in meson\index{meson} 
factories\index{meson factories} or other situations involving entangled states, e.g. 
in quantum optics\index{quantum optics}.  

Clearly much more work, both theoretical and experimental, is needed 
before definite conclusions are reached
on this important research topic, called phenomenology and theory 
of CPT Violation. 
I personally believe that this issue lies at the heart of a 
complete and realistic theory of quantum gravity. For instance, 
CPT and its Violation is certainly an issue 
associated with DSR\index{Doubly-Special Relativity} 
theories, discussed in this
School, and non-commutative geometries\index{non commutative geometry},
which we did not discuss here, but which, 
as I mentioned in the beginning of the lectures,
is also a very active and rich field of research towards 
a theory of quantum gravity\index{quantum gravity}.

In this respect, I believe firmly that theoretical and 
phenomenological 
research on sensitive probes of CPT and quantum mechanics, such as 
photons from extraglactic sources, 
neutrinos and neutral mesons, could soon make important contributions to our 
fundamental quest for understanding the quantum structure of space time. 
Neutrino research 
certainly constitutes a very interesting  
and rapidly developing area of fundamental physics, 
which already provides fruitful collaboration between astrophysics
and particle physics, and which,
apart from the exciting results on 
non-zero neutrino masses which has yielded so far, 
may still hide even further
surprises waiting to be discovered in the near future.  
But other probes, such as photons and neutral mesons, 
may also prove invaluable in this respect, especially if QG effects
discriminate between particle species, a possibility, which as I 
mentioned in these lectures, may not be so unrealistic. 

Let me close, therefore, these lectures with the wish that 
by the year 2015, 
when the physics community will be summoned to celebrate the centennial 
from the development of General Relativity, the dynamical 
theory of curved space-time geometries,
we shall have obtained some concrete 
experimental indications on what is going on in
Physics near the Planck scale.
Let us sincerely hope that this exciting prospect 
will not remain only a wish for the years to come.

\section*{Acknowledgements}

It is a real pleasure to thank J. Kowalski-Glikman and 
G. Amelino-Camelia for 
the invitation to lecture in this School on 
{\it Theory and Phenomenology of Quantum Gravity}, in Poland, 
and for organizing this very successful and 
thought-stimulating meeting. 
I would also like to acknowledge informative discussions with 
G. Barenboim, J. Bernabeu, D. Binosi, G. Gounaris, 
C.N. Ktorides, J. Papavassiliou and A. Waldron-Lauda.
This work is partly supported by 
the European Union (contract HPRN-CT-2000-00152).

\printindex
\end{document}